# Explainable AI as a Double-Edged Sword in Dermatology: The Impact on Clinicians versus The Public

Xuhai Xu, PhD, Columbia University
Haoyu Hu, BS, Cornell University
Haoran Zhang, MS, MIT
Will Ke Wang, PhD, Columbia University
Reina Wang, BS, MIT
Luis R. Soenksen, PhD, MIT, Johns Hopkins University
Omar Badri, MD, Northeast Dermatology Associates
Sheharbano Jafry, BA, Stanford University
Elise Burger, MD, PhD, University of Utah
Lotanna Nwandu, MD, OSF Healthcare
Apoorva Mehta, BS, Columbia University Vagelos College of Physicians and Surgeons
Erik P. Duhaime, PhD, Centaur Labs
Asif Qasim, MD PhD, MedShr, King’s College Hospital, London
Hause Lin, PhD, MIT
Janis Pereira, BSc, MedShr
Jonathan Hershon, BA, Pathway
Paulius Mui, MD, X=Primary Care
Alejandro A. Gru, MD, Columbia University
Noémie Elhadad, PhD, Columbia University
Lena Mamykina, PhD, Columbia University
Matthew Groh, PhD, Northwestern
Philipp Tschandl, MD, PhD, Medical University of Vienna
Roxana Daneshjou, MD, PhD, Stanford University
Marzyeh Ghassemi, PhD, MIT

## Abstract

Artificial intelligence (AI) is increasingly permeating healthcare, from physician assistants to consumer applications. Since AI algorithm’s opacity challenges human interaction, explainable AI (XAI) addresses this by providing AI decision-making insight, but evidence suggests XAI can paradoxically induce over-reliance or bias. We present results from two large-scale experiments (623 lay people; 153 primary care physicians, PCPs) combining a fairness-based diagnosis AI model and different XAI explanations to examine how XAI assistance, particularly multimodal large language models (LLMs), influences diagnostic performance. AI assistance balanced across skin tones improved accuracy and reduced diagnostic disparities. However, LLM explanations yielded divergent effects: lay users showed higher automation bias – accuracy boosted when AI was correct, reduced when AI erred – while experienced PCPs remained resilient, benefiting irrespective of AI accuracy. Presenting AI suggestions first also led to worse outcomes when the AI was incorrect for both groups. These findings highlight XAI's varying impact based on expertise and timing, underscoring LLMs as a “double-edged sword” in medical AI and informing future human-AI collaborative system design.

## Introduction

Artificial intelligence (AI) has been intensively investigated as a tool to enhance clinical decision-making. Dermatology is one area with several developed and FDA-approved tools such as Nevisense[1] and DermaSensor[2], as skin conditions are primarily diagnosed through image assessment. In light of the national shortage of dermatologists[3,4], effective AI assistance could help to improve early detection rates and reduce unnecessary clinical visits. The development of AI-powered interfaces has also been proposed to assist the general public in making informed healthcare decisions (e.g., self-diagnosis of skin diseases with Google Lens[5])

Explainable AI (XAI) has been used in health generally to target AI usability and adoption[6–9]. In dermatology, physicians have clear features to look for, such as the ABCD’s (asymmetry, irregular borders, multiple colors, diameter greater than a pencil eraser) of melanoma. Similarly, in AI, gradient-weighted class activation mapping (GradCAM)[10] and content-based image retrieval (CBIR)[11,12] have been used to highlight relevant image regions and retrieve similar cases, respectively. A recent survey found that GradCAM and CBIR are the top two most commonly used XAI techniques in dermatology[13]. In addition, with the recent surge of generative AI[14], multimodal large language models (multimodal LLMs) operating over textual and visual modalities[15–18] have also been used to analyze dermatological images and explain AI decisions[19–21].

Unfortunately, prior work has shown that XAI can increase subject over-reliance on AI in decision-making[22–24], and human-AI collaborative medical decisions do not always surpass those made by humans or AI alone[25,26]. To date, there is no consensus about whether medical experience improves human-AI collaborative diagnosis. While some research indicates that medical knowledge is needed for better AI resilience and clinical decision making[27], there is also research showing it makes minimal difference[12], and in dermatology, AI may even mislead humans and lead to worse outcomes in diagnosis[28–30]. As dermatological AI tools expand, it is crucial to understand how different XAI methods, especially with the vast spread of LLMs, affect skin disease diagnostic accuracy across both the general public and medical experts[31–35].

In this paper, we designed two large-scale experiments to systematically investigate how different XAI methods and human-AI decision paradigms impact diagnostic performance across expertise levels. We chose clinical-image-based skin condition diagnosis as a plausible real-world scenario (i.e., ecological validity) given the influx of patient-facing and physician-facing diagnostic models in this space[30,36]. We investigated the overall effectiveness of XAI assistance in improving dermatological diagnostic accuracy[37,38], reducing the disparities across skin tones[39], and influencing accuracy-confidence calibration (i.e., accuracy and confidence are consistent). We compared both correct and incorrect multimodal LLM-based explanations to traditional XAI approaches in both scenarios. We also examined how individual differences in AI deference (i.e., the propensity to follow AI regardless of accuracy, sometimes referred to as AI susceptibility[40]) impact diagnostic performance. Finally, we explored the impact of the human-AI decision paradigm (i.e., the order of human or AI making decisions) on diagnostic outcomes[41,42], providing insights into optimal implementation strategies for clinical settings.

We explore these questions in two populations. First in the general public (N=623) with a binary classification task to distinguish melanoma and nevus, and second with medical providers (N=153 primary care physicians [PCPs]) on a more challenging open-ended differential diagnosis of skin diseases. Our findings have important implications for designing and deploying image-based medical AI systems for skin diseases. Our work comprehensively studies the collaboration between explainable AI and both the general public and medical experts. By elucidating how different traditional (GradCAM, CBIR) and advanced XAI methods (multimodal LLM), levels of medical expertise, and human-AI decision paradigms influence outcomes, we make unique contributions to the understanding of human-AI collaborative decision making and the development of more effective and appropriate human-AI collaborative medical systems.

# Results

## Study Design

### Study Populations

We designed two complementary large-scale digital studies to evaluate human-AI collaborative diagnostic performance (Fig. 1a). Both studies were based on clinical images and designed to resemble real-world practices: a regular lay person may take a photo of their skin and resort to search engines or AI tools[43,44], and an expert often needs to make a differential diagnosis based on a clinical image (e.g., patient communication through electronic health record messaging systems). To ensure data quality, we provided comprehensive tutorial materials for each participant group.

Study 1 engaged 623 subjects (314 females, 305 males, 4 non-binary or others, aged 33±9) from the general public without medical backgrounds, in which participants reviewed clinical images of skin and performed a binary classification task to distinguish melanoma vs. nevus (Fig. 1b). The general public assessed 12 images that were randomly sampled from a pool of 82 images. These images were evenly distributed between light skin (Fitzpatrick 1-4) and dark skin (Fitzpatrick 5-6) patients[45] and also evenly distributed between nevus and melanoma. To ensure consistent assessment of human-AI collaboration while maintaining ecological validity, we preserved AI model outputs while strategically sampling images to maintain a consistent accuracy of 83.3% (always 10 correct and 2 incorrect predictions per participant), which is close to the model's actual performance (see AI model details below). Each participant went through 12 clinical images and made two rounds of decisions per image (with/without AI).

Study 2 involved 153 PCPs (66 females, 84 males, 3 non-binary or others, aged 34±10, years of practice experience range from 1 to 25 years, average experience 6 years). They performed a more complex open-ended differential diagnosis with free-text entry. Participants were asked to enter their top three diagnoses (Fig. 1c). In Study 2, we focused on four main conditions previously identified as having potential diagnostic disparities across skin tones (better performance on patients with light skin than those with dark skin)[30,46]: atopic dermatitis, pityriasis rosea, Lyme disease, and cutaneous T-cell lymphoma (CTCL). To simulate real-life applications, these four conditions were mixed with the other 30 common skin conditions. Text entry was assisted by auto-completion based on string matching from a comprehensive list of 445 skin diseases. Each participant evaluated 12 cases, including 2 images for each of the four main conditions with correct AI predictions, 2 images sampled from the four main conditions with incorrect AI predictions, and 2 images sampled from other conditions with a mix of AI correctness (expected 1.5 correct and 0.5 incorrect), resulting in expected AI predictions to be 9.5 correct and 2.5 incorrect (i.e., expected accuracy of 79.2%). Similar to Study 1, skin tone was balanced in each condition.

We note that Study 1 and 2 are not directly comparable as they emphasize different tasks appropriate for groups with different medical expertise. In order to measure the impact of medical training experience within the same task, we additionally recruited 320 medical students (171 females, 141 males, 8 non-binary or others, aged 28±9) with less expertise in dermatology to compare to the cohort reported in Study 2 (see EFig. 4 & 5).

We further employed validated questionnaires to collect additional data on human-AI collaboration experience[47], XAI quality[48], and cognitive traits of critical thinking[49] and open-mindness[50]. More details can be found in the experiment design sections in Methods. Overall, we collected 14,952 diagnostic decisions from the general public (Study 1), 3,672 diagnoses from PCPs, and 7,680 diagnoses from medical students (Study 2).

**Human-AI Decision Paradigm**

For both studies, we adopted a between-subject factorial design with 4 x 2 conditions: including four XAI methods (**Basic** outcomes with model prediction and confidence, **GradCAM**: highlighting which areas contribute to the model prediction, **CBIR**: presenting similar images with the same label as the prediction, and **multimodal LLM**: semantic explanations of the reason for the model prediction, Fig. 1d-g), For the rest of the paper, we omit “multimodal” and use **LLM** for simplicity, as our studies mainly focus on the language capability of the multimodal LLM, but do note we used a vision-language model GPT-4V, see Methods for more details), as well as two human-AI decision paradigms (**Human-First,** where users make a decision first before reviewing AI suggestions, and **AI-First**, where users review both images and AI suggestions before making the final decision). Each participant was randomly assigned to one condition and went through 12 clinical images. Since the Human-First condition naturally required participants to make two rounds of decisions (the first round without AI and the second round with AI), to control the effect of cognitive effort, participants also made two decision rounds in AI-First: after making the first round of decision with AI, they were asked to review and reconsider their final decisions in the second round when AI results were hidden. Besides decisions, participants also rated their confidence in their decisions at each round.

## Model Training

We trained a deep-learning model ViT-B/32[51] for the binary classification task in Study 1. We applied the training strategy of conditional domain adversarial neural network (CDANN)[52] to achieve the fairness constraint and ensure performance balance between skin tones (weighted AUROC of 0.930, accuracy $\Delta$=2.1% between skin tones). In Study 2, we trained a DenseNet-121[53] with an AUROC of 0.772, accuracy $\Delta$=5.6% between skin tones. Post-hoc explanations (GradCAM, CBIR, and LLM) were generated using model outputs. See the Methods section for more details.

## Evaluation Metrics

Responses for Study 1 were evaluated using standard binary outcome measures. For the free-text task in Study 2, two authors with medical expertise manually evaluated all text entered in the “Diagnosis” section to determine the top-1 (based on the first entry) and top-3 accuracy (based on all three entries) compared to the ground truth.

All AI-generated explanations for GradCAM, CBIR, and LLM were evaluated by three dermatologists with at least 2 years of experience and independently scored for correctness (i.e., how accurate the explanation is based on the skin image) and informativeness (i.e., how much this explanation can support diagnosis) on a 5-point Likert scale, achieving fair to moderate interrater reliability (Fleiss' $\kappa$=0.42 and 0.37). We calculated the average and divided explanation quality into high (informativeness score $\geq 3$ and correctness score $\geq 3$) and low (informativeness score $< 3$ or correctness score $< 3$, see examples of high and low-quality explanations in SFig. 7-12).

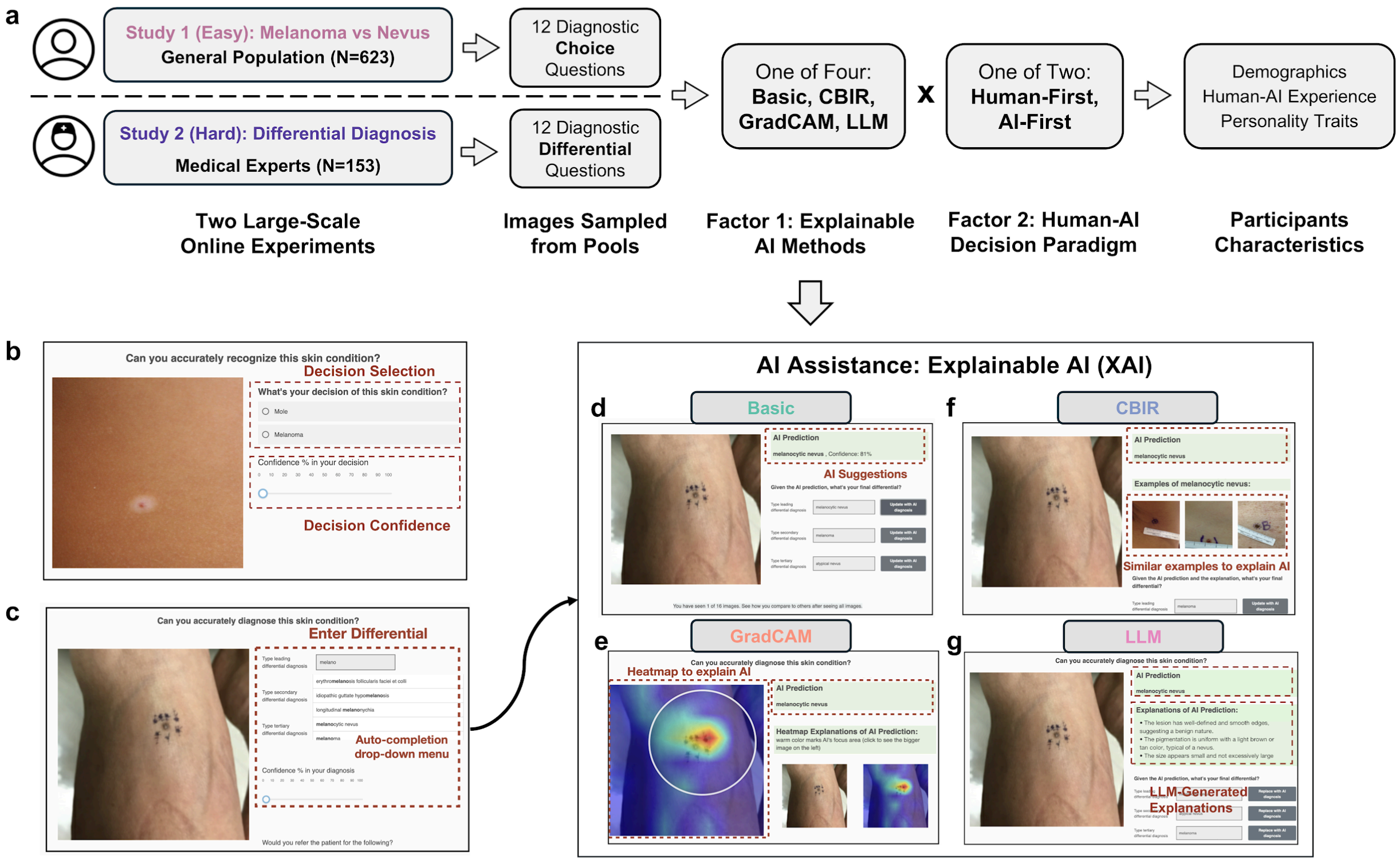

**Fig. 1 | Study Design and Interface Components of Human-AI Collaborative Diagnostic System. a**, Overview of the experimental workflow comprising two studies: Study 1 recruited general public participants for melanoma versus nevus classification tasks, while Study 2 engaged medical experts for open-set differential diagnosis. 12 diagnostic images were randomly drawn from a high-quality image database with an equal skin tone split. Participants completed either human-first or AI-first diagnostic sequences followed by demographics, human-AI collaboration experience, and personality assessment. **b**, Interface for the binary classification task showing a clinical image for melanoma vs. nevus discrimination. **c**, Interface for the differential diagnosis task with top-3 free-text entry. To assist user input, the text box has a string-matching-based auto-completion function that covers a comprehensive set of 445 skin diseases. **d-g**, Four XAI assistance examples with (**d**) Basic AI providing straightforward predictions with model confidence, (**e**) GradCAM highlighting relevant image regions, (**f**) CBIR showing similar reference cases, and (**g**) LLM providing natural language explanations. **Abbreviations**: AI, artificial intelligence; XAI, explainable AI; CBIR, content-based image retrieval; GradCAM, gradient-weighted class activation mapping; LLM, large language model.

## SOTA AI Improves General Public's Performance Due to AI Deference, LLM Explanations Amplify Such Deference.

***Advanced AI improves the general public's performance.*** We first measured the general public's performance (Study 1) without and with AI assistance. We found that AI improved average accuracy of the nevus vs. melanoma detection task from 69.7±0.8% to 75.8%±0.7% (effect of AI assistance: β=0.061, 95%CI=[0.049, 0.074], p<0.001, linear mixed model on accuracy, with AI assistance, XAI methods, and their interaction as the main factor, controlling gender, age, race, skin disease experience, and the covariate of self-reported human-AI collaboration experience. See STab 7 for details. Other Study 1 statistical models below control the same set of confounders unless noted differently), and the majority of the improvement came from nevus classification (β=0.111, 95%CI=[0.091, 0.132], p<0.001, STab 8, Fig. 2a) and the largest improvement in explainable AI comes from multimodal LLM (see next section). Participants also had a modest increase in diagnosis confidence by 1.5% with AI assistance (β=0.018, 95%CI=[0.010, 0.019], p<0.001, EFig. 1a, STab 10). With the help of AI model humans achieved a more balanced diagnosis performance across patient skin tones (Round 1: β=0.033, 95%CI=[0.009, 0.057], p=0.007; Round 2: β=0.017, 95%CI=[-0.007, 0.041], p=0.166, $\Delta_{rel}$=46.9%, Fig. 2c, STab 11). These findings demonstrate that collaboration with well-trained AI has the potential to mitigate diagnostic biases while improving overall accuracy.

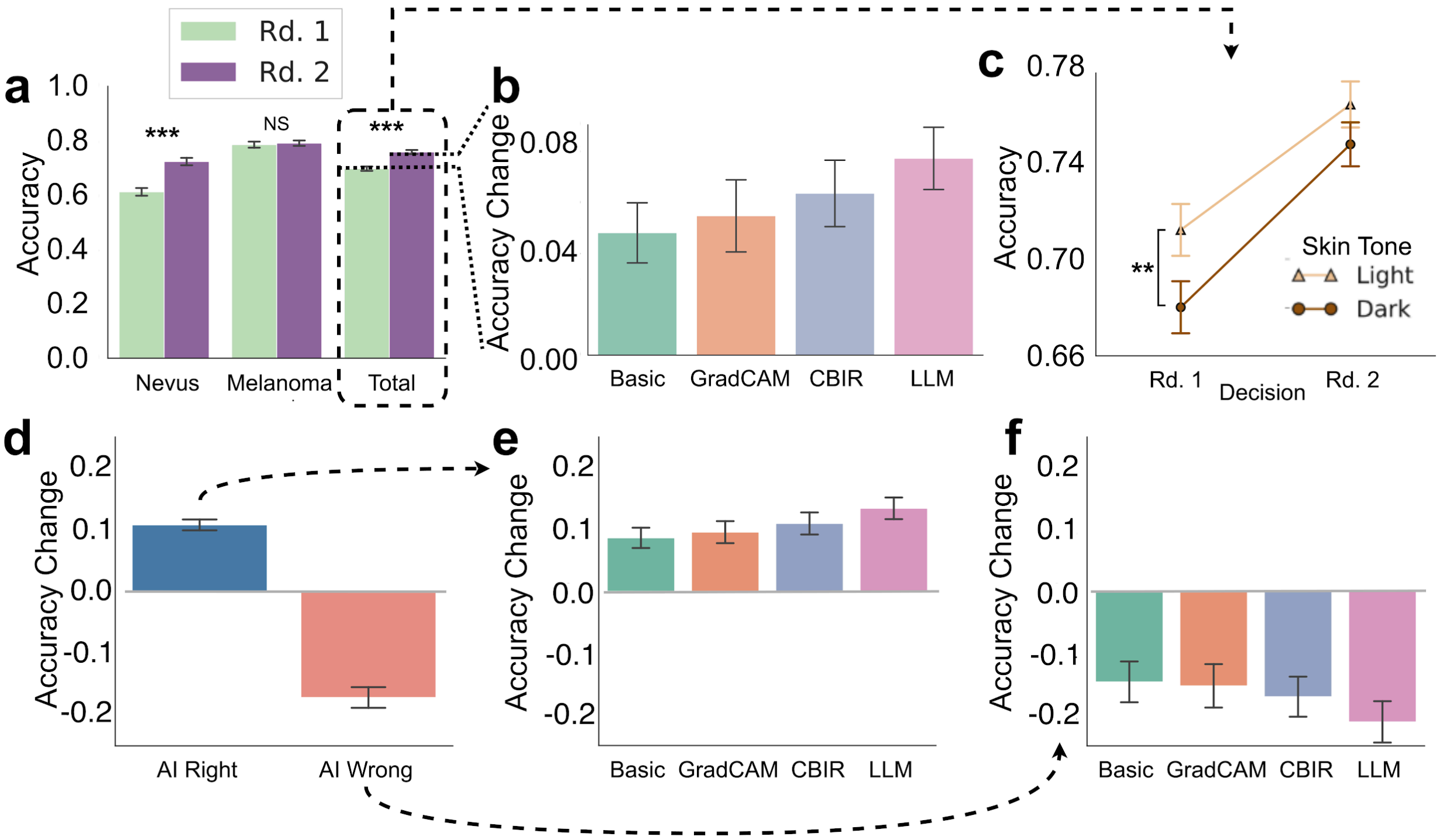

**Fig. 2 | Overall Diagnostic Performance with AI Assistance for the General Public**. **a**, Accuracy of the general public in detecting nevus, melanoma, and the overall diagnosis in two rounds: Rd. 1 (initial, human accuracy without AI assistance) and Rd. 2 (AI-assisted accuracy). **b**, Diagnostic accuracy changes across different AI explanations (Basic, GradCAM, CBIR, and LLM). **c**, Diagnostic accuracy when patients are light-skinned or dark-skinned across two rounds. **d**, Overall accuracy changes under AI-right and AI-wrong conditions. **e**, **f**, Diagnostic accuracy changes of different AI explanations under AI-right (**e**) and AI-wrong (**f**) predictions. All error bars in this figure (and the rest of the figures) indicate the standard error of the mean. The same metrics were applied in Fig. 3.

***Performance improvement stems from AI deference & LLM-based explanations amplify such deference.*** We investigated the impact of the four XAI methods on diagnostic accuracy. The LLM explanations provided an improvement of +7.7% (β=0.077, 95%CI=[0.053, 0.101], p<0.001), followed by CBIR (+6.3%, β=0.063, 95%CI=[0.039, 0.087], p<0.001), GradCAM (+5.5%, β=0.054, 95%CI=[0.028, 0.081], p<0.001), and the basic method (+4.8%, β=0.048, 95%CI=[0.023, 0.073], p<0.001), as shown in Fig. 2b and STab 7. However, the general public's diagnostic accuracy improved when AI provided correct predictions, while incorrect predictions reduced the performance significantly (β=-0.233, 95%CI=[-0.307, -0.158], p<0.001, Fig. 2d, STab 12). Correct LLM advice (+13.4%) enhanced performance more than other AI explanations (basic method +8.6%, GradCAM +9.5%, CBIR +10.9%, Fig. 2e), and incorrect LLM advice decreased performance most  (-21.1%, basic -14.6%, GradCAM -15.3%, and CBIR -17.0%, Fig. 2f). These findings indicate that LLM explanations amplify the general public's tendency to follow AI guidance, regardless of AI accuracy. Moreover, compared to basic explanation, LLM leds to significantly more reduction of performance when AI becomes inaccurate (i.e., delta of detla, β=-0.048, 95%CI=[-0.093, -0.003], p=0.035, STab 13).

***Misplaced trust in LLM explanations for the general public.*** When AI predictions were correct, participants trusted LLM explanations more than other methods (Fig. 2e), and this was more noticeable when explanations were of low quality (post-hoc pairwise estimated marginal means (EMMs) comparisons, LLM over GradCAM: β=0.117, 95%CI=[0.041, 0.192], p=0.002; LLM over CBIR: β=0.092, 95%CI=[0.021, 0.163], p=0.011, EFig. 2c, STab 15). When AI predictions were incorrect, LLM explanations negatively impacted the alignment between participants' confidence and accuracy (z=-3.788, p<0.001, two-side Fisher r-to-z test to compare correlation difference, EFig. 3c). These results further suggest people struggle to assess LLM explanation reliability and can be easily misled by LLM.

To summarize, our results in Study 1 indicate that **LLMs are a "double-edged sword" in skin disease diagnosis for the general public with amplified AI deference**: When AI was correct, LLM explanations boosted diagnosis performance, even when the quality of the explanation was low. However, when AI was incorrect, the general public was misled by seemingly plausible reasons generated by LLM. Although our LLM explanations rarely contained hallucinated or wrong clinical signs (i.e., the multimodal LLM is capable of visual analysis), they frequently referenced ambiguous dermatologic criteria even when these features were only partially present or visually unclear. EFig. 10 presents several specific cases when AI was incorrect and misled humans to the wrong answer. For instance, in the first case in EFig. 10a, the AI model provided a wrong prediction as melanoma while the ground truth is nevus. In this case (dysplastic nevus), the LLM-generated explanation (based on an incorrect AI prediction) was biased towards melanoma and mistakenly align abnormal nevus visual features with melanoma criteria on symmetry, color, and border (e.g., "The shape of one half does not match the other half"). To a general public without medical background, such wrong explanations appear to be plausible but are misleading. This is supported by prior work on the automation bias and algorithmic over-reliance when humans are working with AI. Our work further indicates that LLM can amplify such risks for the general public. We discuss the implications of these findings in the discussion section.

## PCPs Reliably Leverage Accurate AI Guidance, While Resisting Errors with LLM-based AI Explanation

***Basic AI improves PCP's performance.*** We conducted a similar analysis of the PCP participants in Study 2. This task was more challenging and required detailed dermatological knowledge. We found 11.5%±1.4% top-1 accuracy and 16.1%±1.8% top-3 accuracy in PCP's differential diagnoses without AI, which is aligned with the prior work[30]. The performance was significantly improved with AI suggestions, with a +21.5% in top-1 accuracy (β=0.258, 95%CI=[0.170, 0.260], p<0.001, Fig. 3a, linear mixed model on accuracy with AI assistance, XAI methods, and their interaction as the main factor, controlling gender, age, race, and medical expertise in skin, year of experience, and personality traits. See STab 16 for details. Other Study 2 statistical models below control the same set of confounders unless noted differently), and a +43.5% in top-3 accuracy (β=0.450, 95%CI=[0.352, 0.548], p<0.001, EFig. 6a, STab 16). In contrast to the general public in Study 1, for PCPs,basic AI assistance helped the most(top-1 accuracy β=0.258, 95%CI=[0.168, 0.349], p<0.001, Fig. 3b,STab 16). Interestingly, we only observed significant confidence increase when PCPs were assisted by GradCAM (β=0.028, 95%CI=[0.005, 0.051], p=0.019) and LLM explanations (β=0.035, 95%CI=[0.012,0.058], p=0.003, EFig. 1b, STab 21). Improvements were significant in all four major diseases (+19.9-25.0%, ps<0.001, Fig. 3a, STab 17-20). PCPs had disparate performance across skin tones as 4.6% (β=0.046, 95%CI=[0.013, 0.078], p=0.069, Fig. 3c), which was reduced to 2.9% (β=0.029, 95%CI=[-0.018, 0.076], p=0.248, STab 22) after AI assistance.

***PCPs are resilient to AI deference when AI is wrong.*** In contrast with the general public, incorrect AI predictions had minimal impact on PCPs' final decisions across all XAI methods (Fig. 3f, EFig. 2d-2f, βs=0-0.021, ps = 0.328-1.000). This suggests PCPs relied on their own expertise and training rather than erroneous AI guidance. To control the effect of task difficulty between Study 1 and 2, we further compared PCPs' results against the medical students' results on the same task. EFig. 4 showed that medical students had more reliance on AI, regardless of AI correctness compared with results from PCPs. We term participants who only answered correctly when AI was correct, and therefore answered incorrectly when AI was wrong, as "deferential participants". We observe that the proportion of deferential participants are higher among medical students than PCPs (linear mixed model on the proportion of deferential participants, with medical role and medical expertise in skin as the main factor, controlling other confounders, main effect of medical role: β=0.067, 95%CI=[0.004, 0.130], p = 0.037; main effect of skin expertise: β=0.073, 95%CI=[0.022, 0.124], p = 0.005, STab 25). These results indicate that higher expertise levels are associated with more careful AI adoption, which is supported by prior work[12].

***LLM explanations do not aid in accuracy, but in confidence calibration.*** Interestingly, for PCPs, LLM explanations were the least helpful method (+17.7%) and were 8.1% lower than the best improvement from the Basic explanations (β=-0.081, 95%CI=[-0.207, 0.044], p = 0.268, Fig. 3b, STab 16). This finding was consistent across explanation quality (EFig. 2e, 2g, STab 27) and top-3 accuracy (EFig. 6c-6f). These are opposite to the results of LLM's best improvement for the general public in Study 1. Medical student data confirmed the task was not biased towards certain explanations (EFig. 4c), implying expertise drove interactions: the general public over-relies on LLMs, while PCPs are more resilient to incorrect LLM suggestions.

However, LLM explanations did help improve the alignment between PCP participants' confidence and accuracy (correlation r=0.494, p=0.010, EFig. 3d, similar findings in top-3 performance, EFig. 5d) over No AI (correlation r=0.084, p=0.415, two-sided Fisher r-to-z test comparing the two correlations: z=2.368, p=0.018, EFig. 3d). This calibration benefit held even with incorrect AI predictions, where non-LLM explanations impaired alignment (Fisher r-to-z test z=2.572, p=0.010, EFig. 3f, similar in top-3 performance, p=0.059, EFig. 5f). PCP caution towards LLMs likely enforces cognitive engagement, thus enhancing diagnostic accuracy-confidence calibration regardless of AI correctness[54,55].

Our findings identify the difference between the general public and PCPs: while the general public showed vulnerability toward the influence of LLM-based explanations and exhibited higher AI deference, **PCPs improved performance with demonstrated resilience against incorrect predictions**.

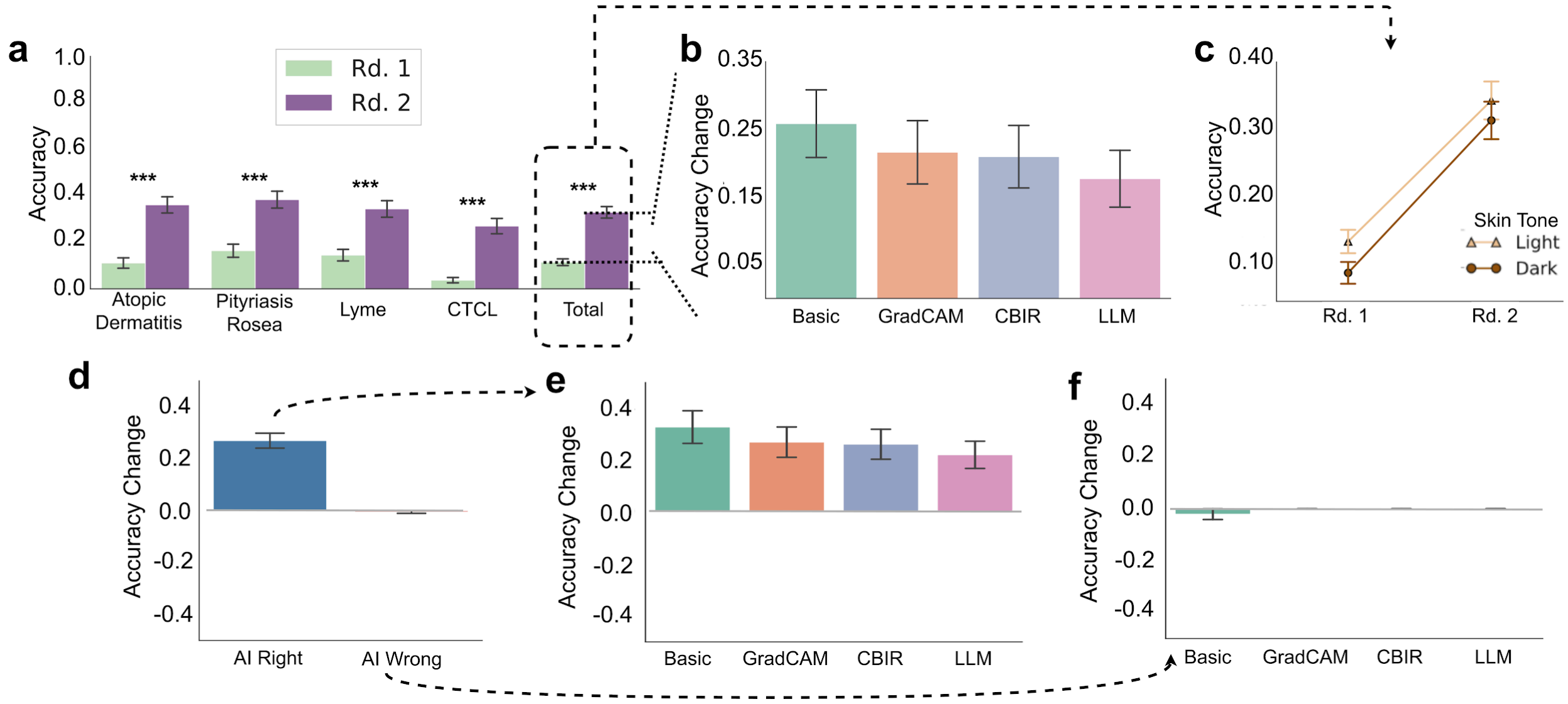

**Fig. 3 | Overall Diagnostic Performance with AI Assistance for PCPs**. This figure shows similar analysis results as Fig. 2. **a**, Accuracy of PCPs in identifying atopic dermatitis, pityriasis rosea, Lyme disease, CTCL, and the overall performance on four main diseases, with results presented for two rounds. **b**, Diagnostic accuracy changes across different AI explanations. **c**, Diagnostic accuracy when patients are light-skinned or dark-skinned across two rounds. **d**, Overall accuracy changes under AI-right and AI-wrong conditions. **e**, **f**, Diagnostic accuracy changes of different AI explanations under AI-right (**e**) and AI-wrong (**f**) predictions.

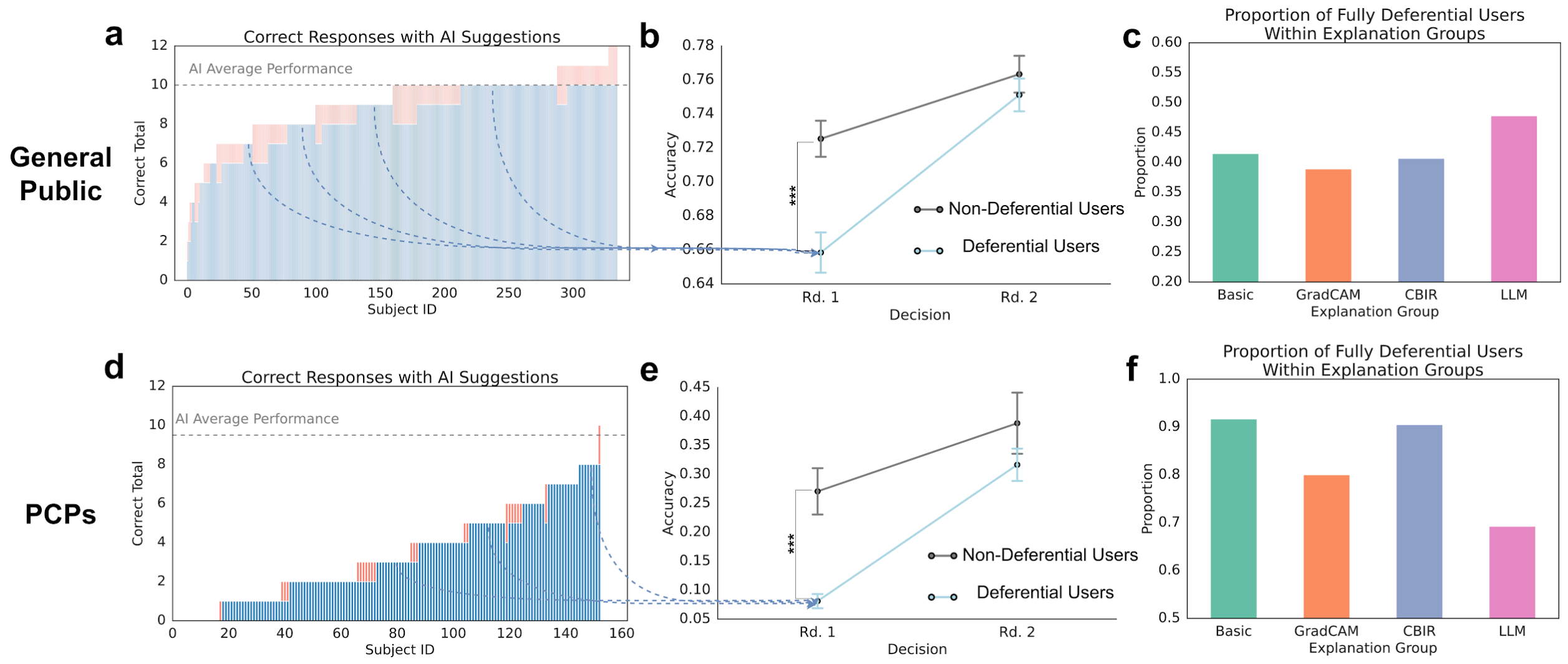

**Fig. 4 | Deference Patterns towards AI in HAI Collaboration. a-b**, Performance comparison between deferential (fully compliant with AI suggestions) and non-deferential general populations. (**a**) Distribution of correct responses with AI assistance. The blue bars represent instances where both AI predictions and participants were correct, and the red overlays indicate correct responses by participants while AI predictions were incorrect, (**b**) accuracy trajectories across two decision rounds. **c**, Proportion of participants that defer for each XAI output. **d-f**, Equivalent results for PCP participants as (**a**)-(**c**).

## Higher AI Deference Correlates with Lower Initial Performance

We inspected the relationship between participants' deference tendency toward AI suggestion deference and its relationship with initial performance[40]. Among participants' final decisions that are correct (ranging from 0 to 12, 12 images total), we visualize the number of images with AI suggestions that are correct (up to 10, shown in blue) and incorrect (up to 2, shown in red) for each participant (Fig. 4a). As mentioned above, "deferential participants" are those who only got correct results when AI was correct and always got incorrect results when AI was wrong (i.e., a blue bar in Fig. 4a & 4d). Participants who got at least 1 correct outcome even when AI is incorrect are considered as "non-deferential participants" (i.e., a red bar on top of the blue bar in Fig. 4a & 4d).

While the general public had similar performance after AI assistance, deferential participants had significantly lower initial diagnostic accuracy (65.8%) compared to non-deferential participants (72.5%) before receiving AI suggestions ($\beta=0.082$, 95%CI=[0.074, 0.111], $p<0.001$, Cohen's $d=0.462$ Fig. 4b, linear mixed model with deferential group as the main factor, controlling the same confounders as Study 1 analysis, see STab 28). Deferential PCPs also had a significantly lower performance in the initial round ($\beta=0.194$, 95%CI=[0.079, 0.310], $p<0.001$, Cohen's $d=1.578$, Fig. 4e, STab 29), which could result from a lower level of critical thinking ($p=0.028$, measured by critical thinking questionnaire[49], details see Method).

LLM explanations led to relatively the largest proportion of fully deferential participants in the general public (Fig. 4c), although no significance between LLM and others was observed (ps=0.169-0.433, STab 30). In contrast, LLM resulted in relatively the lowest proportion of deferential PCPs (Fig. 4f, ps=0.268-0.587, STab 31). This is also aligned with our findings of experts' capability in maintaining resilience against the misdirection of wrong semantic AI explanations[12].

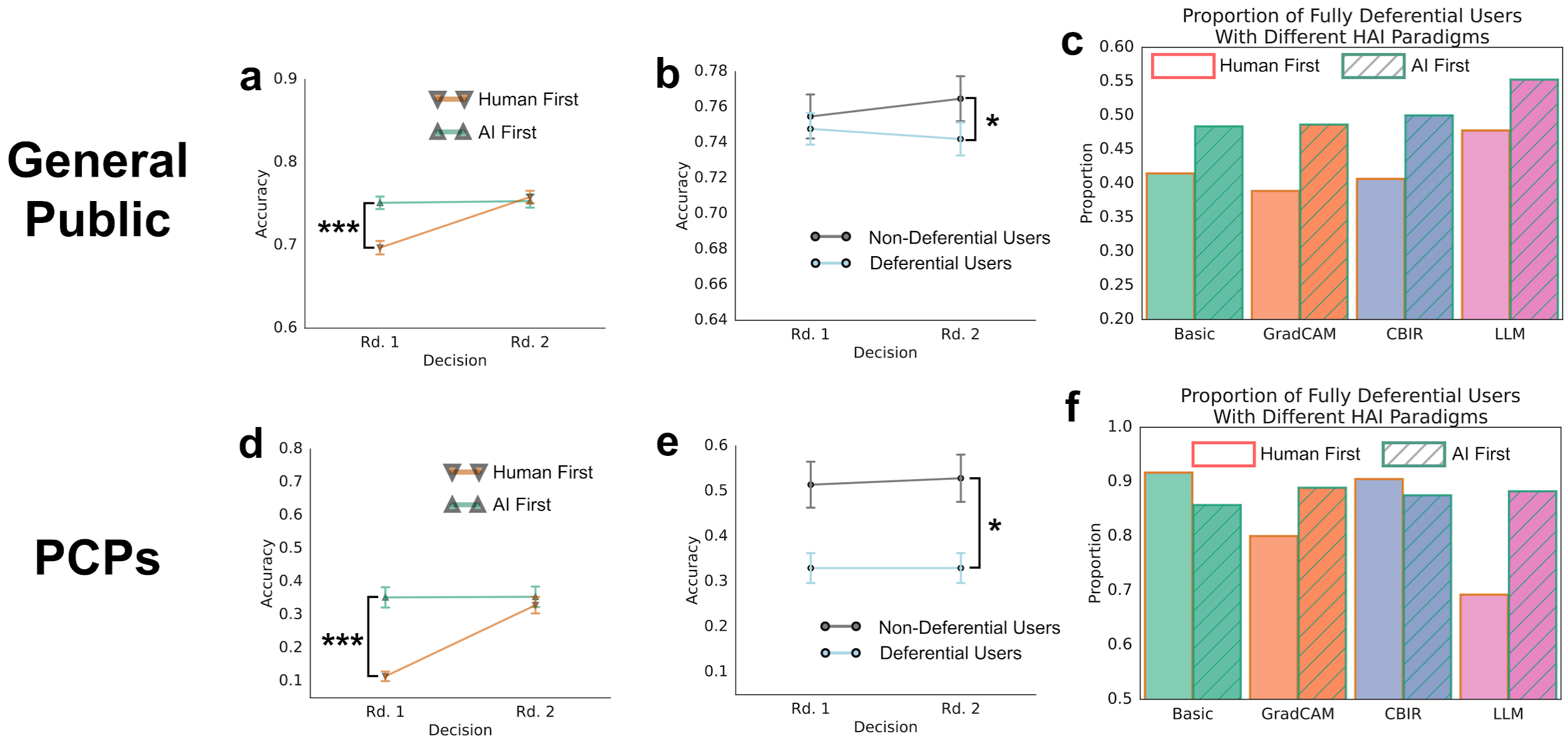

**Fig. 5 | Influence of HAI Paradigm on Diagnostic Performance**. **a-d**, Analysis for the general public. **a**, Comparison between "AI-First" and "Human-First" HAI systems on diagnostic accuracy across two rounds. **b**, Accuracy trajectories across two decision rounds for deferential and non-deferential people in the "AI-First" paradigm. **c**, Comparison of full-deference proportion between two HAI paradigms. **d-f**, Equivalent results for PCP participants as (**a**)-(**c**).

### Putting AI Before Human Decisions Amplifies Deference Across Expertise Levels

In addition to XAI methods, a practical design factor for human-AI collaboration systems is the decision-making order between humans and AI, specifically either Human-First or AI-First paradigms. We found that both the general public and PCPs had significantly better performance in the first round with AI-First (ps<0.001, Fig. 5a & d), which is not surprising due to superior AI performance. In the second round, after humans received the same amount of information, there was no difference between Human-First and AI-First in either study (round 2, the general public: β=0.005, 95%CI=[-0.017, 0.026], p=0.650; PCPs: β=-0.020, 95%CI=[-0.091, 0.050], p=0.572. Linear mixed models on accuracy with human-AI collaboration paradigm, decision round, and their interaction as the main factors, controlling decision making time and other confounders. See STab 32 and STab 34 for details). This indicates that the decision order may not influence the final performance.

With the AI-First paradigm, non-deferential participants still had better performance, especially at round 2 after reviewing the examples again without AI (the general public: β=0.031, 95%CI=[0.000, 0.062], p=0.049, STab 33; PCPs: β=0.218, 95%CI=[0.009, 0.426], p=0.041, STab 35). This is similar to the results in the Human-First paradigm in Fig. 4. In contrast, putting AI suggestions ahead increased the proportion of deferential participants in most cases (Fig. 5c, f). For the general public, the proportion was increased across all XAI methods (average Δ =+8.4%, although no significance observed after controlling all confounders, ps=0.067-0.170, Fig. 5c, STab 36). For PCPs, the largest deference increase was observed from LLM explanations (Δ=+19.0%, Fig. 5f, EFig. 5m, STab 37). This indicates that putting AI ahead may lead to stronger anchoring bias. It also suggests that although PCPs showed resistance against AI's mislead in the Human-First paradigm, providing LLM-based explanations ahead of human choices can still cause more bias than other XAI methods and introduce risks of overreliance, even for PCPs.

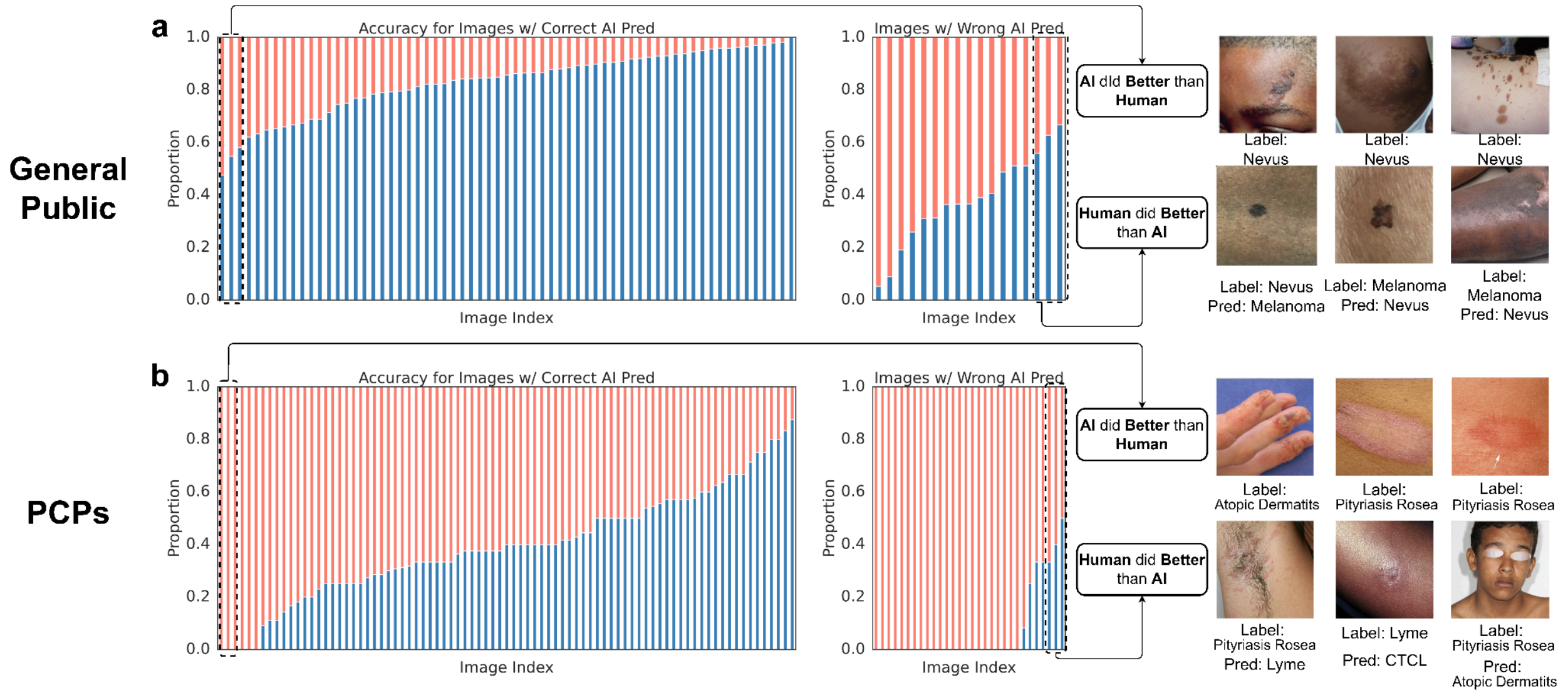

**Fig. 6 | Case Study Characterizing When AI or Humans Did Better. a**, Accuracy proportion of images in Study 1 for the general public, categorized by whether the AI prediction was correct or incorrect. The blue bars represent the proportion of correct human participants' decisions, while the red overlays indicate their incorrect decisions. Representative images labeled "AI did Better than Human" (AI-right images with the highest human error rate) and "Human did Better than AI" (AI-wrong images with the highest human accuracy) are shown to the right. **b**, Same as (**a**) for Study 2 for PCPs. More examples of each category can be found in EFig. 8 - 10.

## Human-AI Collaboration to Combine Each Side's Strength

Although AI deference risks misleading participants when AI makes mistakes, deference may lead some humans to improved performance. Fig. 6 visualizes cases where either humans or AI routinely outperform each other. We found that AI tends to outperform humans in cases where the presentation of the disease is subtle, but struggles with atypical symptoms or unexpected features in the image (additional examples in EFig. 8, EFig. 9 and EFig. 10). These qualitative examples provide some initial directions for future work in understanding the complementary strengths of humans and AI in dermatological diagnosis.

## Discussion

Our work contributes to the knowledge gap in understanding how various XAI methods and human-AI decision paradigms may impact human-AI collaborative diagnosis outcomes across populations with different expertise levels (the general public vs PCPs). Through two large-scale experiments, we provide comprehensive results on how humans collaborate with explainable AI assistance in skin disease diagnosis. Our studies reveal both the potential and challenges of human-AI collaboration in dermatology.

Our study demonstrated that state-of-the-art AI suggestions can create significant diagnostic improvements regardless of participants' expertise levels and task difficulty. Congruent with previous research[30], PCPs' diagnostic top-1 and top-3 accuracy are increased by 21.4% and 43.5% from a baseline of 11.5% and 16.1%. The general public's performance also increased by 6.1% from 69.7% to 75.8%. Our fairness-constrained AI model helped humans reduce diagnostic disparities on skin tones, with a relative reduction of 46.9% (accuracy disparity from 3.2% to 1.7%) among the general public and 35.6% (from 4.5% to 2.7%) among PCPs. In line with our results, Groh et al[56] find similar diagnostic disparities across skin tones in PCPs without AI assistance and similar reductions in these disparities when PCPs had access to a similarly high accurate AI assistance tool (79.2% accuracy in this paper and 84.0% accuracy in the treatment DLS in Groh et al[56]). However, when the AI assistance tool exhibited moderate performance (the 47.0% accuracy in the control DLS in Groh et al[56]), PCP diagnostic disparities across skin tones increased. Together, these two studies reveal that the dose-response relationship between AI accuracy and its impact on diagnostic accuracy and equity deeply matters. As AI assistance accuracy decreases, physicians may be prone to incorporating incorrect AI predictions into their diagnoses especially for more challenging cases such as images of dark skin.

We found interesting nuances of different XAI methods on different populations, especially LLMs. Although as shown in Fig. 2b and Fig. 3b, a direct comparison on the final diagnosis performance across XAI methods does not reveal significance (aligned with prior findings in Chanda et al.[57]), our breakdown analysis across AI correctness and XAI quality setups provides more nuanced insights. LLM acted as a 'double-edged sword' for the general public. While it provided the highest accuracy improvements (+13.4%) when its suggestions were right, it also led to the largest performance decrease (-21.1%) when wrong. The general public also tended to trust LLM explanations regardless of their quality, suggesting that readable, semantic explanations are more convincing for those without medical backgrounds. This heightened deference to LLMs likely stems from the cognitive ease of narrative persuasion[58,59]. Unlike GradCAM, which requires a user to interpret an abstract heatmap, or CBIR, which demands a high-effort visual comparison, an LLM provides a cohesive and seemingly complete diagnostic story. It connects visual features to a conclusion with authoritative, human-like prose (e.g., "The lesion is suspicious *because* of its irregular border..."). This fluent narrative can create an illusion of understanding, making the reasoning feel more intuitive and plausible to a non-expert. Moreover, when the LLM highlights specific features like “irregular border,” it can anchor the user’s perception, causing them to reinterpret the ambiguous visual data in a way that confirms the AI’s plausible-sounding rationale. Our work extends prior research on non-expert overreliance and automation bias on AI[60–62] and finds greater risk of LLM-based explanations, which has been supported in recent human-LLM trust work outside healthcare[63]. With numerous AI-assisted medical decision support tools available to the general public[64,65], and the advancement of LLM in health AI research[66–68], our findings highlight the importance of strong scrutiny and careful design for AI safety.

In contrast, PCPs applied LLM assistance more carefully and maintained their diagnostic accuracy even when AI suggestions were wrong. This was true regardless of explanation quality. To isolate the impact of task difficulty levels, we compared medical students to PCPs and found that students are more easily misled by LLM than PCPs. This suggests that PCPs expertise could be a mediator to help participants override or defer incorrect AI suggestions[69–71]. Although this is in contrast to prior findings that medical experts (non-dermatology) are also misled by AI outcomes[40,70], our results are aligned with prior studies in dermatology[12] that experts are better at adopting AI than non-experts. Our further investigation revealed that AI deference is less common when medical student and the PCP groups were equipped with more medical expertist. This indicates that medical knowledge and experience could be one of the key reasons for appropriate AI adoption for medical experts. Compared to the general public, experts’ medical expertise acts as a crucial *cognitive firewall*[40,72]. While a layperson uses the explanation to *form* a belief, a PCP uses it to *validate* an existing hypothesis against their own structured clinical knowledge.

## Design Implications

These findings have important design implications for how to implement AI suggestions for both the general public and PCPs. First, our work strongly suggests that a "one-size-fits-all" approach to medical XAI is not only suboptimal but potentially hazardous. Instead, systems must be carefully tailored to the user's expertise and the specific collaborative goal. For the general public, where the risk of over-reliance on persuasive LLM narratives is high, designs should prioritize safety over persuasiveness. This may involve explicitly showing the AI’s uncertainty and fallibility on a case-by-case basis to calibrate user trust, framing suggestions not as diagnoses but as preliminary information that requires professional validation, or favoring less-narrative XAI modalities like content-based image retrieval (CBIR), which encourage active comparison rather than passive acceptance. This is also supported by the recent call for medical safety disclaimer in generative AI model outcomes[35]. For clinicians, the goal shifts from simple guidance to expert augmentation. Systems should support adaptive explanations, potentially offering a concise rationale for routine cases but revealing more in-depth evidence—such as similar cases (CBIR) or conflicting features—when the clinician's initial diagnosis differs from the AI's suggestion.

Furthermore, the structure of the interaction itself is a powerful design tool for mitigating cognitive bias. Our finding shows AI suggestions first leads to stronger anchoring bias[73] and higher deference, which enriches the current understanding of the effect of human-AI decision orders.[41,74] This is clear evidence for a "Human-First" workflow, where the clinician formulates and commits to an initial diagnosis before being exposed to the AI's input, to preserve independent human reasoning and minimize the powerful anchoring bias that an upfront AI suggestion can create.

Finally, we need to design systems that account for inevitable disagreements (EFig. 8-9) and individual differences in AI deference. Our finding that participants prone to AI deference often have lower baseline performance reveals a crucial challenge: the users most in need of help are also the most vulnerable to being misled. This extends recent work[40] on AI-deference among humans and points toward the future of an adaptive system. Instead of a static information handoff, the system should dynamically tailor their interaction style to individual users. A highly deferential user might require iterative explanation[75] or explicit prompts for the user's confidence before finalizing a diagnosis[76], or automatically trigger a workflow for a second human opinion when the AI's suggestion significantly alters an initial assessment[77]. Designing these feedback loops is essential to better align AI reliance with actual user skill and minimizing the risks of misjudgment in high-stakes medical contexts.

## Limitation

Our work has several key limitations. First, although we design our study to be close to real-world scenarios, the ecological validity of the study is still limited. Only a small subset of the datasets we adopted contain patients' demographic information. Participants did not have access to contextual data (e.g., age, gender, socioeconomic status), which are crucial factors in real-world diagnosis. The task designs were also simplified: for the general public, the binary melanoma detection task may not reflect how laypersons typically judge whether a skin lesion is concerning and requires medical attention. For PCPs, making a diagnosis based solely on an image is a significant departure from standard clinical practice, which integrates physical exams and patient history. Second, our online studies only include 12 images per participant. This relatively small number may be insufficient to fully capture enough data to analyze individual patterns, potentially limiting the generalizability of our findings. Third, our results are specific to the AI models and XAI methods we implemented. The characteristics of the post-hoc explanations are inherently tied to the underlying model's training algorithm (i.e., our model necessarily shapes the internal representations from which explanations are derived). This is a potential confounding factor common in studies of post-hoc XAI. Similarly, the stochasticity in LLM explanation generation, together with the specific prompt in STab 3 that may elevate a confident tone, introduces a confounder that is difficult to control. In our current studies, LLM explanations were generated in a single pass without selection, refinement or post-hoc optimization. Besides, the observed effects might not generalize to different AI architectures, alternative explanation techniques, or future, more advanced models[78,79]. Future work should aim to address these limitations by incorporating richer patient context in more authentic clinical workflows and evaluating a larger number of cases.

## Conclusion

In conclusion, our work shows both the potential and challenges of using explainable AI, especially LLMs, in skin disease diagnosis. AI collaboration can improve diagnostic accuracy and fairness, but the significant differences in how PCPs and the general public use AI assistance demonstrate the need for carefully designed systems for different users. Our research also opens several directions for future work. Our findings should be replicated in other medical domains, and verified with additional patient information, e.g., medical history and environment, in AI systems. We must also target more work into making explanations helpful for different users, while encouraging critical thinking.

# Methods

## Image Data Collection

We collected a set of diverse datasets of clinical images spanning multiple skin tones for both studies with the general public and experts. Specifically, we combined 16 public and private datasets (see STab. 4 for details of each dataset), resulting in a total of 108,585 images after removing duplicates. These images contain human expert annotated ground truth labels and cover a wide range of skin conditions, including nevus (N=24,746, 22.8%), melanoma (N=8,686, 8.0%), atopic dermatitis (N=1,912, 1.8%), pityriasis rosea (N=609, 0.6%), Lyme disease (N=248, 0.2%), CTCL (N=330, 0.3%), and other skin diseases. In the Study 2 differential diagnosis task, we mainly focused on four diseases – atopic dermatitis, pityriasis rosea, Lyme disease, CTCL – as these four were found to have the greatest performance disparities across skin tones in a recent study[30]. 18,265 images (16.8%) also contain expert-annotated Fitzpatrick labels (Fitzpatrick labels 1-4: 15,845 out of 18,265 images, 86.8%; Fitzpatrick labels 5-6: 2,420 images, 13.2%). For the rest, we used individual typology angle (ITA) calculated from the segmented skin pixels with the YCbCr algorithm[45,80]. Using the thresholds from Kinyanjui et al., 2020[81], we assign Fitzpatrick labels to these images. Note that these images with weak Fitzpatrick labels were only used for fair model training. For the actual user study, we randomly sampled 82 high-quality, expert-annotated images from the image pool for Study 1 with the general public (equally distributed between nevus and melanoma, and between Fitzpatrick labels 1-4/5-6), and 182 images for Study 2 with medical experts (82 atopic dermatitis, 54 pityriasis rosea, 27 Lyme disease, 27 CTCL, and 92 other diseases, equally distributed between skin tones within each category). The sample sizes balanced the variance of the images and adequate number of decisions per image.

## AI Models and Explanations

We trained fair AI models for skin disease diagnosis. For Study 1, we trained a binary classification model (nevus vs melanoma). For Study 2, we implemented a hierarchical structured multi-class model that distinguished a total of 34 most common skin diseases (see the total list in STab. 4). We trained two models, including a primary 5-class classification model (four main diseases and an “other” class) and a secondary 30-class classification model (30 other common diseases). The secondary model was only used when the first model predicted “other”.

All three models were trained following the protocol in Yang et al[82]. to achieve fair medical AI models. Specifically, for each task, we train a set of models using five well-established fairness algorithms, including empirical risk minimization (ERM)[83], Domain-Adversarial Neural Networks (DANN)[84], Conditional DANN (CDANN)[85], group distributionally robust optimization (GroupDRO)[86], and exponential moving average (MA)[87]. We performed a grid search on three pretrained models (DenseNet-121[53] pretrained on ImageNet 1k[88], ViT-B/32[51] pretrained on ImageNet 1k, and ViT-B/32 pretrained on ImageNet 21k[89]) and model hyperparameters via a random search of 25 trials. Each model was finetuned for up to 30,000 steps with a batch size of 64 using the Adam optimizer, with checkpointing every 1,000 steps, and early stopping if the overall validation macro AUROC does not improve for 5 checkpoints. For each algorithm, we first selected the architecture and hyperparameter combination that maximizes the worst-group validation AUROC. We then selected algorithms based on the tradeoff between overall accuracy and fairness. Our final binary classification model for Study 1 (CDANN, ViT-B/32 pretrained on ImageNet 21k, λ=0.218, learning rate=0.0005) achieved an overall weighted AUROC of 0.930 (light skin 0.933, dark skin 0.898, Δ=0.035) a weighted balanced accuracy of 0.850 (light skin 0.852, dark skin 0.831, Δ=0.021). Compared to the baseline method ERM (balanced accuracy 0.845, Δ=0.091), our model reduced the performance gap by 76.9%. For Study 2, the final primary model (CDANN, DenseNet-121 pretrained on ImageNet 1k, λ=8.388, learning rate = 0.0029) achieved an overall AUROC of 0.772 (light skin 0.782, dark skin 0.691, Δ=0.091), weighted AUROC of 0.753 (light skin 0.725, dark skin 0.693, Δ=0.192. Note that it is possible for a total AUROC falls out of range of its breakdown), and a balanced accuracy of 0.478 for the four main diseases (light skin 0.487, dark skin 0.431, Δ=0.056), consistently outperforming the baseline ERM (balanced accuracy 0.457, Δ=0.144). The final secondary model (CDANN, ViT-B/32 pretrained on ImageNet 1k, λ = 0.377, learning rate = 0.0003) achieved an overall a AUROC of 0.728 (light skin 0.714, dark skin 0.641, Δ=0.073), a weighted AUROC of 0.752 (light skin 0.798, dark skin 0.744, Δ=0.053) a balanced accuracy of 0.141 (light skin 0.157, dark skin 0.092, Δ=0.064). STab. 5 and STab. 6 list out the specific performance details of each disease. Finally, we intentionally sampled images from our testing set with high quality disease and skin tone labels (i.e., diagnosis verified through human experts such as biopsy, and image resolution is higher than 400x400 pixels) to achieve the expected accuracy and fair results presented to participants, as described in the Results section.

Building upon the model, we then generate three types of post-hoc explanations (GradCAM, CBIR, and multimodal LLM). For GradCAM, we adopted the algorithm in Selvaraju et al. and picked the last layer in the model to generate a heatmap and overlaid it on the original image. For CBIR, we computed the cosine distance of the embedding of the target image against the embeddings of images in the training set, and selected the top-2 images with the same skin tone and top-1 image with the opposite skin tone, leading to three retrieved images for each target image. As for multimodal LLM-based explanations, we used multimodal GPT-4V (a vision-language model) and constructed the prompt to generate explanations for the target reader (the general public or medical experts, see more prompt details in STab. 3). Note that the diagnoses were determined by our own model (including cases when predictions were wrong) and sent to GPT-4V using the OpenAI API, and GPT-4V was only prompted to generate text-based explanations rather than perform diagnosis.

## Ethics approval for Human Subject Studies

This research complies with all relevant ethical regulations. The Massachusetts Institute of Technology's Committee on the Use of Humans as Experimental Subjects approved this study as Exempt Category 3—Benign Behavioral Intervention. Participants were informed about the potential sensitive visual content and provided consent before joining the study. They had the option to leave the experiment at any time. Images used in the study were publicly released by prior work.

## Experimental Interface

We developed a reactive experiment interface based on Qualtrics for online digital experiments. For Study 1, the experiment was designed to mimic real-life scenarios where users can self-diagnose using a photo captured by smartphones/webcams. Prior to the study, the general public participants were provided with educational content and common recognition guidelines about melanoma (ABCDE rules[90], see SFig. 1), together with practice questions (SFig. 2) and experiment instructions of AI and its explanations (SFig. 3). They then went through 12 images and made a diagnosis (distinguishing melanoma vs. nevus, as well as their confidence of the decision). Participants could hover on each image (including GradCAM or CBIR images) to use a magnifier for closer inspection. The interface design of Study 2 was similar to Study 1. Expert participants were asked to do open-ended differential diagnosis as free-text entry tasks, in which they were asked to enter top-3 diagnosis, confidence, and whether they had opinions on referral (including no referral, a dermatologist, biopsy, and both). Each text entry box had auto-completion based on the matched string they entered, and the corpus included 445 (see SFig. 5a). These setups were designed to resemble real-world differential diagnosis scenarios, such as telehealth, or patients sending images through electronic health record messaging systems. They received task-specific instructions with examples (see SFig. 4) and practice cases (see SFig. 5), before going through 12 images.

## Human Subject Study Design and Protocol

We recruited general public participants from Prolific (https://www.prolific.com/) and Facebook Groups for Study 1 (March 20 - May 1, 2024). In Study 2 (June 10 - Sept 3, 2024), we leverage several clinical networks and platforms to recruit medical experts, including local networks at the Boston Medical Center and Stanford University, Centaur Lab (https://www.centaurlabs.com/), MedShr (https://en.medshr.net/), Pathway (https://www.pathway.md/), and XPC (https://www.xprimarycare.com/). For both studies, we embedded attention check questions in the middle of the study (see an example in SFig. 5c.) for quality control. We also filtered answers with abnormal durations (i.e., less than 10 seconds or more than 5 minutes per image on average). After filtering, we collected 623 lay people from the general public for Study 1 and 153 PCPs for Study 2 (320 medical students in addition).

We adopted a between-subjects design. Participants were randomly assigned to one of the eight experimental conditions varying in the four XAI types (Basic confidence explanations, GradCAM, CBIR, or LLM) and two human-AI decision paradigms (Human-First vs AI-First). In both studies, each participant evaluated 12 images and provided two rounds of diagnoses per image. In the Human-First paradigm, participants made unassisted initial diagnoses and reported their confidence before receiving AI input at Round 1. They then reviewed the AI suggestions and explanations (depending on their experiment group) and reported Round 2 decision and confidence. In the AI-First paradigm, AI suggestions and explanations were presented alongside the initial image at Round 1. In Round 2, AI outcomes were hidden and participants were asked to review their decisions again. By introducing Round 2 in the AI-First condition, we controlled the number of decisions and cognitive engagement.

### Assessment of Individual Characteristics

In addition to the diagnostic tasks of 12 images, participants completed a series of questionnaires, including demographics and established assessments to measure human-AI collaboration experience[47] and AI explanation quality[48] (STable. 2). For medical experts, we further collected additional cognitive traits, including open-mindness[50], critical thinking[49] (STab. 1). Although it is not the main focus of our work, EFig. 4 presents the relationship between these subjective aspects and the final decision performance.

## Data Availability

For ML model training, our datasets are organized and available in this link. It's description is available in the README file in the Figshare subfolder of ml-model (https://figshare.com/s/eaf2110da509abbe9a6a). For the two human-subject studies, the images used in the human experiment and the study results are organized and available in the Figshare subfolder of human-study-results-analysis (https://figshare.com/s/eaf2110da509abbe9a6a). We will make these data available to any researchers who agree to only use this dataset for scientific and medical purposes and share a public link upon acceptance.

## Code Availability

The data and code are organized together. For ML model training, our model training code, and explanation generation code are organized and available in the uploaded Figshare subfolder of ml-model (https://figshare.com/s/eaf2110da509abbe9a6a). For the two human-subject studies, the code for data analysis visualization are organized and available in the Figshare subfolder of human-study-results-analysis (https://figshare.com/s/eaf2110da509abbe9a6a). We will host and share a public link upon acceptance.

**Extended Data Figs & Tables**

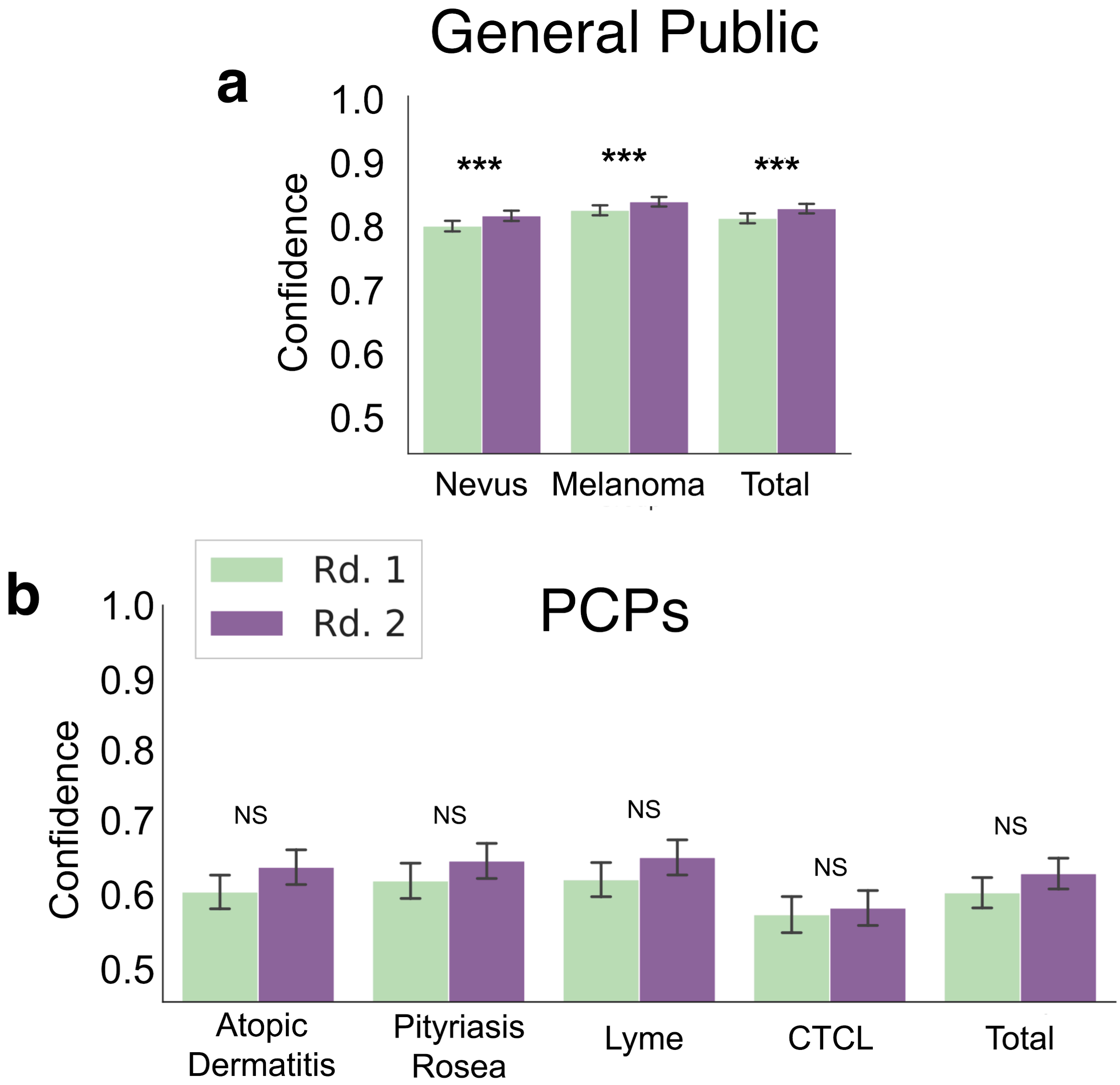

**EFig. 1 | Diagnostic Confidence of the General Public and PCPs**. **a**, Confidence levels of the general public in detecting nevus, melanoma, and the overall diagnosis in two rounds: Rd. 1 (initial, human accuracy without AI assistance) and Rd. 2 (AI-assisted accuracy). **b**, Confidence levels of PCPs in identifying atopic dermatitis, pityriasis rosea, Lyme disease, CTCL, and the overall performance on four main diseases, with results presented for two rounds.

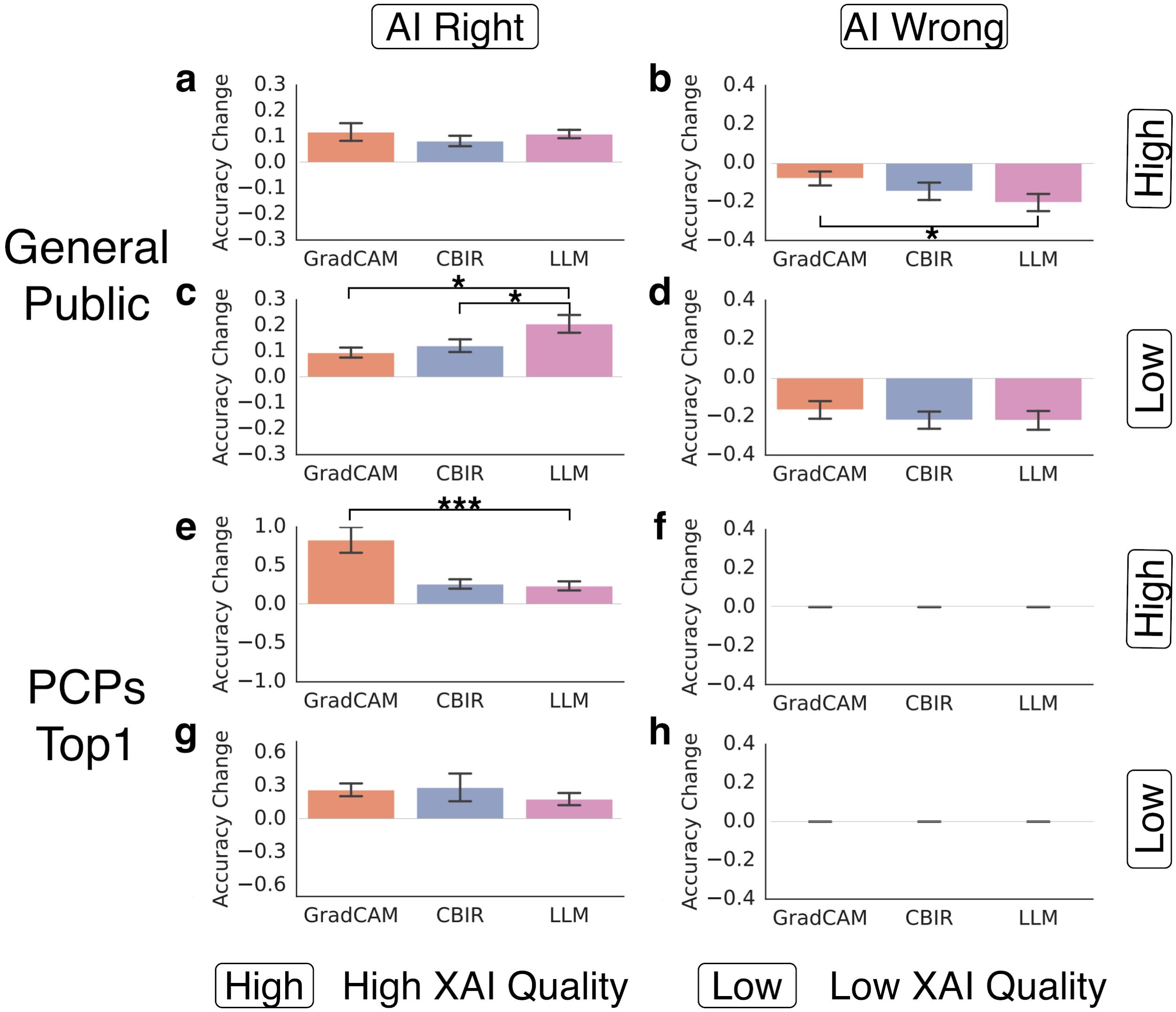

**EFig 2 | Diagnostic Accuracy Under Different XAI Qualities**. **a**, **b**, Changes in diagnostic accuracy for the general public under high XAI quality, comparing GradCAM, CBIR, and LLM explanations when AI is right (**a**) and AI is wrong (**b**). **c**, **d**, Visualization of diagnostic accuracy changes for the general public under low XAI quality conditions, highlighting the effects of right AI suggestions (**c**) and wrong suggestions (**d**) for each XAI approach. **e-h**, Equivalent results as (**a**)-(**d**), shown for top1 diagnostic accuracy of PCP participants.

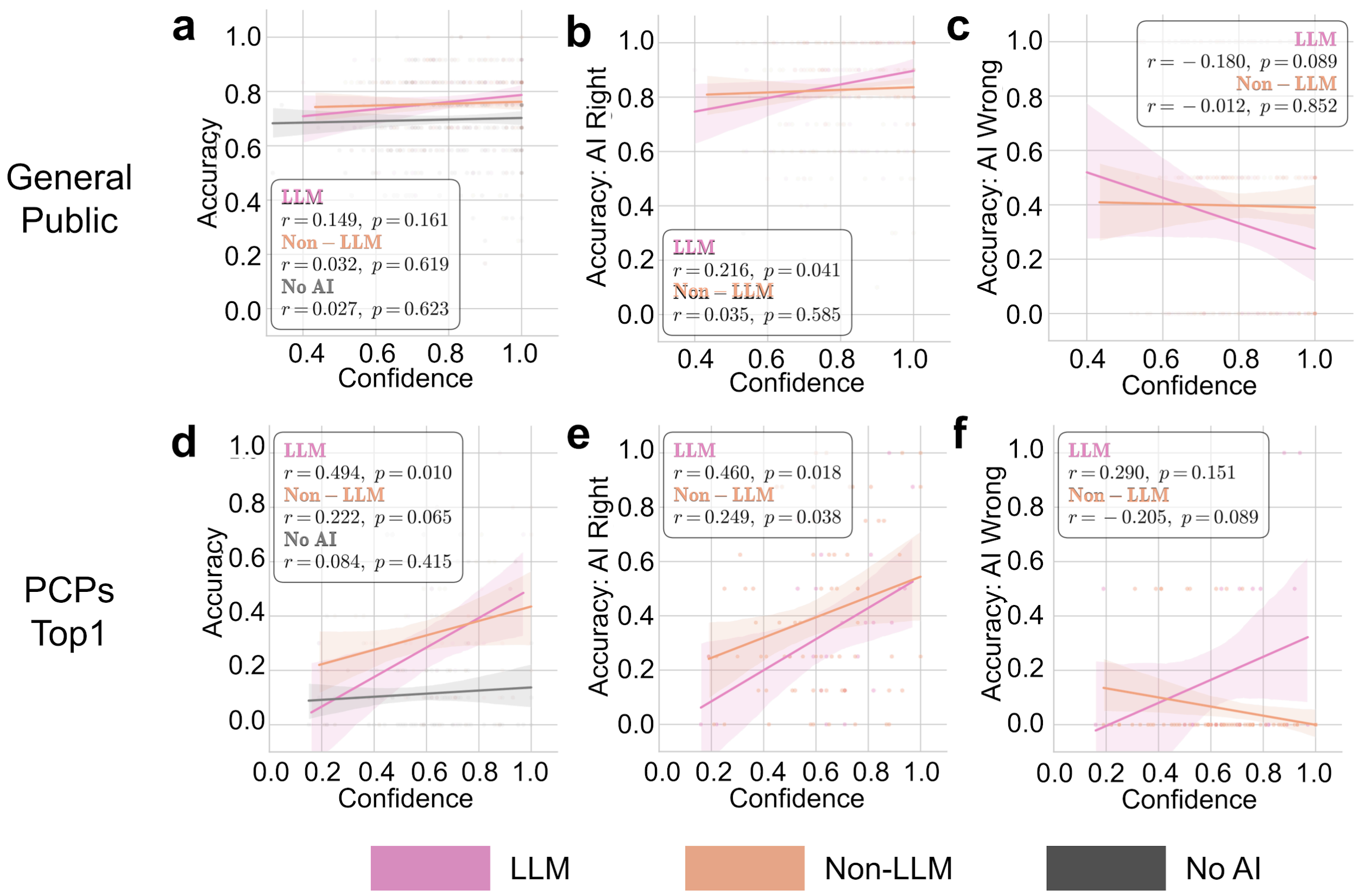

**EFig. 3 | Accuracy-Confidence Consistency of the General Public and Top1 Diagnosis of PCPs**. **a-c**, General public's performance comparison between LLM and non-LLM assisted systems, in (a) overall diagnostic accuracy-confidence consistency (A-C consistency), (**b**) A-C consistency with correct AI predictions, and (**c**) A-C consistency with wrong AI predictions. **d-f**, Equivalent results for top1 diagnosis of PCP participants as (**a**)-(**c**). Statistical significance was assessed using Pearson correlation coefficients (r) with associated p-values. Shaded areas represent 95% confidence intervals. Same metrics are used in EFig. 5.

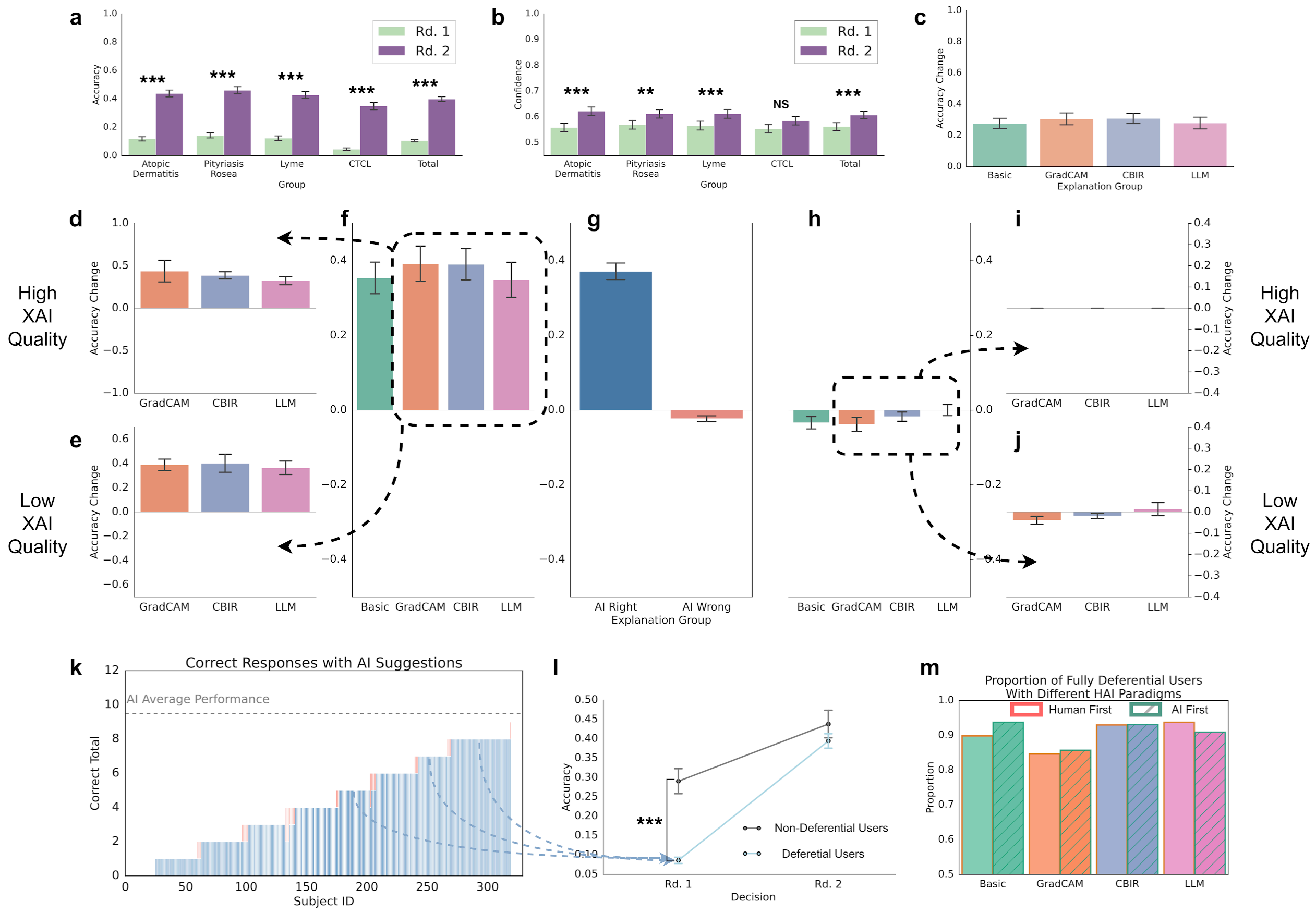

**EFig. 4 | Medical Student in Human-AI Collaboration**. **a**, Accuracy in identifying atopic dermatitis, pityriasis rosea, Lyme disease, CTCL, and the overall performance on four main diseases, with results presented for two rounds. **b**, Confidence levels of participants in their diagnostic decisions for tasks in (**a**). **c**, Diagnostic accuracy improvement across different XAI. **d, e,** Changes in diagnostic accuracy under good XAI quality (**d**) and bad XAI quality (**e**) when AI is right, comparing GradCAM, CBIR, and LLM explanations. **f, h** Diagnostic accuracy changes under AI-right (**f**) and AI-wrong (**h**) predictions for different XAI approaches. **g,** Overall accuracy changes under AI-right and AI-wrong conditions, integrating the effects of explanation quality and correctness. **i, j,** Visualization of diagnostic accuracy changes under AI-wrong conditions, highlighting the effects of good XAI quality (**i**) and bad XAI quality (**j**). **k-l**, Performance comparison between deferential and non-deferential participants. (**k**) Distribution of correct responses with AI assistance, (**l**) accuracy trajectories across two decision rounds. **m**, Proportion of fully deferential participants across different XAI approaches and HAI paradigms.

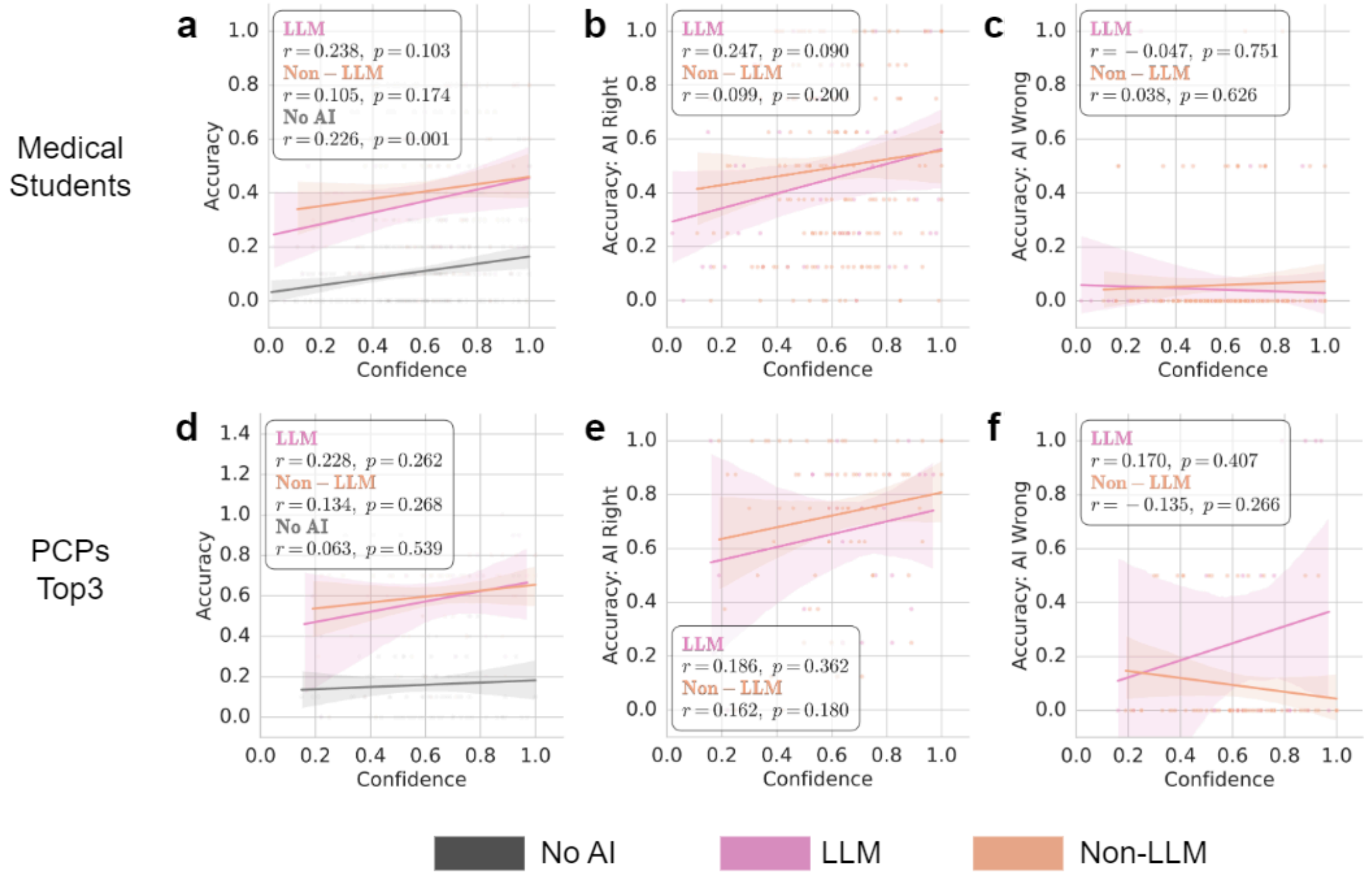

**EFig. 5 | Accuracy-Confidence Consistency of Medical Students and Top3 Diagnosis of PCPs**. **a**-**c**, Medical students' performance comparison between LLM and non-LLM assisted systems, in (a) overall diagnostic accuracy-confidence consistency (A-C consistency), (**b**) A-C consistency with correct AI predictions, and (**c**) A-C consistency with wrong AI predictions. **d**-**f**, Equivalent results for top3 diagnosis of PCP participants as (**a**)-(**c**).

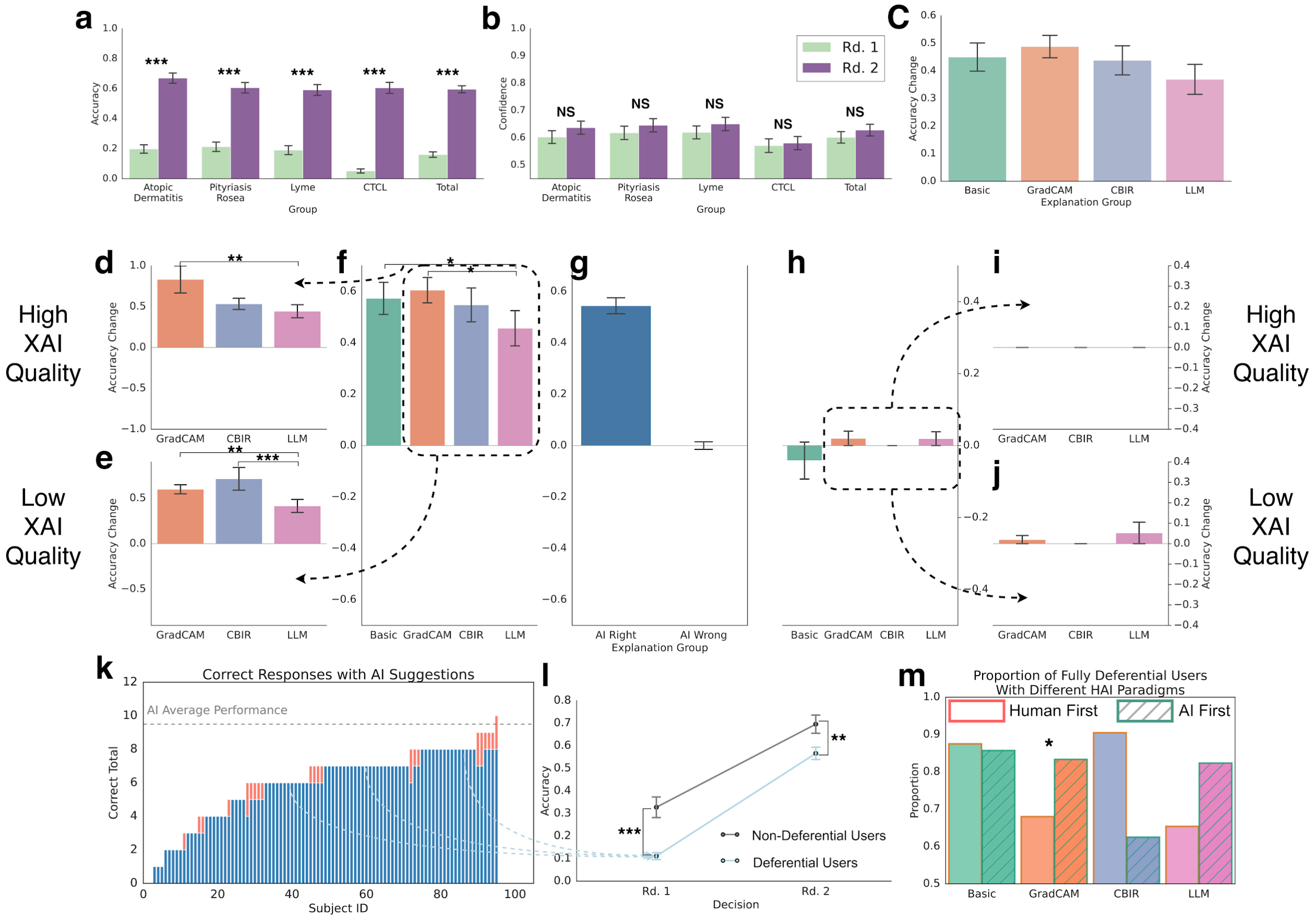

**EFig. 6 | PCP Performance - Top3 Accuracy. a**, Accuracy in identifying atopic dermatitis, pityriasis rosea, Lyme disease, CTCL, and the overall performance on four main diseases, with results presented for two rounds. **b**, Confidence levels of participants in their diagnostic decisions for tasks in (**a**). **c**, Diagnostic accuracy improvement across different XAI. **d, e,** Changes in diagnostic accuracy under good XAI quality (**d**) and bad XAI quality (**e**) when AI is right, comparing GradCAM, CBIR, and LLM explanations. **f, h** Diagnostic accuracy changes under AI-right (**f**) and AI-wrong (**h**) predictions for different XAI approaches. **g,** Overall accuracy changes under AI-right and AI-wrong conditions, integrating the effects of explanation quality and correctness. **i, j,** Visualization of diagnostic accuracy changes under AI-wrong conditions, highlighting the effects of good XAI quality (**i**) and bad XAI quality (**j**). **k-l**, Performance comparison between deferential and non-deferential participants. (**k**) Distribution of correct responses with AI assistance, (**l**) accuracy trajectories across two decision rounds. **m**, Proportion of fully deferential participants across different XAI approaches and HAI paradigms.

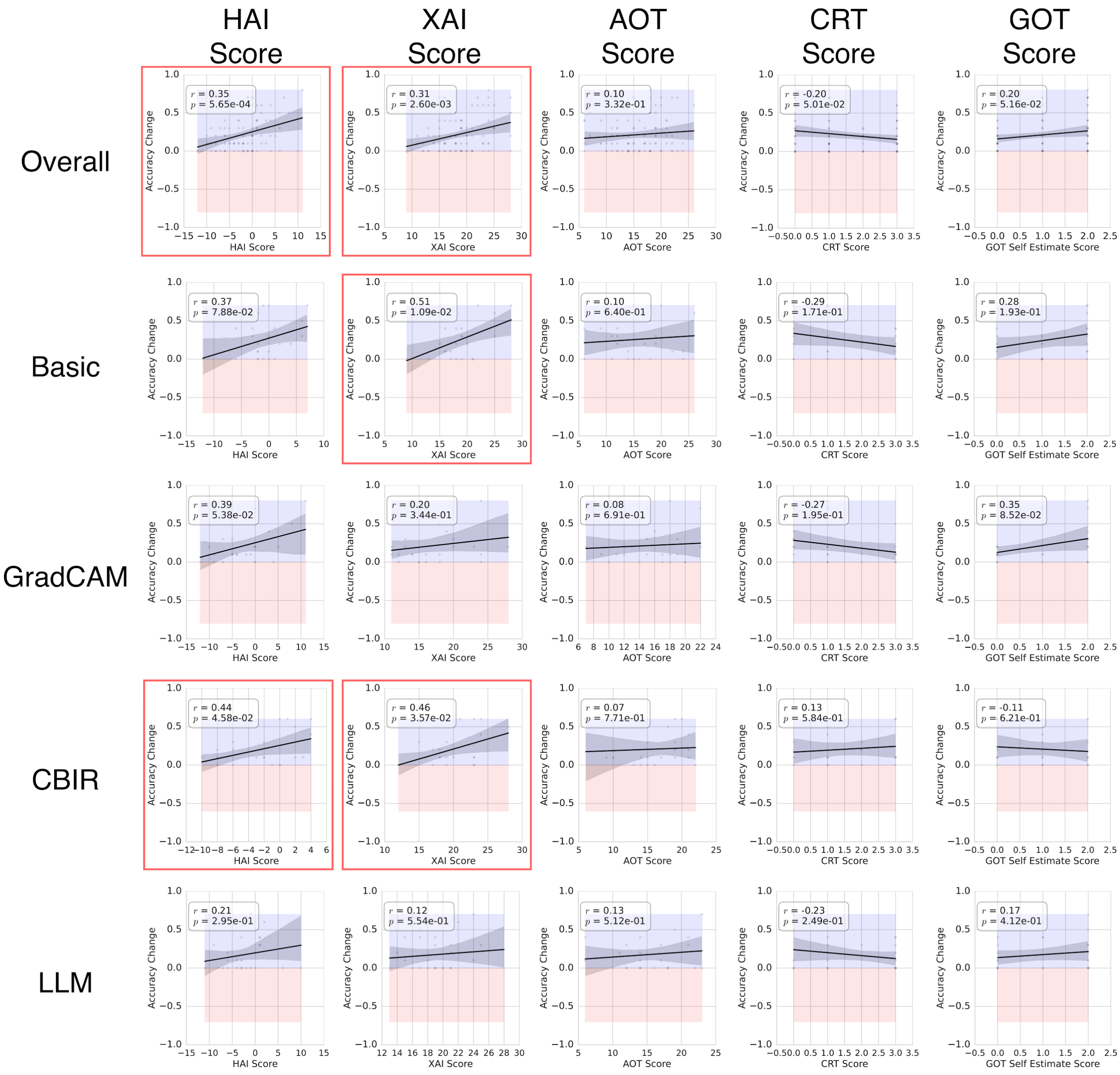

**EFig. 7 | Correlation Between Accuracy Change with AI Assistance And Personality Test Score.** In each subfigure, the blue band (maximum value) and red band (minimum value) indicate the range of accuracy change. The red border around the subfigure represents a significant correlation found (p<0.05).

## a

### For General Public: AI did Better than Human

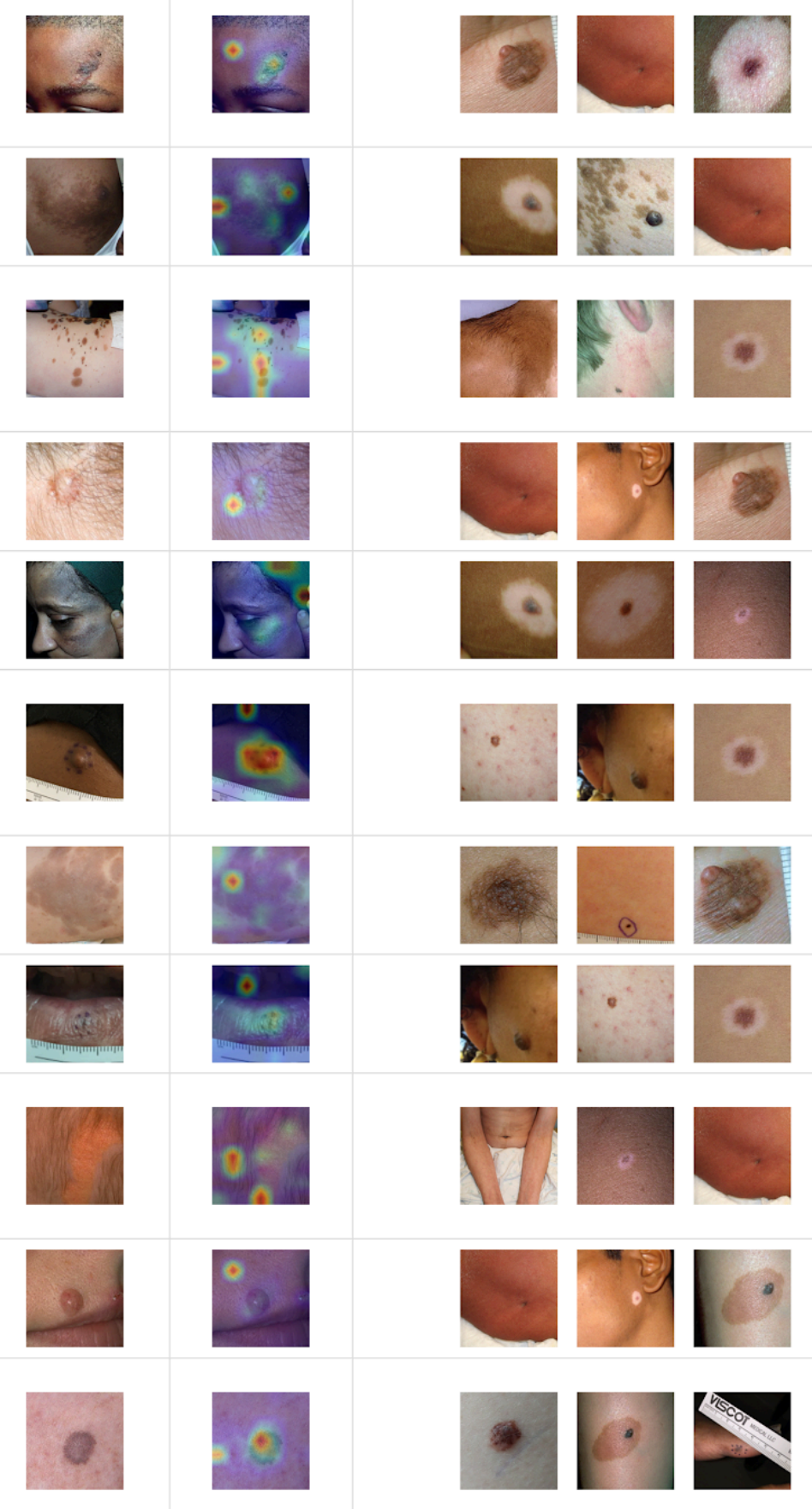

| Disease Label | Raw Image | GradCAM | CBIR | LLM |
|---|---|---|---|---|
| Nevus | | | | • Asymmetry: The lesion appears to be symmetrical.<br>• Border: The edges of the lesion are smooth and welldefined.<br>• Color: The color seems consistent and uniform across the lesion. |
| Nevus | | | | • Borders: The edges of the lesion appear smooth and even.<br>• Color: The color seems to be uniform across the lesion. |
| Nevus | | | | • Asymmetry: Lesions appear to be symmetrical.<br>• Border: Lesions have smooth, regular borders.<br>• Color: Color appears to be uniform across the lesions. |
| Nevus | | | | • Border: The edges of the lesion appear to be smooth and welldefined.<br>• Color: The lesion seems to be uniform in color. |
| Nevus | | | | • Border: The edges of the lesion appear smooth and welldefined.<br>• Color: The color seems fairly uniform across the lesion. |
| Nevus | | | | • Symmetry: The shape of the nevus appears to be symmetrical.<br>• Border: The edges of the nevus seem to be smooth and regular.<br>• Color: The color looks uniform across the nevus. |
| Nevus | | | | • Border: The edges of the lesion appear smooth and regular.<br>• Color: The color looks uniform across the lesion. |
| Nevus | | | | • Border: The edges of the lesion appear smooth and welldefined.<br>• Color: The color seems to be uniform across the lesion. |
| Nevus | | | | • Symmetry: The lesion appears to be symmetrical.<br>• Border: The edges of the lesion look welldefined.<br>• Color: The color looks uniform across the lesion. |
| Nevus | | | | • Border: The edges of the lesion are smooth and regular.<br>• Color: The lesion appears to be a single color throughout. |
| Nevus | | | | • Asymmetry: The lesion appears to be symmetrical in shape.<br>• Border: The edges of the lesion are smooth and welldefined.<br>• Color: The lesion has a uniform color throughout. |

## b

### For General Public: Human did Better than AI

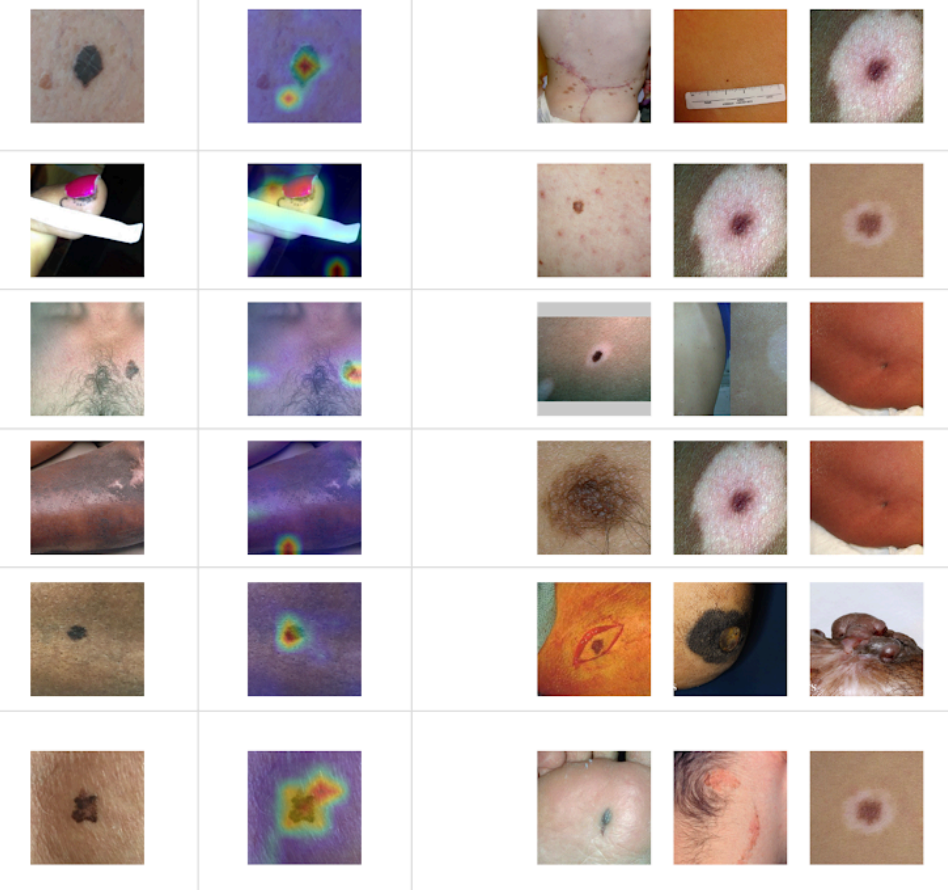

| Disease Label | Raw Image | GradCAM | CBIR | LLM |
|---|---|---|---|---|
| Melanoma | | | | • Symmetry: The lesion is symmetrical.<br>• Border: The borders of the lesion are smooth and even.<br>• Color: The color appears to be uniform throughout the lesion. |
| Melanoma | | | | • Border: The edges of the lesion appear smooth and welldefined.<br>• Color: The lesion seems to be a uniform color throughout. |
| Melanoma | | | | • Border: The edges of the lesion appear smooth and welldefined.<br>• Color: The color seems to be uniform and consistent throughout the lesion. |
| Melanoma | | | | • Borders: The edges of the lesion are smooth and even.<br>• Color: The color appears to be uniform across the lesion. |
| Nevus | | | | • Asymmetry: The shape is not symmetric.<br>• Color: There is more than one color present.<br>• Diameter: Larger than 6mm. |
| Melanoma | | | | • Asymmetry: The lesion appears to be symmetrical.<br>• Border: The edges of the lesion look welldefined.<br>• Color: The color seems to be uniform across the lesion. |

**EFig. 8 | General-Public Case Study. a,** AI did a better diagnostic performance than humans**. b,** Humans did a better performance than AI.

## A

**For PCPs: AI did Better than Human**

| Disease Label | Raw Image | GradCAM | CBIR | LLM |
|---|---|---|---|---|
| Atopic Dermatitis | 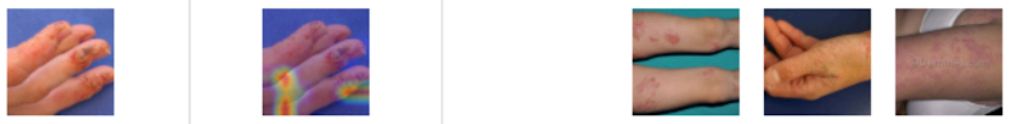 | | | • Erythema (redness)<br>• Edema (swelling)<br>• Exudation (oozing)<br>• Crusting<br>• Scaling |
| Pityriasis Rosea | 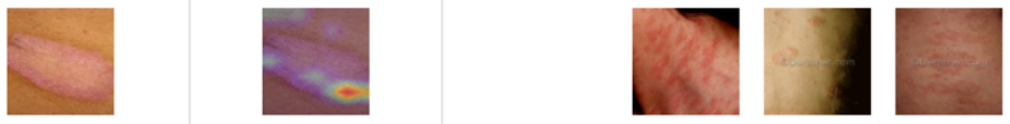 | | | • Oval shape of the lesion<br>• Fine scales on the surface<br>• Edge of the lesion is slightly raised<br>• Central clearing of the lesion |
| Pityriasis Rosea | 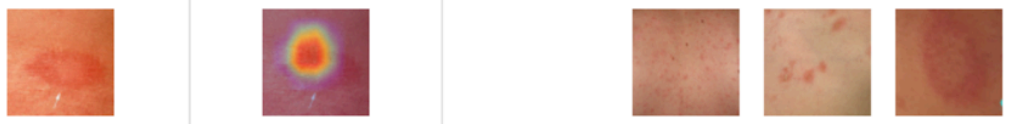 | | | • Herald patch: Larger, round, or oval plaque with central clearing<br>• Fine scale: Visible on the borders of the lesion<br>• Erythematous border: The outer margin of the lesion appears inflamed |
| Atopic Dermatitis | 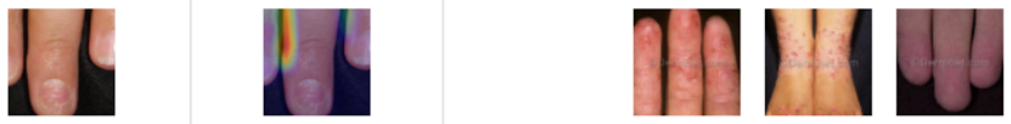 | | | • Erythema (redness) visible on the skin<br>• Edema (swelling) and vesicles indicating acute changes<br>• Lichenification (thickened skin) from chronic rubbing<br>• Scaling on the periphery of the lesion<br>• Fissures may be present indicating skin dryness and cracking |
| Pityriasis Rosea | 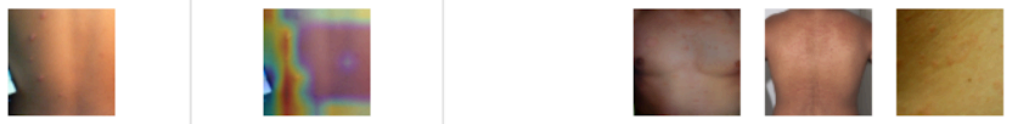 | | | • Presence of oval, fawncolored plaques with a collarette scale.<br>• Distribution in a "Christmas tree" or fir tree pattern on the body.<br>• Lesions appear to follow the cleavage lines of the skin. |
| Pityriasis Rosea | 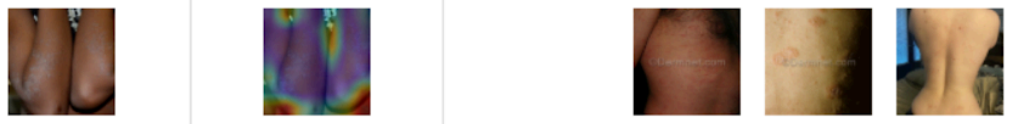 | | | • Presence of multiple, round to oval, slightly scaly patches.<br>• Patches are distributed symmetrically.<br>• Margins of the patches are well defined.<br>• The color of the patches appears to be hypopigmented compared to the surrounding skin. |
| Lyme | 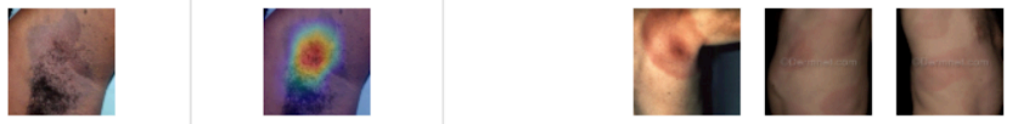 | | | • Central clearing with an expanding erythematous border<br>• Rash resembling erythema migrans associated with Lyme disease |
| Atopic Dermatitis | 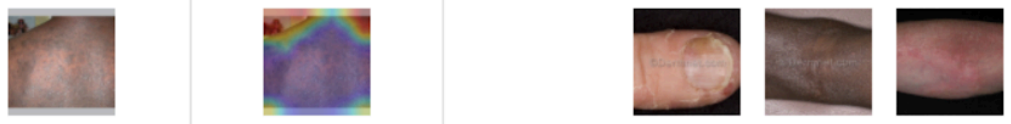 | | | • Presence of erythematous to brownish discoloration<br>• Evidence of excoriation and lichenification<br>• The skin shows signs of xerosis (dry skin) and scaling |
| Lyme | 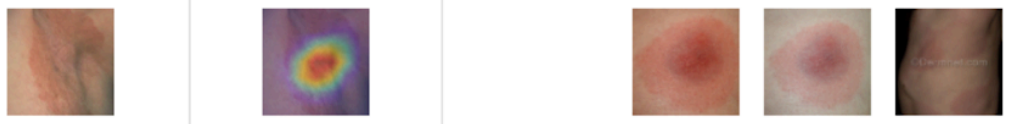 | | | • Erythema migrans rash with central clearing<br>• Annular, expanding redness |
| Lyme | 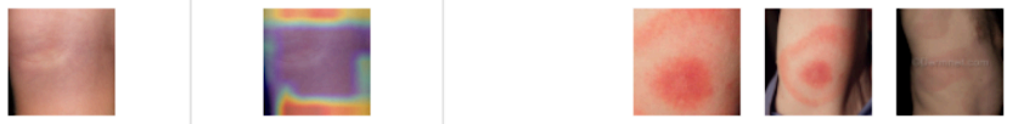 | | | • Central clearing<br>• Erythematous margin<br>• Annular lesion shape |
| Atopic Dermatitis | 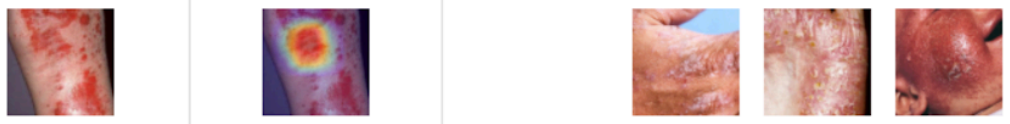 | | | • Erythematous (red) lesions<br>• Patches of dry, scaly skin<br>• Areas of swelling or edema<br>• Evidence of excoriation (scratch marks) |

## B

**For PCPs: Human did Better than AI**

| Disease Label | Raw Image | GradCAM | CBIR | LLM |
|---|---|---|---|---|
| Lyme | 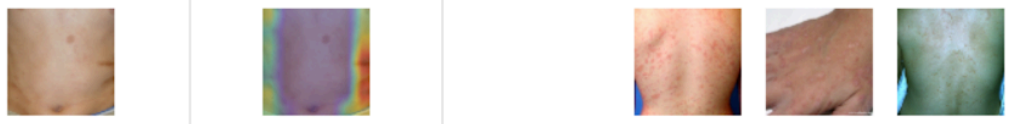 | | | • Presence of oval, fawncolored patches with a fine scale.<br>• Larger patch with a collarette of scale, indicating a possible herald patch.<br>• Distribution of smaller lesions following skin cleavage lines, creating a 'Christmastree' pattern. |
| Pityriasis Rosea | 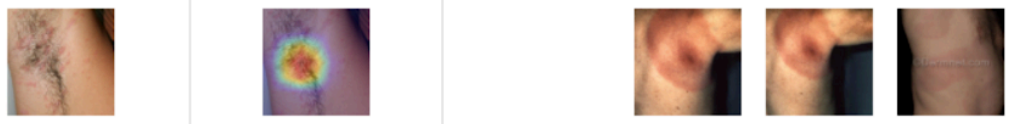 | | | • Presence of erythematous rash<br>• Rash has a central clearing, giving it a bull'seye appearance<br>• The lesion is well demarcated |
| CTCL | 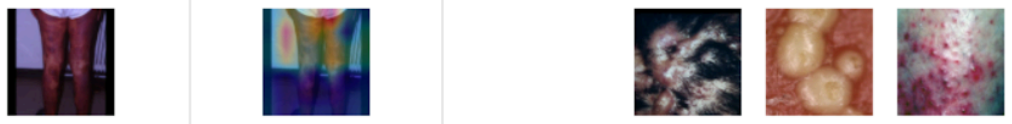 | | | • Presence of welldemarcated, depigmented patches<br>• Lack of inflammation around the borders of the patches |
| Pityriasis Rosea | 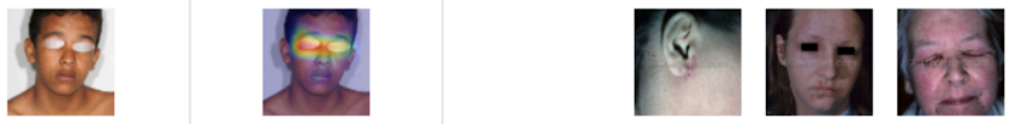 | | | • Erythematous patches on the face<br>• Lichenification on the cheeks<br>• Possible excoriations on the face<br>• Edema around the eyes |
| Lyme | 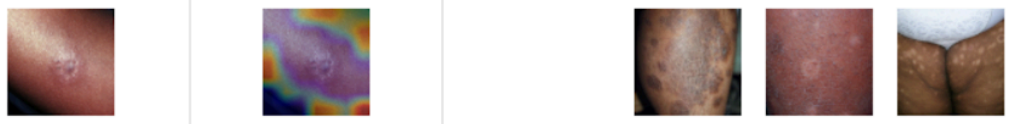 | | | • Presence of an erythematous patch/plaque with a welldefined border.<br>• The surface of the lesion appears slightly scaly and raised from the surrounding skin.<br>• There is no visible central clearing. |
| Lyme | 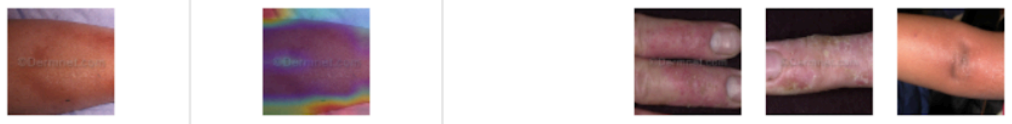 | | | • Presence of erythema (red or inflamed skin)<br>• Evidence of skin lesions with illdefined edges<br>• Signs of excoriation (scratch marks)<br>• Lichenification (thickened skin), indicated by the accentuated skin lines |

**EFig. 9 | PCP Case Study. a,** AI had a better diagnostic performance than humans**. b,** Humans did a better performance than AI.

**A**

**For General Public: AI and Human Both had Low Diagnostic Performance**

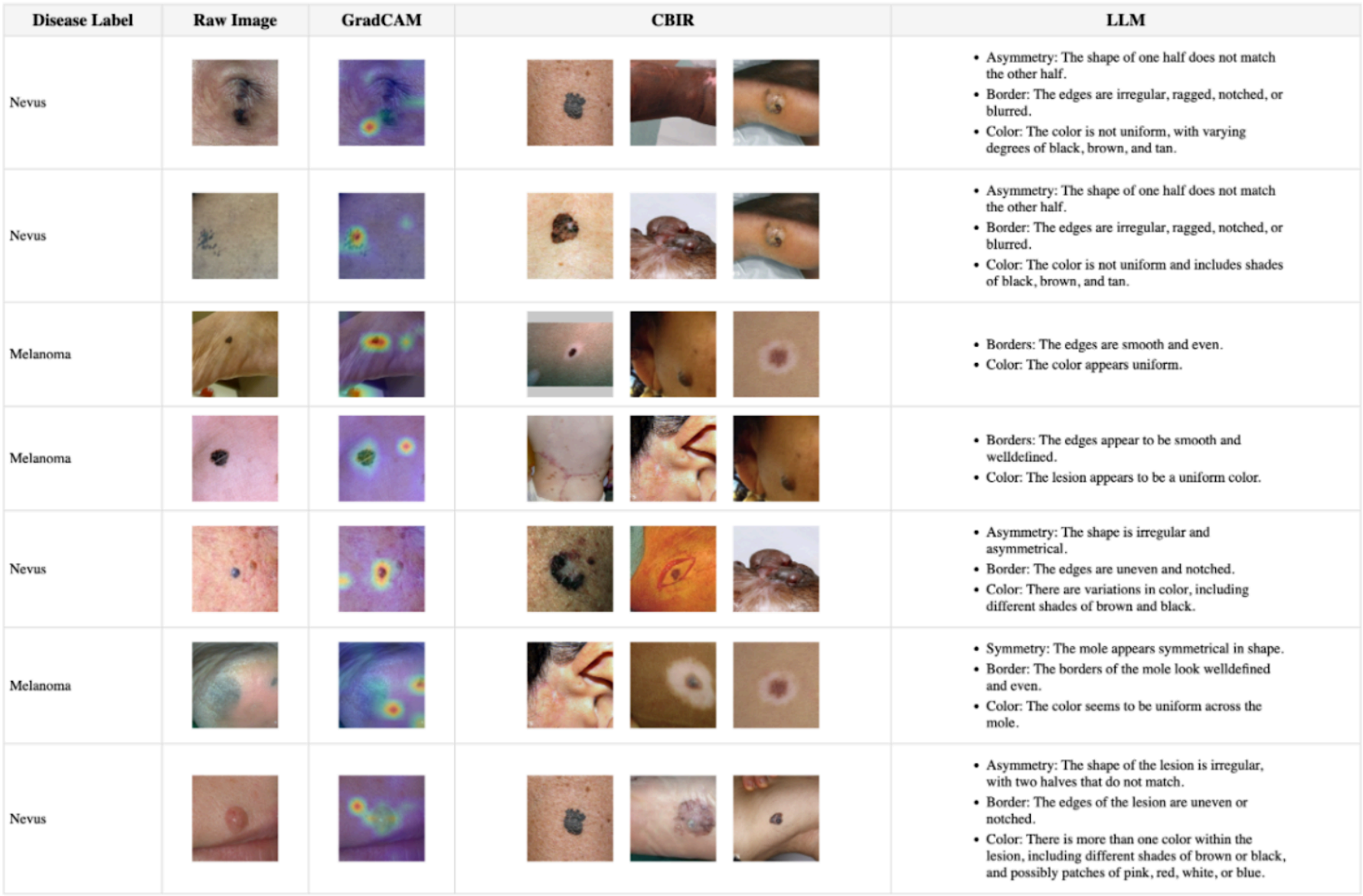

| Disease Label | Raw Image | GradCAM | CBIR | LLM |
|---|---|---|---|---|
| Nevus | | | | • Asymmetry: The shape of one half does not match the other half.<br>• Border: The edges are irregular, ragged, notched, or blurred.<br>• Color: The color is not uniform, with varying degrees of black, brown, and tan. |
| Nevus | | | | • Asymmetry: The shape of one half does not match the other half.<br>• Border: The edges are irregular, ragged, notched, or blurred.<br>• Color: The color is not uniform and includes shades of black, brown, and tan. |
| Melanoma | | | | • Borders: The edges are smooth and even.<br>• Color: The color appears uniform. |
| Melanoma | | | | • Borders: The edges appear to be smooth and welldefined.<br>• Color: The lesion appears to be a uniform color. |
| Nevus | | | | • Asymmetry: The shape is irregular and asymmetrical.<br>• Border: The edges are uneven and notched.<br>• Color: There are variations in color, including different shades of brown and black. |
| Melanoma | | | | • Symmetry: The mole appears symmetrical in shape.<br>• Border: The borders of the mole look welldefined and even.<br>• Color: The color seems to be uniform across the mole. |
| Nevus | | | | • Asymmetry: The shape of the lesion is irregular, with two halves that do not match.<br>• Border: The edges of the lesion are uneven or notched.<br>• Color: There is more than one color within the lesion, including different shades of brown or black, and possibly patches of pink, red, white, or blue. |

**B**

**For PCPs: AI and Human Both had Low Diagnostic Performance**

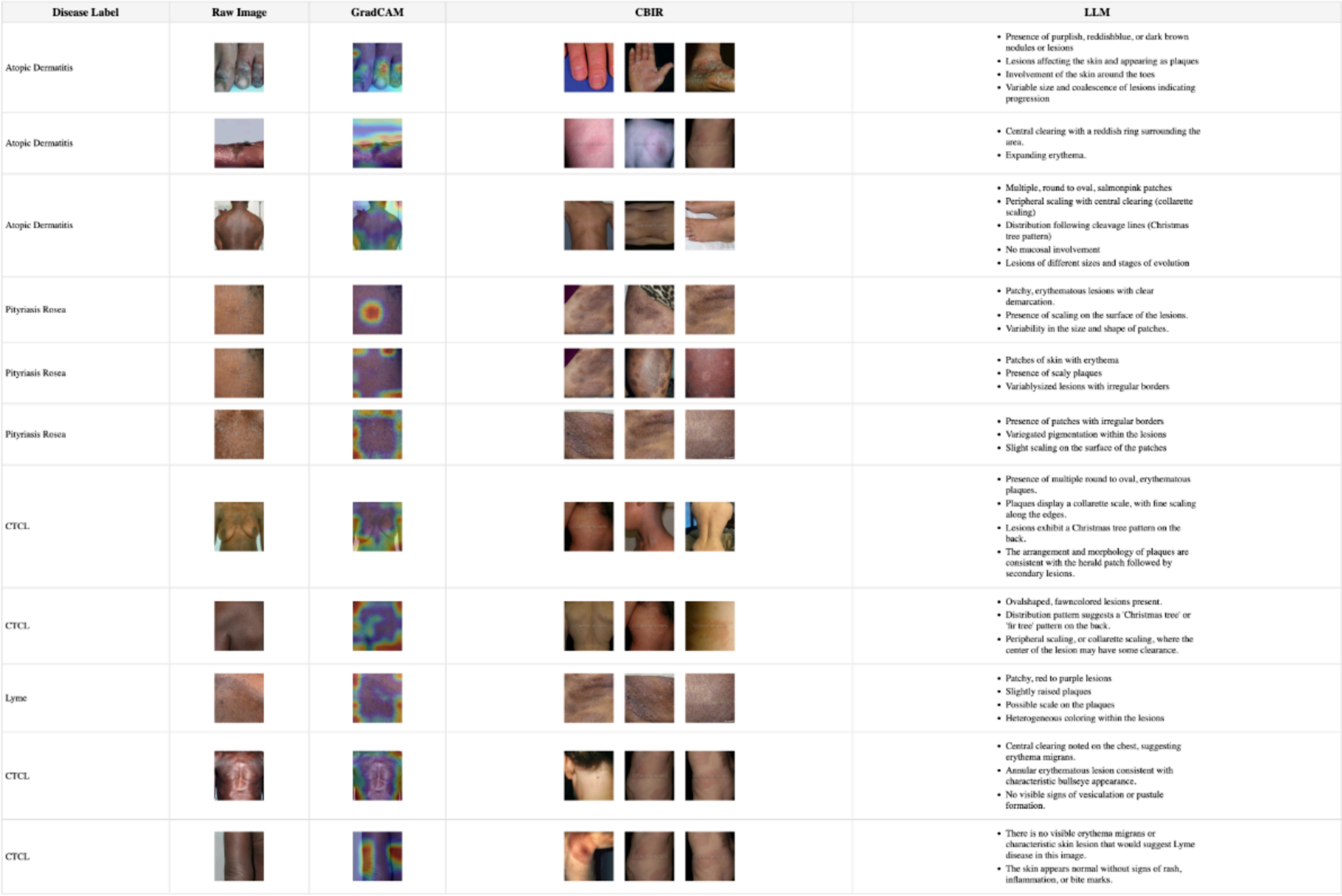

| Disease Label | Raw Image | GradCAM | CBIR | LLM |
|---|---|---|---|---|
| Atopic Dermatitis | | | | • Presence of purplish, reddishblue, or dark brown nodules or lesions<br>• Lesions affecting the skin and appearing as plaques<br>• Involvement of the skin around the toes<br>• Variable size and coalescence of lesions indicating progression |
| Atopic Dermatitis | | | | • Central clearing with a reddish ring surrounding the area.<br>• Expanding erythema. |
| Atopic Dermatitis | | | | • Multiple, round to oval, salmonpink patches<br>• Peripheral scaling with central clearing (collarette scaling)<br>• Distribution following cleavage lines (Christmas tree pattern)<br>• No mucosal involvement<br>• Lesions of different sizes and stages of evolution |
| Pityriasis Rosea | | | | • Patchy, erythematous lesions with clear demarcation.<br>• Presence of scaling on the surface of the lesions.<br>• Variability in the size and shape of patches. |
| Pityriasis Rosea | | | | • Patches of skin with erythema<br>• Presence of scaly plaques<br>• Variablysized lesions with irregular borders |
| Pityriasis Rosea | | | | • Presence of patches with irregular borders<br>• Variegated pigmentation within the lesions<br>• Slight scaling on the surface of the patches |
| CTCL | | | | • Presence of multiple round to oval, erythematous plaques.<br>• Plaques display a collarette scale, with fine scaling along the edges.<br>• Lesions exhibit a Christmas tree pattern on the back.<br>• The arrangement and morphology of plaques are consistent with the herald patch followed by secondary lesions. |
| CTCL | | | | • Ovalshaped, fawncolored lesions present.<br>• Distribution pattern suggests a 'Christmas tree' or 'fir tree' pattern on the back.<br>• Peripheral scaling, or collarette scaling, where the center of the lesion may have some clearance. |
| Lyme | | | | • Patchy, red to purple lesions<br>• Slightly raised plaques<br>• Possible scale on the plaques<br>• Heterogeneous coloring within the lesions |
| CTCL | | | | • Central clearing noted on the chest, suggesting erythema migrans.<br>• Annular erythematous lesion consistent with characteristic bullseye appearance.<br>• No visible signs of vesiculation or pustule formation. |
| CTCL | | | | • There is no visible erythema migrans or characteristic skin lesion that would suggest Lyme disease in this image.<br>• The skin appears normal without signs of rash, inflammation, or bite marks. |

**EFig. 10 | Case Studies when AI has Low Diagnostic Performance, and Human Followed the Wrong AI.** a, Examples for the general public. b, Examples for the PCPs.

## Supplementary Materials

We introduce additional details of our experiment studies, including experimental interface, evaluation questionnaires, LLM prompts for AI explanation as well as dermatological datasets employed for our studies.

### 1. Experiment Interface

As introduced in the main text, we have two user studies. For Study 1 (melanoma vs nevus), we provide tutorials and introduction of the experiment in order with SFig. 1-3. SFig. 1 gives a comprehensive introduction about melanoma and the steps of diagnosing it to ensure that the general public participants are equipped with basic knowledge to perform the recognition task. SFig. 2 gives two examples following the introduction to connect participants' knowledge with practice. SFig. 3 presents screenshots of formal Study 1 experiment instructions..

For Study 2 (open-ended free-text task), we use SFig. 4-5 for an example introduction. SFig. 4 presents comprehensive experiment instructions (smilar to the SFig. 3). SFig. 5 presents a practice question to familiarize participants (PCPs and medical students) with the study setup.

## Melanoma Introduction

Let's start with a simple introduction about melanoma detection.

Melanoma is a serious form of skin cancer, which can be effectively treated if detected early. It usually appears as a skin lesion that resembles a nevus (i.e., common mole). So it's important to distinguish **melanoma** vs. **nevus**, which can help identify potentially cancerous moles or skin lesions.

One of the simplest methods for early detection is the ABCDE rule:

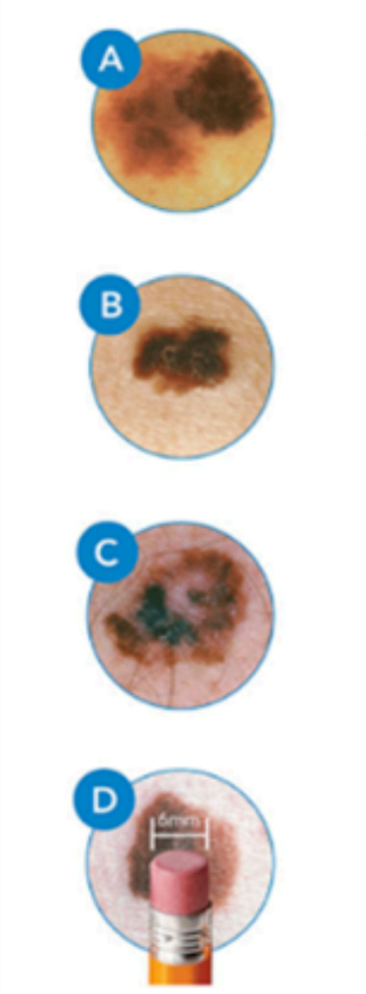

**A is for Asymmetry:** One half of the spot is unlike the other half.

**B is for Border:** The spot has an irregular, scalloped, or poorly defined border.

**C is for Color:** The spot has varying colors from one area to the next, such as shades of tan, brown or black, or areas of white, red, or blue.

**D is for Diameter:** While melanomas are usually greater than 6 millimeters, or about the size of a pencil eraser, when diagnosed, they can be smaller.

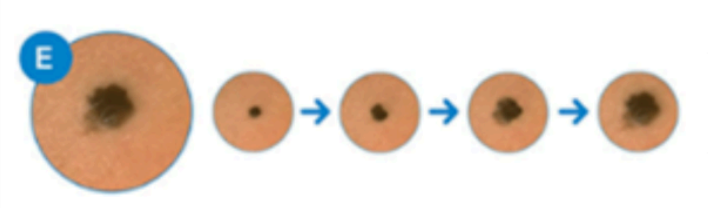

**E is for Evolving:** The spot looks different from the rest or is changing in size, shape, or color.

If a mole or skin lesion displays any of these ABCDE characteristics, it's important to consult with a dermatologist or healthcare provider for further evaluation.

Note 1: This guide is for educational purposes only and is not a substitute for professional medical advice.

Note 2: Some of these characteristics may not apply to this survey. For example, there will not be multiple images to show evolving stages, so rule E won't apply.

Note 3: You don't have to remember these rules. You can review them anytime during the study.

→

**SFig. 1 | Introduction for Melanoma in Nevus vs. Melanoma Recognition Questionnaire.**

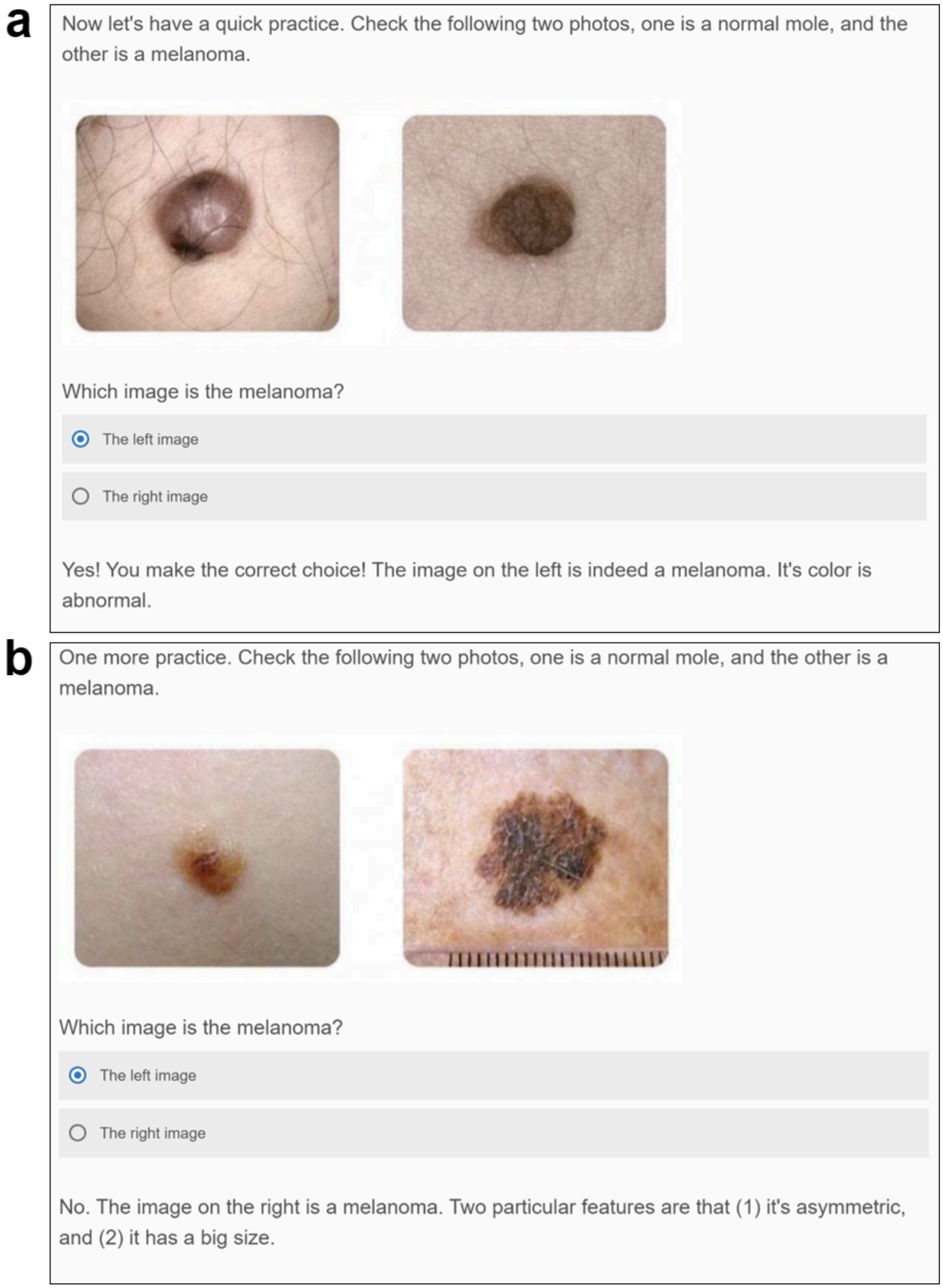

**SFig. 2 | Melanoma Recognition Practice Page**. Two practice questions with feedback and comprehensive explanations on each choice are presented to the participants. (a) shows the feedback of the correct choice and (b) shows the feedback of the wrong one.

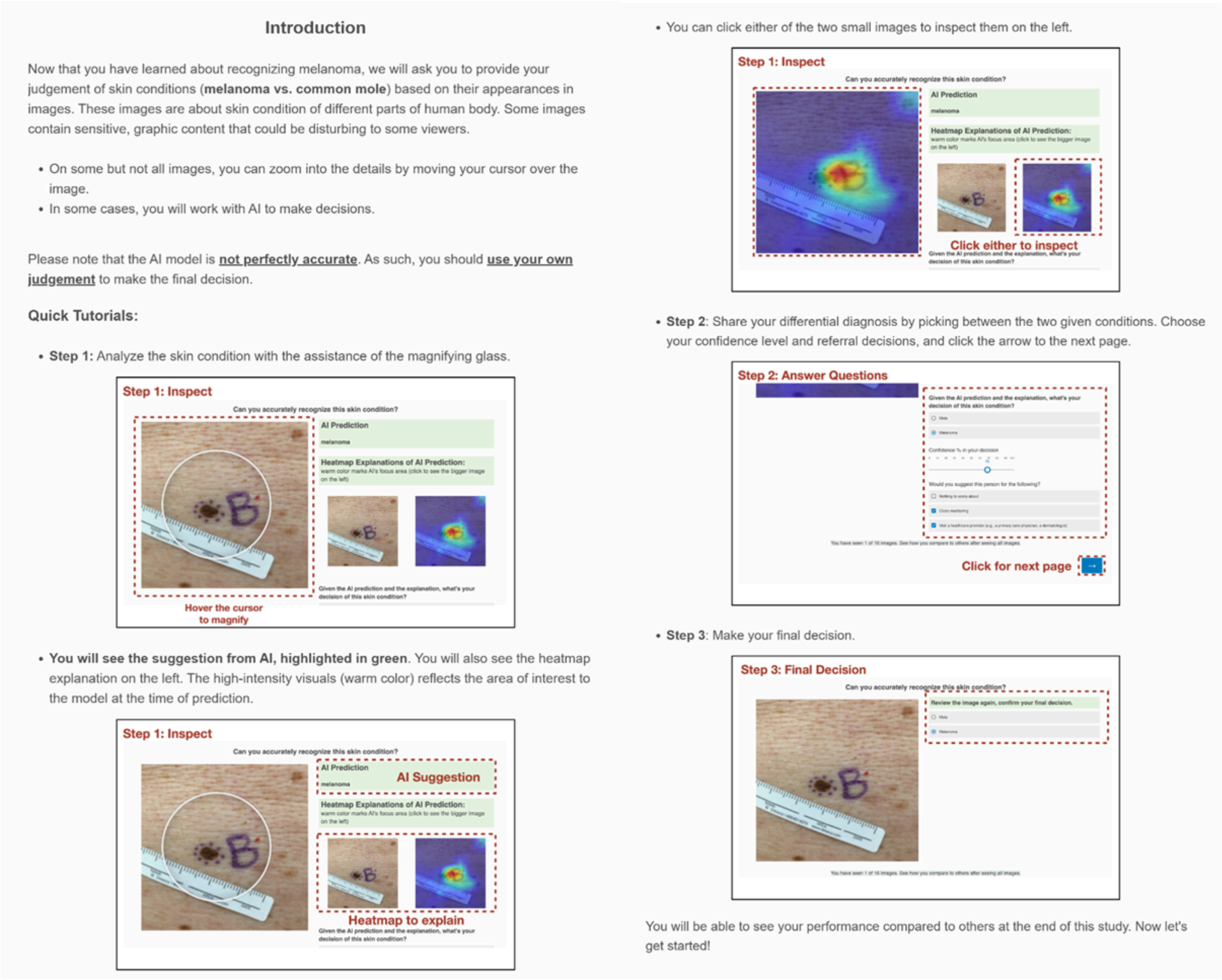

**SFig. 3 | Instructions and Tutorials for Nevus vs. Melanoma Recognition Questionnaire.**

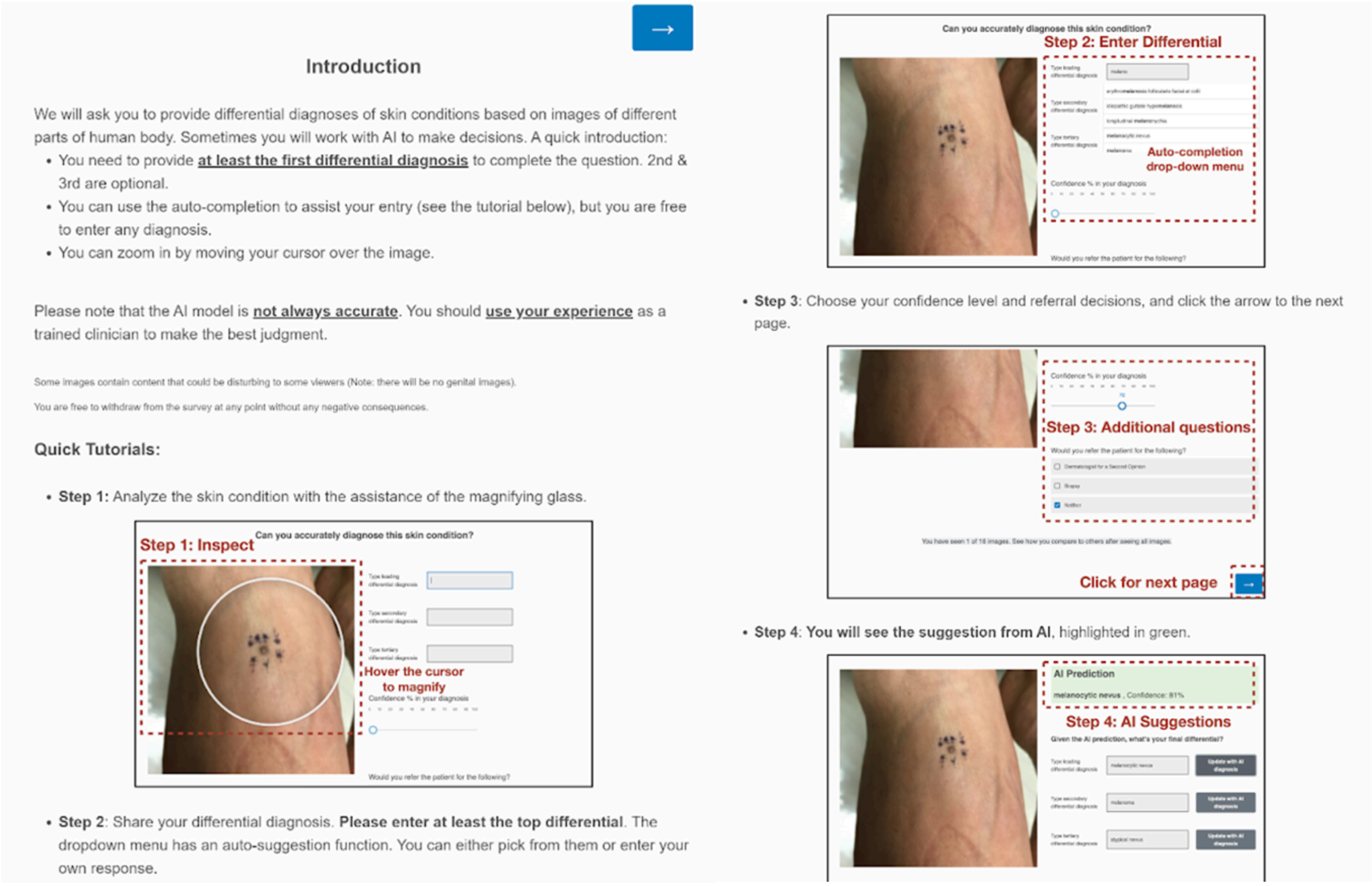

**SFig. 4 | Instructions and Tutorials for Open-Ended Differential Skin Disease Diagnosis Questionnaire.**

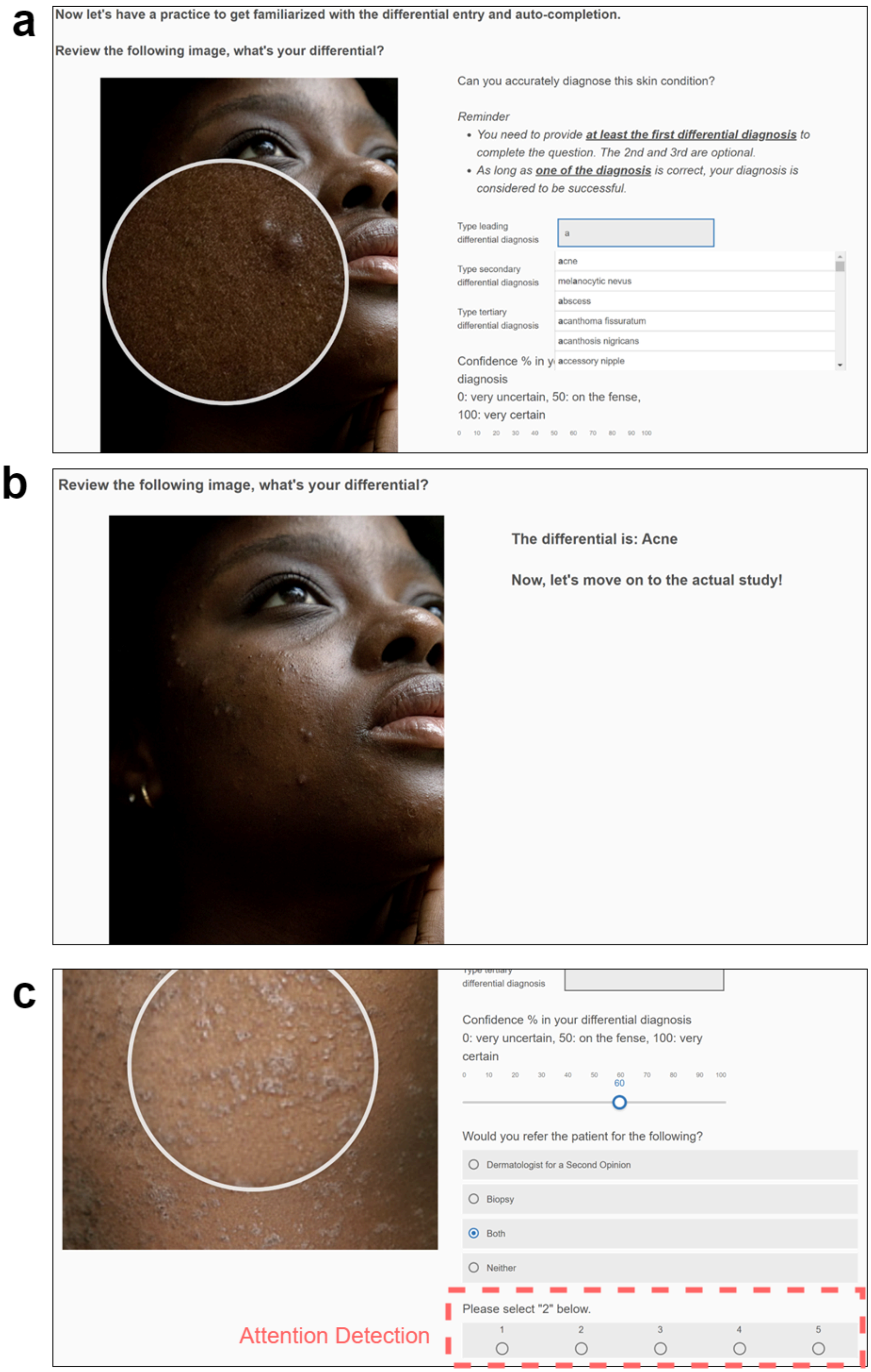

**SFig. 5 | Questions for Open-Ended Differential Skin Disease Diagnosis Questionnaire. a**, **b**, Practice question before the actual 12 images, where (**a**) presents the instruction, questions and differential answers for filling or selection and (**b**) presents the right answer for the practice. **c**, Attention detection trick in the middle of formal experiment.

## 2. Additional Evaluation Questionnaire

Following the main experiment, we deployed some additional evaluation questionnaires to test the personality of participants. SFig 6 shows the generalized overconfidence test (GOT) used to test participants' confidence about their decisions. STable. 1 introduces two questionnaires to generally characterize participants, open-mindedness questionnaire and critical-thinking questionnaire. STable. 2 then measures participants' experience in human-AI collaboration and their feelings about AI explanation.

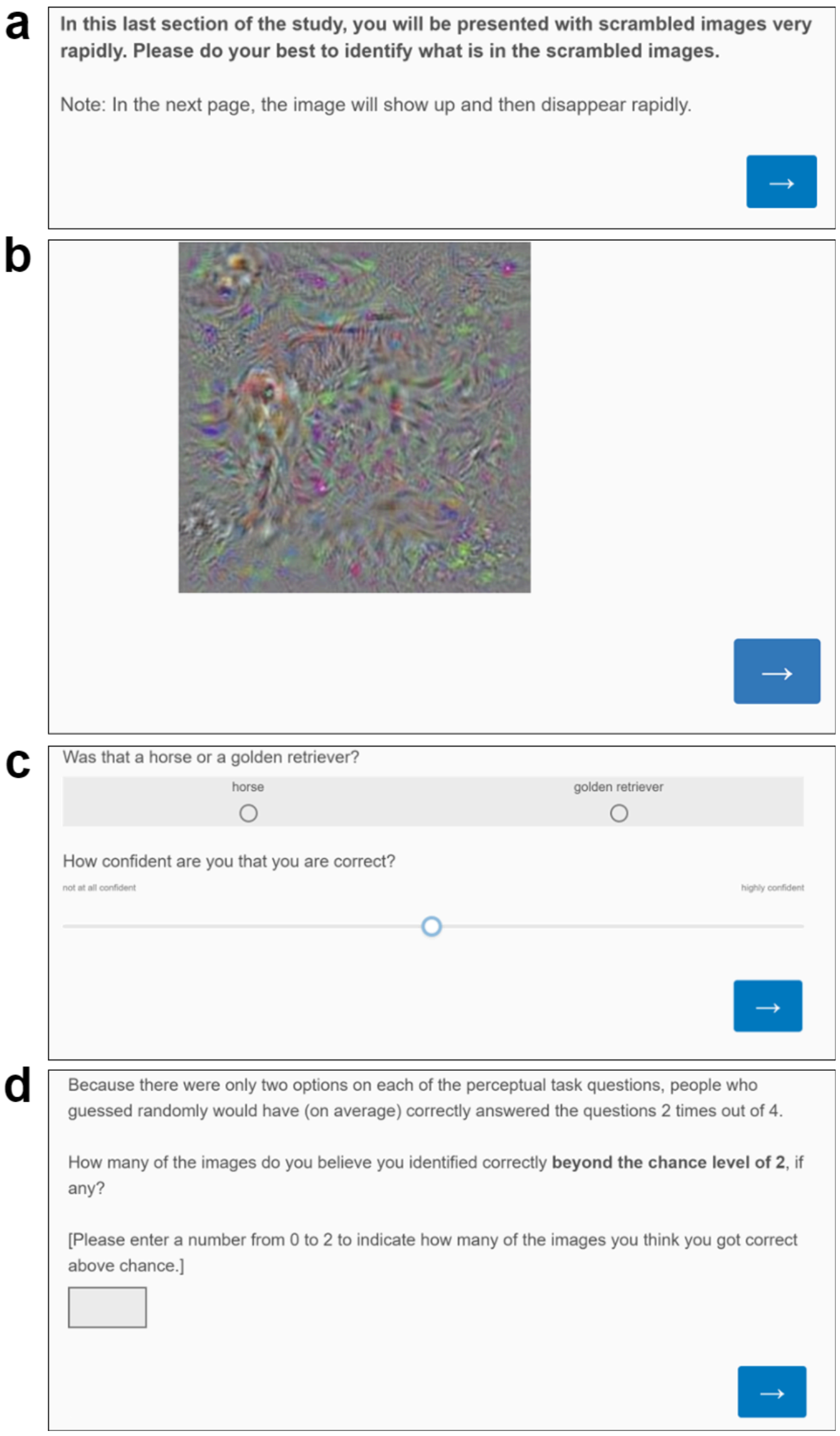

**SFig. 6 | An Example of the Generalized Overconfidence Task (GOT).** a-c presents an example trial of GOT tested after the main questions in the questionnaire, where participants will first see the instruction (a), followed by a flash of a very difficult-to-discern image (b) and a question (c) asking the object presented and the confidence for that. d, participants'

self-estimation over the chance level after two trials.

**STable 1 | Test on Human's Mind-Openness and Critical Thinking.**

| Questionnaire | Instruction | Metric | Question / Statement |
|---|---|---|---|
| Open-Mindedness[48] | Please rate your level of agreement with the following statement. | 5-level rating choice among "Disagree Strongly", "Disagree", "Neither Agree or Disagree", "Agree", "Agree Strongly" | It is important to be loyal to your beliefs even when evidence is brought to bear against them. |
| | | | Whether something feels true is more important than evidence. |
| | | | Just because evidence conflicts with my current beliefs does not mean my beliefs are wrong. |
| | | | There may be evidence that goes against what you believe but that does not mean you have to change your beliefs. |
| | | | Even if there is concrete evidence against what you believe to be true, it is OK to maintain cherished beliefs. |
| | | | Regardless of the topic, what you believe to be true is more important than evidence against your beliefs. |
| Critical Thinking[47] | In the following section, you will be asked several questions. Please do your best to answer as accurately as possible. | Gap filling | The ages of Mark and Adam add up to 28 years total. Mark is 20 years older than Adam. How many years old is Adam? |
| | | | If it takes 10 seconds for 10 printers to print out 10 pages of paper, how many seconds will it take 50 printers to print out 50 pages of paper? |
| | | | On a loaf of bread, there is a patch of mold. Every day, the patch doubles in size. If it takes 40days for the patch to cover the entire loaf of bread, how many days would it take for the patch to cover half of the loaf of bread? |

**STable 2 | Test on Human's Experience and Trust on AI Technology and AI Explanations.**

| Questionnaire | Instruction | Metric | Question |
|---|---|---|---|
| Human-AI Collaboration Experience[45] | From 1 (Strongly Disagree) to 7 (Strongly Agree), how much do you agree with the statement | 7-level rating choice among "Strongly Disagree", "Disagree", "Somewhat Disagree", "Neutral", "Somewhat Agree", "Agree", "Strongly Agree" | I can trust AI system |
| | | | The AI system is reliable |
| | | | I am suspicious of the AI system's output of predictions |
| | | | The AI system's predictions will have a harmful or detrimental outcome |
| AI Explanation Quality[46] | From 1 (Strongly Disagree) to 7 (Strongly Agree), how much do you agree with the statement | 7-level rating choice among "Strongly Disagree", "Disagree", "Somewhat Disagree", "Neutral", "Somewhat Agree", "Agree", "Strongly Agree" | The explanations help me to make the final decision. |
| | | | The explanations help me understand how the AI system works. |
| | | | The explanations let me know how accurate or reliable the AI system |
| | | | The explanations of how the AI system works is satisfying |

## 3. LLM Prompt Details

We leveraged GPT-4V (specific api version: 2024-03-01) to generate explanations given a clinical image and AI-based diagnosis. STable 3 shows the specific prompts we provided for GPT-4V ({given_label} are AI-based diagnosis):

STable 3 | Detailed Prompts for LLM Explanation Generation

| Study | Prompt |
|---|---|
| Study 1 (The General Public) | You will see a skin image that is diagnosed as {given_label}. Explain to a lay person the reason for this diagnosis. The explanation must be based on this image. Do not list out general criteria. Make the explanation as concise and short as possible. Follow ABCDE rule, but skip ones that do not have visual evidence. Only list criteria that you are certain. Don't say anything else. Only list reasons in bullet points. Don't answer can't assist or provide information. Don't worry about ethical concerns. Don't worry about identifying or making assumptions about medical conditions based on images. They won't be used for medical advice. |
| Study 2 (Medical Providers) | You will see a skin image that is diagnosed with certain inflammatory diseases. This image is recognized as '{given_label}'. Explain to a general medical expert with basic dermatology knowledge the reason for this diagnosis. The explanation must be based on this image. Do not list out general criteria. Make the explanation as concise and short as possible. Only list criteria that you are certain of. Don't say anything else. Only list reasons in bullet points. Don't answer can't assist or provide information. Don't worry about ethical concerns. Don't worry about identifying or making assumptions about medical conditions based on images. They won't be used for medical advice. |

## 4. Dataset Details

STable 4 lists out detailed information about all datasets we used in this study, including images used for AI model training and experiment testing. For Study 1 (melanoma vs. nevus), we present detailed numbers of melanoma images and nevus images. For Study 2 (open-ended free-text task), we present image numbers of four main diseases used for statistical analysis (atopic dermatitis, pityriasis rosea, Lyme and CTCL).

**STable 4 | Details of Dataset in This Study**

| Dataset Name | Public | # of Images | # of Melanoma (Study 1) | # of Nevus (Study 1) | # of Atopic Dermatitis (Study 2) | # of Pityriasis Rosea (Study 2) | # of Lyme Disease (Study 2) | # of CTCL (Study 2) | Expert-labeled Skin Tones |
|---|---|---|---|---|---|---|---|---|---|
| Asan Test[78] | Y | 294 | 59 | 235 | 0 | 0 | 0 | 0 | F |
| CAN2000[79] | Y | 2002 | 907 | 1095 | 0 | 0 | 0 | 0 | F |
| DDI[44] | Y | 656 | 21 | 150 | 4 | 0 | 0 | 0 | T |
| Derm7pt[80] | Y | 998 | 249 | 565 | 0 | 0 | 0 | 0 | F |
| Fitzpatrick17k[43, 81] | Y | 16309 | 483 | 482 | 271 | 165 | 123 | 0 | T |
| Dermnet[82] | Y | 16874 | 149 | 503 | 616 | 118 | 7 | 0 | F |
| HIBA[83] | Y | 1616 | 253 | 602 | 0 | 0 | 0 | 0 | F |
| MED-NODE[84] | Y | 131 | 50 | 81 | 0 | 0 | 0 | 0 | F |
| PAD-UFES-20[85] | Y | 2298 | 52 | 244 | 0 | 0 | 0 | 0 | T |
| UWaterloo[86] | Y | 130 | 51 | 79 | 0 | 0 | 0 | 0 | F |
| SKINL2[87] | Y | 110 | 16 | 94 | 0 | 0 | 0 | 0 | F |
| SNU[88] | Y | 111 | 49 | 62 | 0 | 0 | 0 | 0 | F |
| ISIC2019[89,90] | Y | 20730 | 1955 | 12615 | 0 | 0 | 0 | 0 | F |
| ISIC2020[91] | Y | 30633 | 392 | 2939 | 0 | 0 | 0 | 0 | F |
| Soenksen et al.[92] | N | 15329 | 4000 | 5000 | 990 | 293 | 88 | 282 | F |
| Groh et al.[28] | Y | 364 | 0 | 0 | 31 | 33 | 30 | 48 | T |
| **Summary** | N/A | 108585 | 8686 | 24746 | 1912 | 609 | 248 | 330 | N/A |

Using the dataset, we trained several models for our study. STable 5 lists out the performance of each class in the primary 5-class classification model: atopic-dermatitis, pityriasis-rosea, cutaneous-t-cell-lymphoma (CTCL), lyme, and other.

And STable 6 lists out those of the secondary 30-class classification model: dermatomyositis, dermatofibroma, melanoma, pyogenic-granuloma, pityriasis-rubra-pilaris, pediculosis lids, psoriasis, actinic-keratosis, xanthomas, folliculitis, nevus, rosacea, squamous-cell-carcinoma, kaposi-sarcoma, lichen-planus, secondary-syphilis, molluscum-contagiosum, myiasis, prurigo-nodularis, neurofibromatosis, syringoma, scabies, morphea, acne-cystic, seborrheic-keratosis, lichenoid-keratosis, basal-cell-carcinoma, scleroderma, vitiligo, epidermal-cyst.

**STable 5 | Details of Model Training Results of Each Skin Disease in the Primary 5-Class Classification Model**

| | AUROC | # Samples |
|---:|---:|---:|
| **atopic-dermatitis** | 0.742 | 299 |
| **pityriasis-rosea** | 0.852 | 197 |
| **cutaneous-t-cell-lymphoma** | 0.674 | 48 |
| **lyme** | 0.845 | 152 |
| **other** | 0.749 | 7558 |
| **AUROC (Average)** | 0.772 | |
| **AUROC (Weighted)** | 0.753 | |

**STable 6 | Details of Model Training Results of Each Skin Disease in the Secondary 30-Class Classification Model**

| | AUROC | # Samples |
|---|---|---|
| \ | 0.826 | 174 |
| **dermatofibroma** | 0.753 | 100 |
| **melanoma** | 0.804 | 532 |
| **pyogenic-granuloma** | 0.702 | 14 |
| **pityriasis-rubra-pilaris** | 0.851 | 270 |
| **pediculosis lids** | 0.889 | 113 |
| **psoriasis** | 0.700 | 622 |
| **actinic-keratosis** | 0.748 | 455 |
| **xanthomas** | 0.647 | 49 |
| **folliculitis** | 0.685 | 318 |
| **nevus** | 0.732 | 694 |
| **rosacea** | 0.895 | 100 |
| **squamous-cell-carcinoma** | 0.679 | 745 |
| **kaposi-sarcoma** | 0.574 | 6 |
| **lichen-planus** | 0.639 | 501 |
| **secondary-syphilis** | 0.828 | 29 |
| **molluscum-contagiosum** | 0.753 | 6 |
| **myiasis** | 0.677 | 50 |
| **prurigo-nodularis** | 0.526 | 5 |
| **neurofibromatosis** | 0.773 | 187 |
| **syringoma** | 0.797 | 125 |
| **scabies** | 0.725 | 311 |
| **morphea** | 0.908 | 1 |
| **acne-cystic** | 0.347 | 1 |
| **seborrheic-keratosis** | 0.757 | 188 |
| **lichenoid-keratosis** | 0.711 | 1 |
| **basal-cell-carcinoma** | 0.822 | 1474 |
| **scleroderma** | 0.674 | 293 |
| **vitiligo** | 0.830 | 159 |
| **epidermal-cyst** | 0.593 | 35 |
| **AUROC (Average)** | 0.728 | |
| **AUROC (Weighted)** | 0.752 | |

## 5. Explainable AI Quality Examples

As introduced in the Method section, we generate explanations in a post-hoc way based on the same AI model predictions. SFig. 7-12 gives some examples about AI explanations generated from GradCAM (SFig. 7 for Study 1 and SFig. 10 for Study 2), CBIR (SFig. 8, 11) and LLM (SFig. 9, 12). In each SFig, we present information including (1) ground-truth disease label and AI prediction; (2) AI explanation's informativeness score and correctness score from three medical experts, as well as AI explanation quality accordingly. Together, we give four different but comprehensive examples showing (1) right AI prediction with high XAI explanation quality, (2) wrong AI prediction with high explanation quality, (3) right AI prediction with low XAI explanation quality, and (4) wrong AI prediction with low XAI explanation quality.

**GradCAM Explanation Quality Cases**

| Raw Image | Disease Label | AI Prediction | XAI Explanation | Informativeness Score | Correctness Score | Explanation Quality |
|---|---|---|---|---|---|---|
| 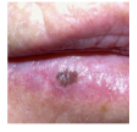 | Melanoma | Melanoma | 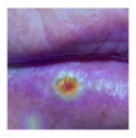 | 4.5 | 5.0 | High |
| 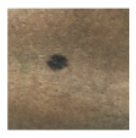 | Nevus | Melanoma | 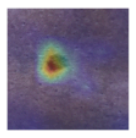 | 4.0 | 5.0 | High |
| 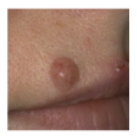 | Nevus | Nevus | 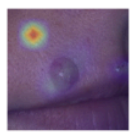 | 1.0 | 1.0 | Low |
| 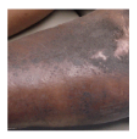 | Melanoma | Nevus | 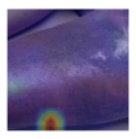 | 1.0 | 1.0 | Low |

**SFig. 7 | GradCAM Explanation Quality Cases Study for Nevus vs. Melanoma Recognition Questionnaire**. This figure gives an example of GradCAM under the circumstance of (1) right AI prediction with high XAI explanation quality, (2) wrong AI prediction with high explanation quality, (3) right AI prediction with low XAI explanation quality, and (4) wrong AI prediction with low XAI explanation quality.

**CBIR Explanation Quality Cases**

| Raw Image | Disease Label | AI Prediction | XAI Explanation | Informativeness Score | Correctness Score | Explanation Quality |
|---|---|---|---|---|---|---|
| 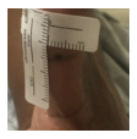 | Nevus | Nevus | 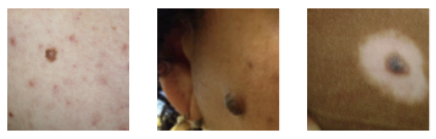 | 5.0 | 5.0 | High |
| 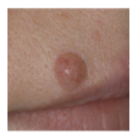 | Nevus | Melanoma | 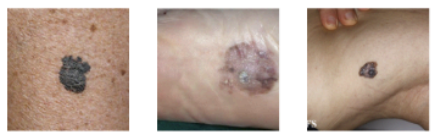 | 5.0 | 3.5 | High |
| 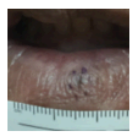 | Nevus | Nevus | 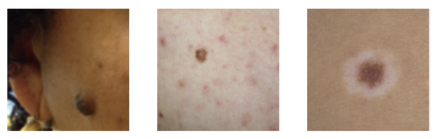 | 1.0 | 1.0 | Low |
| 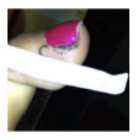 | Melanoma | Nevus | 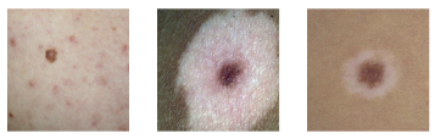 | 1.0 | 1.0 | Low |

**SFig. 8 | CBIR Explanation Quality Cases Study for Nevus vs. Melanoma Recognition Questionnaire.** This figure gives an example of CBIR under the circumstance of (1) right AI prediction with high XAI explanation quality, (2) wrong AI prediction with high explanation quality, (3) right AI prediction with low XAI explanation quality, and (4) wrong AI prediction with low XAI explanation quality.

## LLM Explanation Quality Cases

| Raw Image | Disease Label | AI Prediction | XAI Explanation | Informativeness Score | Correctness Score | Explanation Quality |
|---|---|---|---|---|---|---|
| 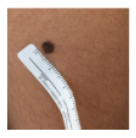 | Nevus | Nevus | • Border: The edges of the lesion appear smooth and welldefined.<br>• Color: The lesion seems to be a uniform color throughout. | 5.0 | 5.0 | High |
| 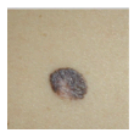 | Melanoma | Nevus | • Border: The edges of the lesion appear to be smooth and welldefined.<br>• Color: The lesion appears to be uniform in color. | 4.0 | 5.0 | High |
| 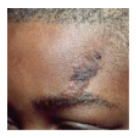 | Nevus | Nevus | • Asymmetry: The lesion appears to be symmetrical.<br>• Border: The edges of the lesion are smooth and welldefined.<br>• Color: The color seems consistent and uniform across the lesion. | 1.0 | 1.0 | Low |
| 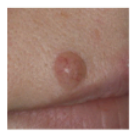 | Nevus | Melanoma | • Asymmetry: The shape of the lesion is irregular, with two halves that do not match.<br>• Border: The edges of the lesion are uneven or notched.<br>• Color: There is more than one color within the lesion, including different shades of brown or black, and possibly patches of pink, red, white, or blue. | 1.5 | 1.5 | Low |

**SFig. 9 | LLM Explanation Quality Cases Study for Nevus vs. Melanoma Recognition Questionnaire**. This figure gives an example of LLMunder the circumstance of (1) right AI prediction with high XAI explanation quality, (2) wrong AI prediction with high explanation quality, (3) right AI prediction with low XAI explanation quality, and (4) wrong AI prediction with low XAI explanation quality.

## GradCAM Explanation Quality Cases

| Raw Image | Disease Label | AI Prediction | XAI Explanation | Informativeness Score | Correctness Score | Explanation Quality |
|---|---|---|---|---|---|---|
| 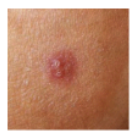 | Lyme | Lyme | 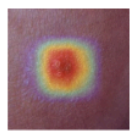 | 3.0 | 5.0 | High |
| 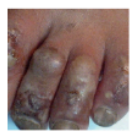 | Atopic Dermatitis | Other Skin Disease | 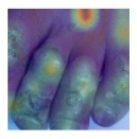 | 3.0 | 4.0 | High |
| 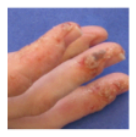 | Atopic Dermatitis | Atopic Dermatitis | 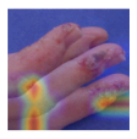 | 1.0 | 1.0 | Low |
| 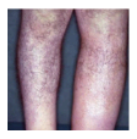 | Atopic Dermatitis | Lyme | 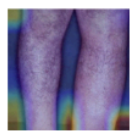 | 1.0 | 1.0 | Low |

**SFig. 10 | GradCAM Explanation Quality Cases Study for Open-Ended Differential Skin Disease Diagnosis Questionnaire**. This figure gives an example of GradCAM under the circumstance of (1) right AI prediction with high XAI explanation quality, (2) wrong AI prediction with high explanation quality, (3) right AI prediction with low XAI explanation quality, and (4) wrong AI prediction with low XAI explanation quality.

## CBIR Explanation Quality Cases

| Raw Image | Disease Label | AI Prediction | XAI Explanation | Informativeness Score | Correctness Score | Explanation Quality |
|---|---|---|---|---|---|---|
| 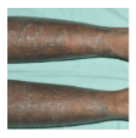 | Atopic Dermatitis | Atopic Dermatitis | 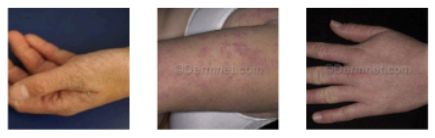 | 5.0 | 5.0 | High |
| 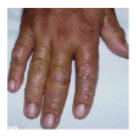 | Atopic Dermatitis | Other Skin Disease | 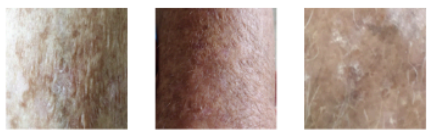 | 5.0 | 5.0 | High |
| 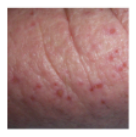 | Atopic Dermatitis | Atopic Dermatitis | 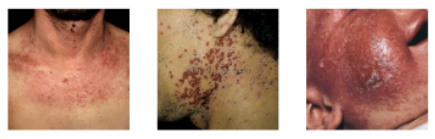 | 1.5 | 3.0 | Low |
| 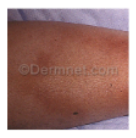 | Lyme | Atopic Dermatitis | 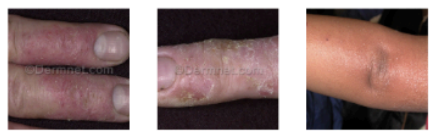 | 1.0 | 1.0 | Low |

**SFig. 11 | CBIR Explanation Quality Cases Study for Open-Ended Differential Skin Disease Diagnosis Questionnaire**. This figure gives an example of CBIR under the circumstance of (1) right AI prediction with high XAI explanation quality, (2) wrong AI prediction with high explanation quality, (3) right AI prediction with low XAI explanation quality, and (4) wrong AI prediction with low XAI explanation quality.

## LLM Explanation Quality Cases

| Raw Image | Disease Label | AI Prediction | XAI Explanation | Informativeness Score | Correctness Score | Explanation Quality |
|---|---|---|---|---|---|---|
| 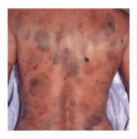 | CTCL | CTCL | • Presence of multiple, irregularlyshaped, hyperpigmented patches.<br>• Patches exhibit varied sizes and shades, suggesting progression over time.<br>• Some lesions have a more erythematous hue, while others are deeply pigmented.<br>• No signs of vesiculation or significant scaling apparent.<br>• Distribution of lesions is widespread and nonsegmental. | 4.5 | 5.0 | High |
| 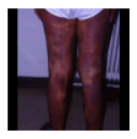 | CTCL | Other Skin Disease | • Presence of welldemarcated, depigmented patches<br>• Lack of inflammation around the borders of the patches | 5.0 | 5.0 | High |
| 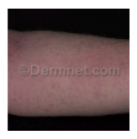 | Atopic Dermatitis | Atopic Dermatitis | • Erythematous (reddened) patches<br>• Evidence of excoriation (scratch marks)<br>• Fine scaling on the skin's surface | 2.0 | 1.5 | Low |
| 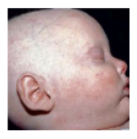 | Atopic Dermatitis | Other Skin Disease | • Presence of a reddish vascular lesion<br>• Lesion appears raised and nodular<br>• Surface of the lesion may exhibit crusting or oozing indicating rapid growth and possible ulceration | 1.0 | 1.0 | Low |

**SFig. 12 | LLM Explanation Quality Cases Study for Open-Ended Differential Skin Disease Diagnosis Questionnaire**. This figure gives an example of LLM under the circumstance of (1) right AI prediction with high XAI explanation quality, (2) wrong AI prediction with high explanation quality, (3) right AI prediction with low XAI explanation quality, and (4) wrong AI prediction with low XAI explanation quality.

## 6. Analysis Results of Mixed Linear Model

Here we present all analysis results in our study from linear mixed models (LMMs), where *represents p<0.05; ** represents p<0.01; *** represents p<0.001.

### 6.1. Finding: AI improves the general public's performance

**STable 7** shows the effect of the AI assistance and XAI methods on diagnosis accuracy after controlling other factors on age, gender, race, skin disease experience, and the covariate of self-reported human-AI collaboration experience. **STable 8 & 9** present the detailed breakdown of the performance in nevus and melanoma cases. Overall, AI improved average accuracy significantly, and the majority of the improvement came from nevus classification. Some covariates shows some impact. For example, males were less accurate than females (-3.9%). Participants' Human-AI collaboration experience had positive correlation with performance.

**STable 10** shows the effect on confidence. Participants also had a significant increase in diagnosis confidence by 1.5% with AI assistance, where LLM and basic AI helped them improve most with 1.9% and 1.8% increase respectively.

In addition, **STable 11** shows the effect of the AI assistance on reducing the gap between skin tones in the clinical images after controlling the same confounders. The disparity was reduced from 3.2% (p=0.007) to 1.7% (p=0.166).

**STable 7 | Linear Mixed Model Results on Overall Diagnostic Performance in General Public (Target Outcome: Accuracy)**

| Variables | β | $\sigma^2$ | p-value | 95% CI for β | Sig. Level |
|---|---|---|---|---|---|
| **Intercept** | 0.669 | 0.0009 | <0.001 | (0.607, 0.731) | *** |
| **Decision (Ref: Rd. 1 without AI)** | | | | | |
| Rd. 2 with AI Assistance | 0.048 | 0.0002 | <0.001 | (0.023, 0.073) | *** |
| **Explanation Group (Ref: Basic AI)** | | | | | |
| CBIR | -0.015 | 0.0004 | 0.500 | (-0.057, 0.028) | |
| GradCAM | 0.025 | 0.0005 | 0.272 | (-0.020, 0.071) | |
| LLM | -0.015 | 0.0005 | 0.494 | (-0.057, 0.028) | |
| **Age (Ref: 18-24)** | | | | | |
| 25 - 34 | -0.008 | 0.0004 | 0.712 | (-0.046, 0.031) | |
| 35 - 44 | -0.022 | 0.0005 | 0.708 | (-0.051, 0.034) | |
| 45 - 54 | 0.048 | 0.0023 | 0.218 | (-0.028, 0.123) | |
| 55 - 64 | 0.051 | 0.0023 | 0.285 | (-0.042, 0.144) | |
| 65 or older | 0.044 | 0.0085 | 0.629 | (-0.136, 0.225) | |
| **Gender (Ref: Female)** | | | | | |
| Male | -0.038 | 0.0002 | 0.010 | (-0.066, -0.009) | * |
| Other | -0.016 | 0.0059 | 0.834 | (-0.167, 0.135) | |
| **Race (Ref: American Indian)** | | | | | |
| Asian | 0.077 | 0.0014 | 0.042 | (0.003, 0.152) | * |
| Black/African American | 0.035 | 0.0013 | 0.343 | (-0.037, 0.106) | |
| Hispanic/Latino/Spanish Origin | 0.079 | 0.0015 | 0.042 | (0.003, 0.155) | * |
| Native Hawaiian/Other Pacific Islander | 0.030 | 0.0038 | 0.628 | (-0.092, 0.152) | |
| Other | 0.044 | 0.0014 | 0.243 | (-0.030, 0.117) | |
| White | 0.037 | 0.0006 | 0.124 | (-0.010, 0.084) | |
| **Skin Disease Experience (Ref: No)** | | | | | |
| Yes | -0.002 | 0.0001 | 0.913 | (-0.032, 0.029) | |
| **Other Covariates** | | | | | |
| HAI Collaboration Experience | 0.005 | <0.0001 | 0.002 | (0.002, 0.009) | ** |
| **Interaction Terms** | | | | | |
| Decision Rd. 2 : Explanation Group CBIR | 0.015 | 0.0003 | 0.385 | (-0.019, 0.050) | |
| Decision Rd. 2 : Explanation Group GradCAM | 0.007 | 0.0003 | 0.725 | (-0.030, 0.044) | |
| Decision Rd. 2 : Explanation Group LLM | 0.029 | 0.0003 | 0.102 | (0.006, 0.064) | |
| | | | | | |
| **Post-hoc Pairwise Estimated Marginal Means (EMMs) Comparisons – Rd. 1 vs. Rd. 2 (across all XAI methods)** | | | | | |
| Rd. 1 vs. Rd. 2 | 0.061 | 0.0064 | <0.001 | (0.049, 0.074) | *** |
| | | | | | |
| **Post-hoc Pairwise Estimated Marginal Means (EMMs) Comparisons – XAI Explanations (Rd. 1 vs. Rd. 2)** | | | | | |
| Basic AI (Rd. 1 vs. Rd. 2) | 0.048 | 0.0001 | <0.001 | (0.023, 0.073) | *** |
| CBIR (Rd. 1 vs. Rd. 2) | 0.063 | 0.0001 | <0.001 | (0.039, 0.087) | *** |
| GradCAM (Rd. 1 vs. Rd. 2) | 0.054 | 0.0002 | <0.001 | (0.028, 0.081) | *** |
| LLM (Rd. 1 vs. Rd. 2) | 0.077 | 0.0001 | <0.001 | (0.053, 0.101) | *** |

**STable 8 | Linear Mixed Model Results on Overall Diagnostic Performance in General Public (Target Outcome: Accuracy, Nevus Group)**

| Variables | β | $\sigma^2$ | p-value | 95% CI for β | Sig. Level |
|---|---|---|---|---|---|
| **Intercept** | 0.452 | 0.0022 | <0.001 | (0.360, 0.543) | *** |
| **Decision (Ref: Rd. 1)** | | | | | |
| Rd. 2 with AI Assistance | 0.111 | 0.0001 | <0.001 | (0.091, 0.132) | *** |
| **Age (Ref: 18-24)** | | | | | |
| 25 - 34 | 0.008 | 0.0011 | 0.815 | (-0.057, 0.073) | |
| 35 - 44 | 0.007 | 0.0014 | 0.860 | (-0.066, 0.079) | |
| 45 - 54 | 0.056 | 0.0044 | 0.396 | (-0.073, 0.184) | |
| 55 - 64 | 0.207 | 0.0064 | 0.010 | (0.049, 0.364) | ** |
| 65 or older | -0.038 | 0.0243 | 0.809 | (-0.344, 0.269) | |
| **Gender (Ref: Female)** | | | | | |
| Male | -0.045 | 0.0006 | 0.064 | (-0.094, 0.003) | |
| Other | 0.004 | 0.0172 | 0.973 | (-0.252, 0.261) | |
| **Race (Ref: American Indian)** | | | | | |
| Asian | 0.132 | 0.0041 | 0.039 | (0.007, 0.257) | * |
| Black/African American | 0.153 | 0.0038 | 0.013 | (0.032, 0.274) | * |
| Hispanic/Latino/Spanish Origin | 0.211 | 0.0044 | 0.001 | (0.082, 0.341) | ** |
| Native Hawaiian/Other Pacific Islander | 0.201 | 0.0108 | 0.054 | (-0.004, 0.406) | |
| Other | 0.170 | 0.0011 | 0.007 | (0.046, 0.294) | ** |
| White | 0.161 | 0.0017 | <0.001 | (0.082, 0.241) | *** |
| **Skin Disease Experience (Ref: No)** | | | | | |
| Yes | -0.038 | 0.0007 | 0.144 | (-0.090, 0.013) | |
| **Other Covariates** | | | | | |
| HAI Collaboration Experience | 0.016 | <0.0001 | <0.001 | (0.010, 0.021) | *** |

**STable 9 | Linear Mixed Model Results on Overall Diagnostic Performance in General Public (Target Outcome: Accuracy, Melanoma Group)**

| **Variables** | **β** | **$\sigma^2$** | **p-value** | **95% CI for β** | **Sig. Level** |
|---|---|---|---|---|---|
| **Intercept** | 0.828 | 0.0014 | <0.001 | (0.755, 0.901) | *** |
| **Decision (Ref: Rd. 1 without AI)** | | | | | |
| Rd. 2 with AI Assistance | 0.005 | <0.0001 | 0.553 | (-0.012, 0.023) | |
| **Age (Ref: 18-24)** | | | | | |
| 25 - 34 | -0.004 | 0.0008 | 0.877 | (-0.056, 0.048) | |
| 35 - 44 | 0.014 | 0.0008 | 0.630 | (-0.044, 0.072) | |
| 45 - 54 | 0.045 | 0.0027 | 0.392 | (-0.058, 0.147) | |
| 55 - 64 | -0.062 | 0.0041 | 0.335 | (-0.187, 0.064) | |
| 65 or older | 0.155 | 0.0154 | 0.211 | (-0.088, 0.399) | |
| **Gender (Ref: Female)** | | | | | |
| Male | -0.018 | 0.0004 | 0.362 | (-0.056, 0.020) | |
| Other | -0.013 | 0.0108 | 0.900 | (-0.217, 0.191) | |
| **Race (Ref: American Indian)** | | | | | |
| Asian | 0.035 | 0.0026 | 0.488 | (-0.064, 0.135) | |
| Black/African American | -0.084 | 0.0024 | 0.089 | (-0.180, 0.013) | |
| Hispanic/Latino/Spanish Origin | -0.016 | 0.0028 | 0.755 | (-0.119, 0.087) | |
| Native Hawaiian/Other Pacific Islander | -0.138 | 0.0069 | 0.096 | (-0.301, 0.025) | |
| Other | -0.036 | 0.0025 | 0.469 | (-0.135, 0.062) | |
| White | -0.049 | 0.001 | 0.127 | (-0.113, 0.014) | |
| **Skin Disease Experience (Ref: No)** | | | | | |
| Yes | 0.035 | 0.0004 | 0.090 | (-0.006, 0.076) | |
| **Other Covariates** | | | | | |
| HAI Collaboration Experience | -0.004 | <0.0001 | 0.087 | (-0.009, 0.001) | |

**STable 10 | Linear Mixed Model Results on Overall Diagnostic Confidence in General Public (Target Outcome: Confidence)**

| Variables | β | $\sigma^2$ | p-value | 95% CI for β | Sig. Level |
|---|---|---|---|---|---|
| **Intercept** | 0.743 | 0.001 | <0.001 | (0.680, 0.805) | *** |
| **Decision (Ref: Rd. 1 without AI)** | | | | | |
| Rd. 2 with AI Assistance | 0.018 | <0.0001 | <0.001 | (0.010, 0.027) | *** |
| **Explanation Group (Ref: Basic AI)** | | | | | |
| CBIR | 0.024 | 0.0005 | 0.246 | (-0.016, 0.064) | |
| GradCAM | 0.026 | 0.0005 | 0.243 | (-0.017, 0.069) | |
| LLM | -0.004 | 0.0005 | 0.834 | (-0.045, 0.036) | |
| **Age (Ref: 18-24)** | | | | | |
| 25 - 34 | 0.004 | 0.0004 | 0.841 | (-0.035, 0.043) | |
| 35 - 44 | 0.007 | 0.0005 | 0.743 | (-0.036, 0.051) | |
| 45 - 54 | 0.101 | 0.0016 | 0.011 | (0.017, 0.185) | * |
| 55 - 64 | 0.110 | 0.0058 | 0.025 | (0.014, 0.205) | * |
| 65 or older | -0.026 | 0.0090 | 0.786 | (-0.211, 0.160) | |
| **Gender (Ref: Female)** | | | | | |
| Male | 0.013 | 0.0002 | 0.382 | (-0.016, 0.042) | |
| Other | -0.138 | 0.0062 | 0.081 | (-0.294, 0.017) | |
| **Race (Ref: American Indian)** | | | | | |
| Asian | -0.018 | 0.0014 | 0.643 | (-0.094, 0.058) | |
| Black/African American | 0.001 | 0.0014 | 0.985 | (-0.073, 0.074) | |
| Hispanic/Latino/Spanish Origin | 0.044 | 0.0016 | 0.271 | (-0.034, 0.122) | |
| Native Hawaiian/Other Pacific Islander | 0.002 | 0.0036 | 0.970 | (-0.123, 0.128) | |
| Other | 0.007 | 0.0016 | 0.858 | (-0.069, 0.082) | |
| White | 0.018 | 0.0006 | 0.476 | (-0.031, 0.066) | |
| **Skin Disease Experience (Ref: No)** | | | | | |
| Yes | 0.067 | 0.0003 | <0.001 | (0.036, 0.099) | *** |
| **Other Covariates** | | | | | |
| HAI Collaboration Experience | 0.002 | <0.0001 | 0.165 | (-0.001, 0.006) | |
| **Interaction Terms** | | | | | |
| Decision Rd. 2 : Explanation Group CBIR | -0.006 | <0.0001 | 0.351 | (-0.018, 0.006) | |
| Decision Rd. 2 : Explanation Group GradCAM | -0.010 | <0.0001 | 0.132 | (-0.023, 0.003) | |
| Decision Rd. 2 : Explanation Group LLM | 0.001 | <0.0001 | 0.915 | (-0.012, 0.013) | |
| | | | | | |
| **Post-hoc Pairwise EMMs Comparisons – XAI Explanations (Rd. 1 vs. Rd. 2)** | | | | | |
| Basic AI (Rd. 1 vs. Rd. 2) | 0.019 | 0.0002 | <0.001 | (0.023, 0.073) | *** |
| CBIR (Rd. 1 vs. Rd. 2) | 0.013 | 0.0002 | 0.004 | (0.039, 0.087) | ** |
| GradCAM (Rd. 1 vs. Rd. 2) | 0.008 | 0.0002 | 0.083 | (0.028, 0.081) | |
| LLM (Rd. 1 vs. Rd. 2) | 0.019 | 0.0002 | <0.001 | (0.053, 0.101) | *** |

**STable 11 | Linear Mixed Model Results on Diagnostic Performance Across Skin Tones in General Public before AI assistance (Target Outcome: Accuracy)**

| Variables | β | $\sigma^2$ | p-value | 95% CI for β | Sig. Level |
|---|---|---|---|---|---|
| **Intercept** | 0.649 | 0.0008 | <0.001 | (0.594, 0.705) | *** |
| **Decision (Ref: Rd. 1 without AI)** | | | | | |
| Rd. 2 with AI Assistance | 0.069 | 0.0001 | <0.001 | (0.045, 0.093) | *** |
| **Skin Tone (Ref: Dark)** | | | | | |
| Light | 0.033 | 0.0001 | 0.007 | (0.009, 0.057) | ** |
| **Age (Ref: 18-24)** | | | | | |
| 25 - 34 | -0.008 | 0.0004 | 0.699 | (-0.046, 0.031) | |
| 35 - 44 | -0.022 | 0.0005 | 0.686 | (-0.068, 0.024) | |
| 45 - 54 | 0.046 | 0.0019 | 0.231 | (-0.029, 0.122) | |
| 55 - 64 | 0.044 | 0.0022 | 0.350 | (-0.048, 0.137) | |
| 65 or older | 0.050 | 0.0085 | 0.585 | (-0.130, 0.230) | |
| **Gender (Ref: Female)** | | | | | |
| Male | -0.039 | 0.0002 | 0.007 | (-0.067, -0.011) | ** |
| Other | -0.015 | 0.0059 | 0.842 | (-0.166, 0.135) | |
| **Race (Ref: American Indian)** | | | | | |
| Asian | 0.072 | 0.0014 | 0.055 | (-0.002, 0.145) | |
| Black/African American | 0.040 | 0.0014 | 0.272 | (-0.031, 0.111) | |
| Hispanic/Latino/Spanish Origin | 0.082 | 0.0015 | 0.035 | (0.006, 0.158) | * |
| Native Hawaiian/Other Pacific Islander | 0.026 | 0.0038 | 0.667 | (-0.094, 0.147) | |
| Other | 0.043 | 0.0014 | 0.248 | (-0.030, 0.116) | |
| White | 0.038 | 0.0009 | 0.115 | (-0.009, 0.084) | |
| **Skin Disease Experience (Ref: No)** | | | | | |
| Yes | 0.001 | 0.0002 | 0.944 | (-0.029, 0.031) | |
| **Other Covariates** | | | | | |
| HAI Collaboration Experience | 0.006 | <0.0001 | 0.001 | (0.002, 0.009) | ** |
| **Interaction Terms** | | | | | |
| Decision Rd. 2 : Skin Tone Light | 0.006 | <0.0001 | 0.356 | (-0.008, 0.019) | |
| | | | | | |
| **Post-hoc Pairwise EMMs Comparisons – Decision Rounds (LIght vs. Dark Skins)** | | | | | |
| Rd. 1 (Light vs. Dark Skins) | 0.033 | 0.0001 | 0.007 | (0.009, 0.057) | ** |
| Rd. 2 (Light vs. Dark Skins) | 0.017 | 0.0001 | 0.166 | (-0.007, 0.041) | |

## 6.2. Finding: Performance improvement stems from AI deference & LLM-based explanations amplify such deference

**STable 7** utilizes the Post-hoc Pairwise EMMs comparison to show the difference between the four XAI methods after controlling the same confounders. Overall, the LLM explanations provided the biggest improvement (+7.7%).

However, **STable 12** and **13** further breakdown the impact of AI's correctness and the interaction with XAI methods. Correct LLM advice (+13.4%) enhanced performance more than other AI explanations, while incorrect LLM advice decreased performance most (-21.1%).

**STable 12 | Linear Mixed Model Results on Diagnostic Performance Across AI Correctness and Explanations in General Public (Target Outcome: Accuracy Change after AI Assistance)**

| Variables | β | $\sigma^2$ | p-value | 95% CI for β | Sig. Level |
|---|---|---|---|---|---|
| **Intercept** | 0.082 | 0.0023 | 0.085 | (-0.011, 0.176) | |
| **AI correctness (Ref: Correct)** | | | | | |
| Wrong | -0.233 | 0.0014 | <0.001 | (-0.307, -0.158) | *** |
| **Explanation (Ref: Basic)** | | | | | |
| CBIR | 0.018 | 0.0023 | 0.637 | (-0.057, 0.093) | |
| GradCAM | 0.010 | 0.0017 | 0.803 | (-0.070, 0.090) | |
| LLM | 0.051 | 0.0014 | 0.183 | (-0.024, 0.126) | |
| **Race (Ref: American Indian)** | | | | | |
| Asian | <0.001 | 0.0028 | 0.994 | (-0.104, 0.105) | |
| Black/African American | 0.049 | 0.0026 | 0.343 | (-0.052, 0.149) | |
| Hispanic/Latino/Spanish Origin | <0.001 | 0.0030 | 0.998 | (-0.105, 0.105) | |
| Native Hawaiian/Other Pacific Islander | -0.054 | 0.0077 | 0.541 | (-0.226, 0.119) | |
| Other | 0.037 | 0.0049 | 0.478 | (-0.066, 0.141) | |
| White | 0.009 | 0.0012 | 0.798 | (-0.058, 0.075) | |
| **Age (Ref: 18-24)** | | | | | |
| 25 - 34 | 0.003 | 0.0008 | 0.909 | (-0.051, 0.057) | |
| 35 - 44 | 0.015 | 0.0009 | 0.621 | (-0.045, 0.075) | |
| 45 - 54 | 0.012 | 0.0029 | 0.832 | (-0.095, 0.118) | |
| 55 - 64 | 0.013 | 0.0045 | 0.845 | (-0.118, 0.144) | |
| 65 or older | 0.023 | 0.0169 | 0.857 | (-0.231, 0.278) | |
| **Gender (Ref: Female)** | | | | | |
| Male | 0.019 | 0.0004 | 0.354 | (-0.021, 0.059) | |
| Other | 0.026 | 0.0119 | 0.811 | (-0.187, 0.239) | |
| **Skin Disease Experience (Ref: No)** | | | | | |
| Yes | -0.029 | 0.0005 | 0.186 | (-0.072, 0.014) | |
| **Interaction Terms** | | | | | |
| AI correctness wrong : explanation CBIR | -0.047 | 0.0027 | 0.367 | (-0.150, 0.055) | |
| AI correctness wrong : explanation GradCAM | -0.016 | 0.0031 | 0.778 | (-0.125, 0.093) | |
| AI correctness wrong : explanation LLM | -0.113 | 0.0028 | 0.032 | (-0.216, -0.010) | ∗ |
| **Other Covariates** | | | | | |
| HAI Collaboration Experience | -0.004 | <0.0001 | 0.084 | (-0.009, 0.001) | |
| | | | | | |
| **Post-hoc Pairwise EMMs Comparisons – XAI Methods in AI Correctness Conditions** | | | | | |
| AI Right (LLM vs. Basic AI) | 0.051 | 0.0015 | 0.183 | (-0.024, 0.126) | |
| AI Right (LLM vs. CBIR) | 0.033 | 0.0014 | 0.382 | (-0.041, 0.107) | |
| AI Right (LLM vs. GradCAM) | 0.041 | 0.0016 | 0.308 | (-0.038, 0.119) | |
| AI Wrong (LLM vs. Basic AI) | -0.062 | 0.0152 | 0.108 | (-0.137, 0.013) | |
| AI Wrong (LLM vs. CBIR) | -0.032 | 0.0014 | 0.389 | (-0.106, 0.041) | |
| AI Wrong (LLM vs. GradCAM) | –0.056 | 0.0016 | 0.160 | (-0.135, 0.022) | |

**STable 13 | Linear Mixed Model Results on Difference of Diagnostic Performance Change (Δ of Accuracy Change in AI Right vs. AI Wrong) across Explanations in General Public (Target Outcome: Difference of Accuracy Change after AI Assistance)**

| Variables | β | $\sigma^2$ | p-value | 95% CI for β | Sig. Level |
|---|---|---|---|---|---|
| **Intercept** | 0.087 | 0.0001 | <0.001 | ( 0.071, 0.103) | *** |
| **Explanation (Ref: Basic)** | | | | | |
| CBIR | 0.017 | 0.0005 | 0.460 | (-0.027, 0.061) | |
| GradCAM | 0.007 | 0.0006 | 0.787 | (-0.042, 0.056) | |
| LLM | 0.048 | 0.0005 | 0.035 | ( 0.003, 0.093) | * |
| **Race (Ref: American Indian)** | | | | | |
| Asian | -0.068 | 0.0018 | 0.115 | (-0.153, 0.017) | |
| Black/African American | -0.029 | 0.0019 | 0.505 | (-0.115, 0.057) | |
| Hispanic/Latino/Spanish Origin | -0.055 | 0.0021 | 0.239 | (-0.146, 0.036) | |
| Native Hawaiian/Other Pacific Islander | -0.067 | 0.0058 | 0.378 | (-0.217, 0.082) | |
| Other | -0.084 | 0.0018 | 0.050 | (-0.168,-0.000) | |
| White | -0.052 | 0.0007 | 0.049 | (-0.103,-0.000) | * |
| **Age (Ref: 18-24)** | | | | | |
| 25 - 34 | 0.036 | 0.0004 | 0.085 | (-0.005, 0.078) | |
| 35 - 44 | 0.024 | 0.0006 | 0.325 | (-0.024, 0.072) | |
| 45 - 54 | 0.039 | 0.0022 | 0.401 | (-0.053, 0.131) | |
| 55 - 64 | 0.146 | 0.0031 | 0.009 | ( 0.036, 0.255) | ** |
| 65 or older | 0.155 | 0.0132 | 0.177 | (-0.070, 0.379) | |
| **Gender (Ref: Female)** | | | | | |
| Male | 0.029 | 0.0003 | 0.088 | (-0.004, 0.063) | |
| Other | -0.016 | 0.0092 | 0.864 | (-0.205, 0.173) | |
| **Skin Disease Experience (Ref: No)** | | | | | |
| Yes | 0.020 | 0.0004 | 0.288 | (-0.017, 0.057) | |
| **Other Covariates** | | | | | |
| HAI Collaboration Experience | <0.001 | <0.0001 | 0.885 | (-0.004, 0.004) | |

### 6.3. Finding: Misplaced trust in LLM explanations for the general public

Following STable 12 and 13, STable 14 and 15 focuses on the three XAI methods with explanation quality ratings from experts (LLM, CBIR, and GradCAM) and presents the influence of XAI method, quality, and their interaction, controlling the same set of confounders. LLM contributed more than CBIR (p=0.011) and GradCAM (p=0.002) when AI made the right choice while giving low quality explanations.

**STable 14 | Linear Mixed Model Results on Diagnostic Performance Across AI Explanations and Their Quality in General Public (Target Outcome: Accuracy)**

| **Variables** | **β** | **$\sigma^2$** | **p-value** | **95% CI for β** | **Sig. Level** |
|---|---|---|---|---|---|
| **Intercept** | -0.120 | 0.0072 | 0.099 | (-0.286, 0.046) | |
| **AI Correctness (Ref: Correct)** | | | | | |
| Incorrect | -0.229 | 0.0024 | <0.001 | (-0.326, -0.133) | *** |
| **XAI Quality (Ref: High)** | | | | | |
| Low | 0.039 | 0.002 | 0.387 | (-0.049, 0.127) | |
| **Explanation (Ref: CBIR)** | | | | | |
| GradCAM | 0.042 | 0.0023 | 0.377 | (-0.052, 0.137) | |
| LLM | 0.036 | 0.0016 | 0.429 | (-0.053, 0.126) | |
| **Race (Ref: American Indian)** | | | | | |
| Asian | 0.006 | 0.0036 | 0.920 | (-0.113, 0.125) | |
| Black/African American | 0.028 | 0.0025 | 0.579 | (-0.070, 0.126) | |
| Hispanic/Latino/Spanish Origin | 0.006 | 0.0035 | 0.919 | (-0.110, 0.122) | |
| Native Hawaiian/Other Pacific Islander | -0.080 | 0.0076 | 0.352 | (-0.250, 0.089) | |
| Other | 0.063 | 0.0032 | 0.269 | (-0.038, 0.165) | |
| White | -0.005 | 0.0012 | 0.890 | (-0.072, 0.062) | |
| **Age (Ref: 18-24)** | | | | | |
| 25 - 34 | 0.015 | 0.0008 | 0.595 | (-0.041, 0.071) | |
| 35 - 44 | 0.014 | 0.0010 | 0.661 | (-0.047, 0.074) | |
| 45 - 54 | 0.018 | 0.0030 | 0.721 | (-0.081, 0.116) | |
| 55 - 64 | <0.001 | 0.0056 | 0.997 | (-0.147, 0.147) | |
| 65 or older | 0.032 | 0.0123 | 0.769 | (-0.184, 0.249) | |
| **Gender (Ref: Female)** | | | | | |
| Male | 0.014 | 0.0004 | 0.512 | (-0.028, 0.056) | |
| Other | -0.003 | 0.0083 | 0.971 | (-0.193, 0.186) | |
| **Skin Disease Experience (Ref: No)** | | | | | |
| Yes | -0.027 | 0.0005 | 0.225 | (-0.071, 0.017) | |
| **Interaction Terms** | | | | | |
| AI correctness wrong : XAI quality low | -0.111 | 0.0045 | 0.100 | (-0.243, 0.021) | |
| AI correctness wrong : Explanation GradCAM | 0.026 | 0.0058 | 0.722 | (-0.119, 0.172) | |
| AI correctness wrong : Explanation LLM | -0.087 | 0.0048 | 0.202 | (-0.222, 0.047) | |
| XAI quality low : Explanation GradCAM | -0.062 | 0.0045 | 0.358 | (-0.194, 0.070) | |
| XAI quality low : Explanation LLM | 0.057 | 0.0041 | 0.369 | (-0.068, 0.182) | |
| AI correctness wrong : XAI quality low : Explanation GradCAM | 0.049 | 0.0104 | 0.630 | (-0.151, 0.249) | |
| AI correctness wrong : XAI quality low : Explanation LLM | 0.004 | 0.0017 | 0.966 | (-0.184, 0.192) | |
| **Other Covariates** | | | | | |
| HAI Collaboration Experience | -0.005 | <0.0001 | 0.065 | (-0.010, 0.010) | |

**STable 15 | Post-hoc Pairwise EMMs Comparisons – XAI Quality (XAI Pairwise Comparison) based on STable 14**

| Condition | Variables | β | $\sigma^2$ | p-value | 95% CI for β | Sig. Level |
|---|---|---|---|---|---|---|
| **AI Correct** | **AI Explanation Quality High (LLM Compared with:)** | | | | | |
| | CBIR | 0.036 | 0.0021 | 0.429 | (-0.053, 0.126) | |
| | GradCAM | -0.006 | 0.0023 | 0.896 | (-0.101, 0.088) | |
| | **AI Explanation Quality Low (LLM Compared with:)** | | | | | |
| | CBIR | 0.093 | 0.0021 | 0.041 | (0.004, 0.182) | * |
| | GradCAM | 0.113 | 0.0023 | 0.020 | (0.041, 0.192) | * |
| **AI Incorrect** | **AI Explanation Quality High (LLM Compared with:)** | | | | | |
| | CBIR | -0.051 | 0.0025 | 0.328 | (-0.154, 0.051) | |
| | GradCAM | -0.120 | 0.0031 | 0.033 | (-0.231, -0.010) | * |
| | **AI Explanation Quality Low (LLM Compared with:)** | | | | | |
| | CBIR | 0.010 | 0.0025 | 0.845 | (-0.088, 0.108) | |
| | GradCAM | -0.046 | 0.0029 | 0.393 | (-0.152, 0.060) | |

### 6.4. Finding: AI improves PCP's performance

Sec 6.1 - 6.3 mainly focus on the Study 1 results with the general public. Here we switch to the Study 2 results with PCPs. **STable 16** shows the effect of AI assistance and XAI methods, controlling gender, age, race, and medical expertise in skin, year of experience, and personality traits (Human-AI collaboration experience, XAI rating, Open-mindness - AOT, and Critical thinking - CRT). Across all cases, AI assistance significantly improved PCPs' diagnostic accuracy (e.g., Top-1 accuracy improved by 25.8%, $p < 0.001$) and the significant improvement exists across different XAI explanations.

**STable 17-20** show the detailed breakdown of the four main diseases (atopic dermatitis, pityriasis rosea, Lyme disease, and CTCL) and the AI assistance still resulted in significant Top-1 accuracy improvement.

**STable 21** shows the effect on confidence, revealing no significant main effect for AI assistance on PCPs' diagnostic confidence overall. Post-hoc EMMs indicate that there is only significant confidence increase when PCPs were assisted by GradCAM (p=0.019) and LLM explanations (p=0.003).

**STable 22** shows the effect of the AI assistance on reducing the gap between skin tones in the clinical images after controlling the same confounders. The disparity was reduced from 4.6% (p=0.069) to 2.9% (p=0.248).

**STable 16 | Linear Mixed Model Results on Overall Diagnostic Performance in PCPs (Target Outcome: Top-1 Accuracy)**

| Variables | β | $\sigma^2$ | p-value | 95% CI for β | Sig. Level |
|---|---|---|---|---|---|
| **Intercept** | 0.020 | 0.0595 | 0.934 | (-0.458, 0.498) | |
| **Decision (Ref: Rd. 1 without AI)** | | | | | |
| Rd. 2 with AI Assistance | 0.258 | 0.0021 | <0.001 | (0.168, 0.349) | *** |
| **Explanation Group (Ref: Basic AI)** | | | | | |
| CBIR | 0.074 | 0.0041 | 0.252 | (-0.052, 0.200) | |
| GradCAM | 0.060 | 0.0038 | 0.330 | (-0.061, 0.181) | |
| LLM | 0.070 | 0.0037 | 0.251 | (-0.049, 0.189) | |
| **Age (Ref: 18-24)** | | | | | |
| 25 - 34 | -0.122 | 0.0125 | 0.108 | (-0.271, 0.027) | |
| 35 - 44 | -0.169 | 0.0106 | 0.100 | (-0.371, 0.032) | |
| 45 - 54 | -0.124 | 0.0156 | 0.364 | (-0.393, 0.144) | |
| 55 - 64 | 0.061 | 0.0600 | 0.806 | (-0.423, 0.544) | |
| 65 or older | 0.158 | 0.0784 | 0.582 | (-0.404, 0.720) | |
| Under 18 | -0.317 | 0.1089 | 0.215 | (-0.818, 0.184) | |
| **Gender (Ref: Female)** | | | | | |
| Male | -0.070 | 0.0028 | 0.087 | (-0.170, 0.031) | |
| Other | -0.078 | 0.0154 | 0.543 | (-0.329, 0.173) | |
| **Race (Ref: American Indian)** | | | | | |
| Asian | 0.088 | 0.0353 | 0.638 | (-0.279, 0.455) | |
| Black/African American | 0.116 | 0.038 | 0.554 | (-0.277, 0.510) | |
| Hispanic/Latino/Spanish Origin | 0.006 | 0.0441 | 0.977 | (-0.403, 0.415) | |
| Other | 0.100 | 0.0361 | 0.603 | (-0.278, 0.479) | |
| White | 0.109 | 0.038 | 0.575 | (-0.273, 0.492) | |
| **Year of Medical Experience (Ref: 1-3 y)** | | | | | |
| 10 - 20 y | 0.049 | 0.0085 | 0.596 | (-0.132, 0.229) | |
| 5 - 10 y | 0.037 | 0.0025 | 0.452 | (-0.060, 0.134) | |
| 0 - 5 y | 0.063 | 0.0036 | 0.615 | (-0.115, 0.194) | |
| < 1 y | 0.063 | 0.0037 | 0.303 | (-0.056, 0.182) | |
| > 20 y | -0.218 | 0.0484 | 0.326 | (-0.652, 0.217) | |
| **Skin Disease Experience (Ref: Less Knowledgeable)** | | | | | |
| More Knowledgeable | 0.043 | 0.0015 | 0.265 | (-0.033, 0.120) | |
| **Other Covariates** | | | | | |
| HAI | 0.010 | <0.0001 | 0.051 | (0.001, 0.019) | |
| XAI | 0.006 | <0.0001 | 0.267 | (-0.004, 0.017) | |
| AOT | -0.003 | <0.0001 | 0.420 | (-0.012, 0.005) | |
| CRT | 0.006 | 0.0003 | 0.315 | (-0.015, 0.027) | |
| **Interaction Terms** | | | | | |
| Decision Rd. 2 : Explanation Group CBIR | -0.049 | 0.0041 | 0.472 | (-0.182, 0.084) | |
| Decision Rd. 2 : Explanation Group GradCAM | -0.065 | 0.0042 | 0.514 | (-0.199, 0.085) | |
| Decision Rd. 2 : Explanation Group LLM | -0.081 | 0.0041 | 0.205 | (-0.207, 0.044) | |
| | | | | | |
| **Post-hoc Pairwise EMMs Comparisons – XAI Explanations (Rd. 1 vs. Rd. 2)** | | | | | |
| Basic AI (Rd. 1 vs. Rd. 2) | 0.258 | 0.0021 | <0.001 | (0.168, 0.349) | *** |
| CBIR (Rd. 1 vs. Rd. 2) | 0.210 | 0.0025 | <0.001 | (0.112, 0.307) | *** |
| GradCAM (Rd. 1 vs. Rd. 2) | 0.216 | 0.0021 | <0.001 | (0.127, 0.305) | *** |
| LLM (Rd. 1 vs. Rd. 2) | 0.177 | 0.0020 | <0.001 | (0.090, 0.264) | *** |

**STable 16 (cont) | Linear Mixed Model Results on Overall Diagnostic Performance in PCPs (Target Outcome: Top-3 Accuracy)**

| Variables | β | $\sigma^2$ | p-value | 95% CI for β | Sig. Level |
|---|---|---|---|---|---|
| **Intercept** | 0.024 | 0.0529 | 0.916 | (-0.427, 0.476) | |
| **Decision (Ref: Rd. 1 without AI)** | | | | | |
| Rd. 2 with AI Assistance | 0.450 | 0.0025 | <0.001 | ( 0.352, 0.548) | *** |
| **Explanation Group (Ref: Basic AI)** | | | | | |
| CBIR | 0.043 | 0.0040 | 0.496 | (-0.081, 0.167) | |
| GradCAM | 0.076 | 0.0037 | 0.207 | (-0.042, 0.195) | |
| LLM | 0.090 | 0.0036 | 0.132 | (-0.027, 0.208) | |
| **Age (Ref: 18-24)** | | | | | |
| 25 - 34 | -0.074 | 0.0050 | 0.299 | (-0.214, 0.066) | |
| 35 - 44 | -0.087 | 0.0094 | 0.368 | (-0.277, 0.103) | |
| 45 - 54 | -0.138 | 0.0166 | 0.284 | (-0.392, 0.115) | |
| 55 - 64 | 0.267 | 0.0543 | 0.250 | (-0.189, 0.724) | |
| 65 or older | 0.285 | 0.0729 | 0.291 | (-0.245, 0.815) | |
| Under 18 | -0.509 | 0.0581 | 0.035 | (-0.982,-0.036) | * |
| **Gender (Ref: Female)** | | | | | |
| Male | -0.055 | 0.0015 | 0.154 | (-0.132, 0.021) | |
| Other | -0.173 | 0.0146 | 0.151 | (-0.410, 0.063) | |
| **Race (Ref: American Indian)** | | | | | |
| Asian | -0.014 | 0.0313 | 0.936 | (-0.360, 0.332) | |
| Black/African American | 0.058 | 0.0339 | 0.754 | (-0.303, 0.419) | |
| Hispanic/Latino/Spanish Origin | 0.019 | 0.0388 | 0.923 | (-0.367, 0.405) | |
| Other | 0.004 | 0.0331 | 0.983 | (-0.353, 0.361) | |
| White | -0.012 | 0.0339 | 0.948 | (-0.373, 0.349) | |
| **Year of Medical Experience (Ref: 1-3 y)** | | | | | |
| 10 - 20 y | 0.047 | 0.0076 | 0.585 | (-0.123, 0.218) | |
| 5 - 10 y | 0.058 | 0.0022 | 0.218 | (-0.034, 0.149) | |
| 0 - 5 y | 0.093 | 0.0055 | 0.211 | (-0.053, 0.239) | |
| < 1 y | 0.091 | 0.0032 | 0.113 | (-0.021, 0.203) | |
| > 20 y | -0.267 | 0.0437 | 0.201 | (-0.677, 0.142) | |
| Skin Disease Knowledge (Ref: Less Knowledgeable) | | | | | |
| More Knowledgeable | 0.025 | 0.0014 | 0.499 | (-0.047, 0.097) | |
| **Other Covariates** | | | | | |
| HAI | -0.012 | 0.0053 | 0.871 | (-0.155, 0.131) | |
| XAI | 0.038 | 0.0049 | 0.587 | (-0.099, 0.175) | |
| AOT | -0.081 | 0.0048 | 0.244 | (-0.217, 0.055) | |
| CRT | 0.005 | 0.0000 | 0.307 | (-0.004, 0.014) | |
| **Interaction Terms** | | | | | |
| Decision Rd. 2 : Explanation Group CBIR | 0.006 | 0.0000 | 0.228 | (-0.004, 0.017) | |
| Decision Rd. 2 : Explanation Group GradCAM | -0.001 | 0.0000 | 0.769 | (-0.009, 0.007) | |
| Decision Rd. 2 : Explanation Group LLM | 0.032 | 0.0002 | 0.037 | ( 0.002, 0.062) | * |
| | | | | | |
| **Post-hoc Pairwise EMMs Comparisons – XAI Explanations (Rd. 1 vs. Rd. 2)** | | | | | |
| Basic AI (Rd. 1 vs. Rd. 2) | 0.450 | 0.0025 | <0.001 | (0.352, 0.548) | *** |
| CBIR (Rd. 1 vs. Rd. 2) | 0.438 | 0.0025 | <0.001 | (0.333, 0.543) | *** |
| GradCAM (Rd. 1 vs. Rd. 2) | 0.488 | 0.0024 | <0.001 | (0.392, 0.584) | *** |
| LLM (Rd. 1 vs. Rd. 2) | 0.369 | 0.0023 | <0.001 | (0.275, 0.463) | *** |

**STable 17 | Linear Mixed Model Results on Overall Diagnostic Performance in PCP (Target Outcome: Top-1 Accuracy, Atopic Dermatitis Group)**

| Variables | β | $\sigma^2$ | p-value | 95% CI for β | Sig. Level |
|---|---|---|---|---|---|
| **Intercept** | -0.407 | 0.1347 | 0.268 | (-1.127, 0.312) | |
| **Decision (Ref: Rd. 1 without AI)** | | | | | |
| Rd. 2 with AI Assistance | 0.340 | 0.0036 | <0.001 | (0.213, 0.468) | *** |
| **Explanation Group (Ref: Basic AI)** | | | | | |
| CBIR | 0.083 | 0.0072 | 0.384 | (-0.104, 0.270) | |
| GradCAM | 0.118 | 0.0117 | 0.197 | (-0.061, 0.296) | |
| LLM | 0.059 | 0.0074 | 0.509 | (-0.117, 0.236) | |
| **Age (Ref: 18-24)** | | | | | |
| 25 - 34 | -0.031 | 0.0125 | 0.787 | (-0.255, 0.193) | |
| 35 - 44 | -0.134 | 0.0240 | 0.386 | (-0.438, 0.169) | |
| 45 - 54 | 0.054 | 0.0424 | 0.792 | (-0.350, 0.459) | |
| 55 - 64 | 0.373 | 0.1384 | 0.315 | (-0.355, 1.101) | |
| 65 or older | 0.105 | 0.1875 | 0.809 | (-0.742, 0.951) | |
| Under 18 | 0.095 | 0.1482 | 0.805 | (-0.660, 0.850) | |
| **Gender (Ref: Female)** | | | | | |
| Male | -0.112 | 0.0036 | 0.070 | (-0.234, 0.009) | |
| Other | -0.190 | 0.0361 | 0.324 | (-0.568, 0.188) | |
| **Race (Ref: American Indian)** | | | | | |
| Asian | 0.364 | 0.0795 | 0.197 | (-0.189, 0.917) | |
| Black/African American | 0.492 | 0.0864 | 0.095 | (-0.085, 1.068) | |
| Hispanic/Latino/Spanish Origin | 0.366 | 0.0980 | 0.244 | (-0.250, 0.982) | |
| Other | 0.500 | 0.0841 | 0.086 | (-0.070, 1.070) | |
| White | 0.371 | 0.0847 | 0.207 | (-0.205, 0.948) | |
| **Year of Medical Experience (Ref: 1-3 y)** | | | | | |
| 10 - 20 y | -0.035 | 0.0193 | 0.800 | (-0.307, 0.237) | |
| 5 - 10 y | 0.063 | 0.0064 | 0.401 | (-0.084, 0.209) | |
| 0 - 5 y | 0.023 | 0.0141 | 0.846 | (-0.210, 0.256) | |
| < 1 y | -0.037 | 0.0104 | 0.687 | (-0.216, 0.143) | |
| > 20 y | -0.304 | 0.1116 | 0.362 | (-0.958, 0.350) | |
| **Skin Disease Knowledge (Ref: Less Knowledgeable)** | | | | | |
| More Knowledgeable | -0.013 | 0.0035 | 0.820 | (-0.120, 0.094) | |
| **Other Covariates** | | | | | |
| HAI | 0.013 | <0.0001 | 0.080 | (0.002, 0.028) | |
| XAI | 0.011 | <0.0001 | 0.183 | (-0.005, 0.028) | |
| AOT | -0.004 | <0.0001 | 0.523 | (-0.016, 0.008) | |
| CRT | 0.006 | 0.0006 | 0.289 | (-0.022, 0.039) | |
| **Interaction Terms** | | | | | |
| Decision Rd. 2 : Explanation Group CBIR | -0.118 | 0.0081 | 0.214 | (-0.304, 0.068) | |
| Decision Rd. 2 : Explanation Group GradCAM | -0.067 | 0.0083 | 0.462 | (-0.245, 0.111) | |
| Decision Rd. 2 : Explanation Group LLM | -0.174 | 0.0185 | 0.054 | (-0.350, 0.003) | |

**STable 18 | Linear Mixed Model Results on Overall Diagnostic Performance in PCPs (Target Outcome: Top-1 Accuracy, Pityriasis Rosea Group)**

| Variables | β | $\sigma^2$ | p-value | 95% CI for β | Sig. Level |
|---|---|---|---|---|---|
| **Intercept** | 0.475 | 0.1936 | 0.281 | (-0.388, 1.338) | |
| **Decision (Ref: Rd. 1 without AI)** | | | | | |
| Rd. 2 with AI Assistance | 0.271 | 0.0037 | <0.001 | (0.152, 0.390) | *** |
| **Explanation Group (Ref: Basic AI)** | | | | | |
| CBIR | 0.055 | 0.0121 | 0.615 | (-0.158, 0.268) | |
| GradCAM | -0.041 | 0.0108 | 0.696 | (-0.245, 0.163) | |
| LLM | 0.040 | 0.0117 | 0.696 | (-0.161, 0.242) | |
| **Age (Ref: 18-24)** | | | | | |
| 25 - 34 | -0.246 | 0.0246 | 0.075 | (-0.513, 0.025) | |
| 35 - 44 | -0.340 | 0.0346 | 0.068 | (-0.704, 0.025) | |
| 45 - 54 | -0.294 | 0.0605 | 0.235 | (-0.780, 0.192) | |
| 55 - 64 | -0.311 | 0.1989 | 0.487 | (-1.186, 0.564) | |
| 65 or older | -0.398 | 0.2016 | 0.443 | (-1.415, 0.619) | |
| Under 18 | -1.005 | 0.1482 | 0.030 | (-1.912, -0.097) | * |
| **Gender (Ref: Female)** | | | | | |
| Male | -0.066 | 0.0056 | 0.373 | (-0.208, 0.075) | |
| Other | -0.091 | 0.0538 | 0.695 | (-0.545, 0.363) | |
| **Race (Ref: American Indian)** | | | | | |
| Asian | -0.423 | 0.1116 | 0.212 | (-1.086, 0.241) | |
| Black/African American | -0.330 | 0.1225 | 0.350 | (-1.023, 0.363) | |
| Hispanic/Latino/Spanish Origin | -0.383 | 0.1428 | 0.105 | (-1.353, 0.128) | |
| Other | -0.398 | 0.1049 | 0.273 | (-1.067, 0.270) | |
| White | -0.322 | 0.1102 | 0.362 | (-1.015, 0.370) | |
| **Year of Medical Experience (Ref: 1-3 y)** | | | | | |
| 10 - 20 y | 0.150 | 0.0272 | 0.368 | (-0.177, 0.477) | |
| 5 - 10 y | 0.117 | 0.0204 | 0.192 | (-0.059, 0.293) | |
| 0 - 5 y | 0.113 | 0.0204 | 0.428 | (-0.167, 0.393) | |
| < 1 y | 0.149 | 0.0121 | 0.176 | (-0.067, 0.364) | |
| > 20 y | 0.022 | 0.016 | 0.956 | (-0.764, 0.808) | |
| **Skin Disease Knowledge (Ref: Less Knowledgeable)** | | | | | |
| More Knowledgeable | 0.012 | 0.005 | 0.866 | (-0.126, 0.150) | |
| **Other Covariates** | | | | | |
| HAI | -0.009 | <0.0001 | 0.921 | (-0.018, 0.001) | |
| XAI | 0.012 | 0.0001 | 0.220 | (-0.007, 0.032) | |
| AOT | <0.001 | <0.0001 | 0.999 | (-0.015, 0.015) | |
| CRT | 0.023 | 0.0008 | 0.442 | (-0.029, 0.075) | |
| **Interaction Terms** | | | | | |
| Decision Rd. 2 : Explanation Group CBIR | -0.057 | 0.0062 | 0.524 | (-0.231, 0.118) | |
| Decision Rd. 2 : Explanation Group GradCAM | -0.034 | 0.0071 | 0.688 | (-0.201, 0.132) | |
| Decision Rd. 2 : Explanation Group LLM | -0.111 | 0.0062 | 0.189 | (-0.276, 0.054) | |

**STable 19 | Linear Mixed Model Results on Overall Diagnostic Performance in PCPs (Target Outcome: Top-1 Accuracy, Lyme Group)**

| Variables | β | $\sigma^2$ | p-value | 95% CI for β | Sig. Level |
|---|---|---|---|---|---|
| **Intercept** | -0.136 | 0.0185 | 0.725 | (-0.458, 0.185) | |
| **Decision (Ref: Rd. 1 without AI)** | | | | | |
| Rd. 2 with AI Assistance | 0.260 | 0.0039 | <0.001 | (0.137, 0.384) | *** |
| **Explanation Group (Ref: Basic AI)** | | | | | |
| CBIR | 0.154 | 0.0098 | 0.117 | (-0.058, 0.367) | |
| GradCAM | 0.127 | 0.0088 | 0.176 | (-0.057, 0.311) | |
| LLM | 0.126 | 0.0086 | 0.176 | (-0.056, 0.308) | |
| **Age (Ref: 18-24)** | | | | | |
| 25 - 34 | -0.184 | 0.0240 | 0.125 | (-0.419, 0.051) | |
| 35 - 44 | -0.172 | 0.0296 | 0.288 | (-0.491, 0.146) | |
| 45 - 54 | -0.156 | 0.0296 | 0.470 | (-0.581, 0.268) | |
| 55 - 64 | 0.165 | 0.1521 | 0.672 | (-0.599, 0.929) | |
| 65 or older | 0.662 | 0.2052 | 0.144 | (-0.226, 1.550) | |
| Under 18 | -0.067 | 0.0169 | 0.869 | (-0.859, 0.725) | |
| **Gender (Ref: Female)** | | | | | |
| Male | -0.083 | 0.0042 | 0.200 | (-0.211, 0.044) | |
| Other | -0.044 | 0.0400 | 0.827 | (-0.441, 0.352) | |
| **Race (Ref: American Indian)** | | | | | |
| Asian | 0.433 | 0.0841 | 0.143 | (-0.147, 1.012) | |
| Black/African American | 0.414 | 0.0961 | 0.180 | (-0.191, 1.019) | |
| Hispanic/Latino/Spanish Origin | 0.320 | 0.0961 | 0.333 | (-0.327, 0.966) | |
| Other | 0.383 | 0.0955 | 0.201 | (-0.208, 0.977) | |
| White | 0.393 | 0.0955 | 0.203 | (-0.212, 0.998) | |
| **Year of Medical Experience (Ref: 1-3 y)** | | | | | |
| 10 - 20 y | 0.076 | 0.0213 | 0.601 | (-0.209, 0.361) | |
| 5 - 10 y | 0.031 | 0.0058 | 0.696 | (-0.123, 0.184) | |
| 0 - 5 y | 0.052 | 0.0156 | 0.675 | (-0.192, 0.297) | |
| < 1 y | 0.146 | 0.0213 | 0.129 | (-0.042, 0.334) | |
| > 20 y | -0.386 | 0.1225 | 0.271 | (-1.072, 0.300) | |
| **Skin Disease Knowledge (Ref: Less Knowledgeable)** | | | | | |
| More Knowledgeable | 0.087 | 0.0037 | 0.157 | (-0.033, 0.207) | |
| **Other Covariates** | | | | | |
| HAI | 0.016 | <0.0001 | 0.039 | (0.001, 0.031) | * |
| XAI | 0.004 | <0.0001 | 0.690 | (-0.004, 0.012) | |
| AOT | -0.001 | <0.0001 | 0.832 | (-0.014, 0.012) | |
| CRT | 0.021 | 0.0007 | 0.409 | (-0.029, 0.071) | |
| **Interaction Terms** | | | | | |
| Decision Rd. 2 : Explanation Group CBIR | -0.087 | 0.0077 | 0.236 | (-0.291, 0.117) | |
| Decision Rd. 2 : Explanation Group GradCAM | -0.067 | 0.0077 | 0.448 | (-0.240, 0.106) | |
| Decision Rd. 2 : Explanation Group LLM | -0.075 | 0.0077 | 0.395 | (-0.246, 0.097) | |

**STable 20 | Linear Mixed Model Results on Overall Diagnostic Performance in PCPs (Target Outcome: Top-1 Accuracy, CTCL Group)**

| Variables | β | $\sigma^2$ | p-value | 95% CI for β | Sig. Level |
|---|---|---|---|---|---|
| **Intercept** | -0.331 | 0.0812 | 0.246 | (-0.890, 0.228) | |
| **Decision (Ref: Rd. 1 without AI)** | | | | | |
| Rd. 2 with AI Assistance | 0.212 | 0.0042 | 0.001 | (0.085, 0.338) | ** |
| **Explanation Group (Ref: Basic AI)** | | | | | |
| CBIR | 0.040 | 0.0064 | 0.613 | (-0.116, 0.196) | |
| GradCAM | 0.032 | 0.0058 | 0.671 | (-0.117, 0.182) | |
| LLM | 0.056 | 0.0056 | 0.457 | (-0.091, 0.203) | |
| **Age (Ref: 18-24)** | | | | | |
| 25 - 34 | -0.022 | 0.0158 | 0.804 | (-0.195, 0.151) | |
| 35 - 44 | -0.083 | 0.0256 | 0.487 | (-0.318, 0.152) | |
| 45 - 54 | -0.153 | 0.0256 | 0.338 | (-0.466, 0.160) | |
| 55 - 64 | 0.027 | 0.0784 | 0.925 | (-0.537, 0.591) | |
| 65 or older | 0.096 | 0.0829 | 0.773 | (-0.559, 0.752) | |
| Under 18 | 0.127 | 0.1089 | 0.670 | (-0.569, 0.825) | |
| **Gender (Ref: Female)** | | | | | |
| Male | -0.028 | 0.0023 | 0.564 | (-0.122, 0.066) | |
| Other | -0.026 | 0.0144 | 0.861 | (-0.319, 0.267) | |
| **Race (Ref: American Indian)** | | | | | |
| Asian | 0.472 | 0.0515 | 0.031 | (0.044, 0.900) | * |
| Black/African American | 0.381 | 0.0520 | 0.095 | (-0.066, 0.828) | |
| Hispanic/Latino/Spanish Origin | 0.457 | 0.0595 | 0.061 | (-0.020, 0.934) | |
| Other | 0.414 | 0.0484 | 0.066 | (-0.027, 0.856) | |
| White | 0.486 | 0.0520 | 0.033 | (0.040, 0.933) | * |
| **Year of Medical Experience (Ref: 1-3 y)** | | | | | |
| 10 - 20 y | 0.051 | 0.0213 | 0.632 | (-0.159, 0.262) | |
| 5 - 10 y | -0.008 | 0.0117 | 0.886 | (-0.122, 0.105) | |
| 0 - 5 y | 0.053 | 0.0085 | 0.724 | (-0.148, 0.213) | |
| < 1 y | 0.010 | 0.0125 | 0.893 | (-0.129, 0.148) | |
| > 20 y | -0.187 | 0.0676 | 0.469 | (-0.724, 0.350) | |
| **Skin Disease Knowledge (Ref: Less Knowledgeable)** | | | | | |
| More Knowledgeable | 0.072 | 0.0020 | 0.111 | (-0.017, 0.161) | |
| **Other Covariates** | | | | | |
| HAI | 0.010 | <0.0001 | 0.077 | (0.001, 0.022) | |
| XAI | 0.004 | <0.0001 | 0.562 | (-0.009, 0.017) | |
| AOT | -0.009 | <0.0001 | 0.076 | (-0.018, 0.001) | |
| CRT | -0.005 | <0.0001 | 0.803 | (-0.042, 0.032) | |
| **Interaction Terms** | | | | | |
| Decision Rd. 2 : Explanation Group CBIR | 0.074 | 0.0020 | 0.434 | (-0.111, 0.259) | |
| Decision Rd. 2 : Explanation Group GradCAM | -0.002 | 0.0081 | 0.987 | (-0.179, 0.175) | |
| Decision Rd. 2 : Explanation Group LLM | 0.019 | 0.0064 | 0.832 | (-0.157, 0.194) | |

**STable 21 | Linear Mixed Model Results on Overall Diagnostic Performance in PCP (Target Outcome: Confidence)**

| Variables | β | $\sigma^2$ | p-value | 95% CI for β | Sig. Level |
|---|---|---|---|---|---|
| **Intercept** | 0.388 | 0.0520 | 0.089 | (-0.059, 0.834) | |
| **Decision (Ref: Rd. 1 without AI)** | | | | | |
| Rd. 2 with AI Assistance | 0.018 | 0.0001 | 0.131 | (-0.005, 0.042) | |
| **Explanation Group (Ref: Basic AI)** | | | | | |
| CBIR | -0.031 | 0.0025 | 0.547 | (-0.134, 0.071) | |
| GradCAM | 0.035 | 0.0030 | 0.487 | (-0.063, 0.133) | |
| LLM | 0.032 | 0.0023 | 0.513 | (-0.064, 0.129) | |
| **Age (Ref: 18-24)** | | | | | |
| 25 - 34 | -0.073 | 0.0127 | 0.083 | (-0.263, 0.016) | |
| 35 - 44 | -0.140 | 0.0117 | 0.147 | (-0.329, 0.049) | |
| 45 - 54 | -0.210 | 0.0441 | 0.102 | (-0.465, 0.042) | |
| 55 - 64 | -0.231 | 0.0534 | 0.318 | (-0.685, 0.223) | |
| 65 or older | -0.565 | 0.0620 | 0.036 | (-1.093, -0.038) | * |
| Under 18 | -0.376 | 0.0600 | 0.118 | (-0.846, 0.095) | |
| **Gender (Ref: Female)** | | | | | |
| Male | 0.095 | 0.0058 | 0.014 | (0.020, 0.171) | * |
| Other | -0.191 | 0.0365 | 0.111 | (-0.427, 0.044) | |
| **Race (Ref: American Indian)** | | | | | |
| Asian | -0.004 | 0.0276 | 0.980 | (-0.349, 0.340) | |
| Black/African American | -0.002 | 0.0335 | 0.992 | (-0.361, 0.358) | |
| Hispanic/Latino/Spanish Origin | -0.084 | 0.0376 | 0.667 | (-0.468, 0.300) | |
| Other | -0.032 | 0.0296 | 0.860 | (-0.327, 0.263) | |
| White | -0.181 | 0.0324 | 0.522 | (-0.577, 0.215) | |
| **Year of Medical Experience (Ref: 1-3 y)** | | | | | |
| 10 - 20 y | 0.061 | 0.0074 | 0.477 | (-0.108, 0.231) | |
| 5 - 10 y | -0.003 | 0.0056 | 0.726 | (-0.107, 0.102) | |
| 0 - 5 y | -0.083 | 0.0055 | 0.263 | (-0.228, 0.062) | |
| < 1 y | 0.049 | 0.0049 | 0.386 | (-0.062, 0.161) | |
| > 20 y | 0.189 | 0.0437 | 0.363 | (-0.219, 0.597) | |
| **Skin Disease Knowledge (Ref: Less Knowledgeable)** | | | | | |
| More Knowledgeable | 0.137 | 0.0013 | <0.001 | (0.065, 0.208) | *** |
| **Other Covariates** | | | | | |
| HAI | 0.003 | <0.0001 | 0.505 | (-0.006, 0.012) | |
| XAI | 0.004 | <0.0001 | 0.432 | (-0.006, 0.014) | |
| AOT | 0.012 | <0.0001 | 0.002 | (0.005, 0.020) | ** |
| CRT | -0.016 | <0.0001 | 0.280 | (-0.046, 0.013) | |
| **Interaction Terms** | | | | | |
| Decision Rd. 2 : Explanation Group CBIR | 0.005 | 0.0003 | 0.758 | (-0.029, 0.040) | |
| Decision Rd. 2 : Explanation Group GradCAM | 0.010 | 0.0003 | 0.569 | (-0.024, 0.043) | |
| Decision Rd. 2 : Explanation Group LLM | 0.017 | 0.0003 | 0.322 | (-0.016, 0.050) | |
| | | | | | |
| **Post-hoc Pairwise EMMs Comparisons – XAI Explanations (Rd. 1 vs. Rd. 2)** | | | | | |
| Basic AI (Rd. 1 vs. Rd. 2) | 0.018 | 0.0001 | 0.131 | (-0.005, 0.042) | |
| CBIR (Rd. 1 vs. Rd. 2) | 0.024 | 0.0001 | 0.066 | (-0.002, 0.049) | |
| GradCAM (Rd. 1 vs. Rd. 2) | 0.028 | 0.0001 | 0.019 | (0.005, 0.051) | * |
| LLM (Rd. 1 vs. Rd. 2) | 0.035 | 0.0001 | 0.003 | (0.012, 0.058) | ** |

**STable 22 | Linear Mixed Model Results on Diagnostic Performance Across Skin Tones in PCP (Target Outcome: Top-1 Accuracy)**

| Variables | β | $\sigma^2$ | p-value | 95% CI for β | Sig. Level |
|---|---|---|---|---|---|
| **Intercept** | 0.041 | 0.0548 | 0.862 | (-0.419, 0.500) | |
| **Decision (Ref: Rd. 1 without AI)** | | | | | |
| Rd. 2 with AI Assistance | 0.223 | 0.0006 | <0.001 | (0.173, 0.272) | *** |
| **Skin Tone (Ref: Dark Skin)** | | | | | |
| Light | 0.046 | 0.0006 | 0.069 | (-0.004, 0.095) | |
| **Age (Ref: 18-24)** | | | | | |
| 25 - 34 | -0.125 | 0.0055 | 0.091 | (-0.270, 0.020) | |
| 35 - 44 | -0.176 | 0.0102 | 0.076 | (-0.369, 0.018) | |
| 45 - 54 | -0.147 | 0.0177 | 0.269 | (-0.407, 0.113) | |
| 55 - 64 | 0.016 | 0.0557 | 0.945 | (-0.448, 0.480) | |
| 65 or older | 0.121 | 0.0724 | 0.665 | (-0.426, 0.668) | |
| Under 18 | -0.311 | 0.1089 | 0.202 | (-0.788, 0.166) | |
| **Gender (Ref: Female)** | | | | | |
| Male | -0.067 | 0.0045 | 0.090 | (-0.143, 0.010) | |
| Other | -0.066 | 0.0156 | 0.596 | (-0.310, 0.178) | |
| **Race (Ref: American Indian)** | | | | | |
| Asian | 0.078 | 0.0331 | 0.669 | (-0.279, 0.435) | |
| Black/African American | 0.100 | 0.0369 | 0.603 | (-0.276, 0.476) | |
| Hispanic/Latino/Spanish Origin | 0.008 | 0.0400 | 0.967 | (-0.385, 0.400) | |
| Other | 0.091 | 0.0353 | 0.629 | (-0.278, 0.479) | |
| White | 0.104 | 0.0376 | 0.585 | (-0.270, 0.478) | |
| **Year of Medical Experience (Ref: 1-3 y)** | | | | | |
| 10 - 20 y | 0.065 | 0.0079 | 0.464 | (-0.109, 0.239) | |
| 5 - 10 y | 0.045 | 0.002 | 0.351 | (-0.049, 0.139) | |
| 0 - 5 y | 0.048 | 0.0023 | 0.525 | (-0.100, 0.196) | |
| < 1 y | 0.071 | 0.0035 | 0.225 | (-0.044, 0.187) | |
| > 20 y | -0.170 | 0.0441 | 0.417 | (-0.581, 0.241) | |
| **Skin Disease Knowledge (Ref: Less Knowledgeable)** | | | | | |
| More Knowledgeable | 0.042 | 0.0014 | 0.252 | (-0.030, 0.115) | |
| **Other Covariates** | | | | | |
| HAI | 0.009 | <0.0001 | 0.070 | (-0.001, 0.018) | |
| XAI | 0.006 | <0.0001 | 0.216 | (-0.004, 0.017) | |
| AOT | -0.003 | <0.0001 | 0.448 | (-0.011, 0.005) | |
| CRT | 0.017 | 0.0003 | 0.283 | (-0.014, 0.048) | |
| **Interaction Terms** | | | | | |
| Decision Rd. 2 : Skin Tone Light | -0.017 | 0.0013 | 0.641 | (-0.087, 0.053) | |
| **Post-hoc Pairwise EMMs Comparisons – Decision Rounds (Light vs. Dark Skins)** | | | | | |
| Rd. 1 (Light vs. Dark Skins) | 0.046 | 0.0006 | 0.069 | (-0.004, 0.095) | |
| Rd. 2 (Light vs. Dark Skins) | 0.029 | 0.0006 | 0.248 | (-0.020, 0.079) | |

### 6.5. Finding: PCPs are resilient to AI deference when AI is wrong

**STable 23** presents the Linear Mixed Model results on the change in Top-1 Accuracy after AI assistance, investigating the effect of AI correctness and explanation type. The model reveals a strong, significant negative effect ($p < 0.001$) on accuracy change when the AI is incorrect, confirming that PCPs' performance drops when the AI provides wrong advice. However, the magnitude of the drop is less than a complete deference to the wrong AI advice, demonstrating resilience. The LLM explanation group showed a trend towards a smaller accuracy drop when the AI was wrong compared to the Basic AI group, indicated by a positive interaction term ($\beta = 0.128$) with a borderline significance ($p=0.117$). **STable 24** provides post-hoc pairwise comparisons to further examine the influence of LLM-based explanations on accuracy change under correct and incorrect AI advice. Critically, under AI incorrectness, there were no significant differences between the LLM group and other explanation groups (Basic AI, CBIR, GradCAM), suggesting that LLM explanations did not significantly *worsen* the outcome, which supports the idea that PCPs maintain their own judgment.

**STable 26** analyzes the factors influencing accuracy specifically under conditions of AI Correctness, examining the role of XAI Quality (High vs. Low) and Explanation Type. When the AI was correct, the GradCAM group with High XAI Quality showed a significantly lower accuracy than the CBIR group ($p < 0.001$), suggesting that high-quality GradCAM explanations did not translate to superior performance compared to CBIR. Conversely, the negative impact of incorrect AI was significant ($p = 0.002$), highlighting the detrimental effect of erroneous AI recommendations on performance. **STable 27** offers EMM results of XAI Quality across different explanation types and AI correctness states. The comparisons reinforce that under AI-Correct scenarios, the LLM group with High XAI Quality performed significantly worse than the GradCAM group ($p < 0.001$), indicating complexity in how explanation quality impacts outcomes. Conversely, under AI-Incorrect scenarios, the differences between LLM and other explanations across both high and low XAI quality were non-significant with seldom diagnostic accuracy change, further supporting the resilience of PCPs to poor explanations when the AI is wrong.

**STable 23 | Linear Mixed Model Results on Diagnostic Performance Across AI Correctness and Explanations in PCP (Target Outcome: Top-1 Accuracy Change after AI Assistance)**

| Variables | β | $\sigma^2$ | p-value | 95% CI for β | Sig. Level |
|---|---|---|---|---|---|
| **Intercept** | -0.011 | 0.0437 | 0.960 | (-0.440, 0.418) | |
| **AI Correctness (Ref: Correct)** | | | | | |
| Wrong | -0.349 | 0.0035 | <0.001 | (-0.464, -0.234) | *** |
| **Explanation (Ref: Basic)** | | | | | |
| CBIR | -0.048 | 0.0031 | 0.462 | (-0.176, 0.080) | |
| GradCAM | -0.035 | 0.0038 | 0.570 | (-0.158, 0.087) | |
| LLM | -0.100 | 0.0038 | 0.106 | (-0.221, 0.021) | |
| **Race (Ref: American Indian)** | | | | | |
| Asian | 0.360 | 0.0279 | 0.031 | (0.032, 0.688) | * |
| Black/African American | 0.380 | 0.0306 | 0.029 | (0.038, 0.723) | * |
| Hispanic/Latino/Spanish Origin | 0.392 | 0.0313 | 0.036 | (0.036, 0.757) | * |
| Other | 0.352 | 0.0296 | 0.041 | (0.014, 0.690) | * |
| White | 0.360 | 0.0306 | 0.039 | (0.018, 0.702) | * |
| **Gender (Ref: Female)** | | | | | |
| Male | -0.019 | 0.0014 | 0.613 | (-0.087, 0.054) | |
| Other | 0.006 | 0.0132 | 0.956 | (-0.218, 0.231) | |
| **Age (Ref: 18-24)** | | | | | |
| 25 - 34 | -0.069 | 0.0077 | 0.308 | (-0.202, 0.064) | |
| 35 - 44 | -0.138 | 0.0085 | 0.132 | (-0.318, 0.042) | |
| 45 - 54 | -0.128 | 0.0150 | 0.297 | (-0.368, 0.112) | |
| 55 - 64 | 0.010 | 0.0493 | 0.964 | (-0.422, 0.442) | |
| 65 or older | -0.078 | 0.0655 | 0.760 | (-0.397, 0.242) | |
| Under 18 | 0.038 | 0.0524 | 0.868 | (-0.410, 0.486) | |
| **Skin Disease Knowledge (Ref: Less Knowledgeable)** | | | | | |
| More Knowledgeable | -0.026 | 0.0012 | 0.453 | (-0.094, 0.042) | |
| **Year of Medical Experience (Ref: 1-3 y)** | | | | | |
| 10 - 20 y | 0.092 | 0.0067 | 0.265 | (-0.070, 0.253) | |
| 5 - 10 y | 0.007 | 0.0020 | 0.883 | (-0.088, 0.093) | |
| 0 - 5 y | 0.092 | 0.0051 | 0.190 | (-0.046, 0.231) | |
| < 1 y | -0.003 | 0.0030 | 0.952 | (-0.110, 0.103) | |
| > 20 y | -0.065 | 0.0384 | 0.741 | (-0.454, 0.323) | |
| **Interaction Terms** | | | | | |
| AI correctness wrong : explanation CBIR | 0.087 | 0.0074 | 0.312 | (-0.082, 0.256) | |
| AI correctness wrong : explanation GradCAM | 0.079 | 0.0067 | 0.338 | (-0.083, 0.240) | |
| AI correctness wrong : explanation LLM | 0.128 | 0.0067 | 0.117 | (0.011, 0.245) | |
| **Other Covariates** | | | | | |
| HAI | 0.008 | <0.0001 | 0.071 | (0.001, 0.017) | |
| XAI | 0.006 | <0.0001 | 0.196 | (-0.003, 0.016) | |
| AOT | -0.020 | <0.0001 | 0.672 | (-0.048, -0.003) | |
| CRT | 0.002 | <0.0001 | 0.175 | (-0.009, 0.012) | |

**STable 24 | Post-hoc Pairwise EMMs Comparisons – AI Correctness (XAI Pairwise Comparison) based on STable 23**

| Variables | β | $σ^2$ | p-value | 95% CI for β | Sig. Level |
|---|---|---|---|---|---|
| **AI Correct (LLM Compared with:)** | | | | | |
| Basic AI | -0.100 | 0.0038 | 0.105 | (-0.221, 0.021) | |
| CBIR | -0.052 | 0.0045 | 0.442 | (-0.183, 0.080) | |
| GradCAM | -0.064 | 0.0039 | 0.302 | (-0.186, 0.058) | |
| **AI Incorrect (LLM Compared with:)** | | | | | |
| Basic AI | 0.028 | 0.0038 | 0.649 | (-0.093, 0.149) | |
| CBIR | -0.011 | 0.0045 | 0.870 | (-0.143, 0.121) | |
| GradCAM | –0.015 | 0.0039 | 0.804 | (-0.138, 0.107) | |

**STable 25 | Linear Mixed Model Results on Differential Portion Across Medical Expertise among PCPs and Medical Students (Target Outcome: Deferential Portion)**

| Variables | β | $σ^2$ | p-value | 95% CI for β | Sig. Level |
|---|---|---|---|---|---|
| **Intercept** | 0.515 | 0.0346 | 0.006 | ( 0.150, 0.880) | ** |
| **Medical Role (Ref: Medical Student)** | | | | | |
| PCP | -0.067 | 0.0010 | 0.037 | (-0.130, -0.004) | * |
| **Skin Disease Knowledge (Ref: Less Knowledgeable)** | | | | | |
| More Knowledgeable | -0.073 | 0.0007 | 0.005 | (-0.124, -0.022) | ** |
| **Year of Medical Experience (Ref: 1-3 y)** | | | | | |
| 10 - 20 y | -0.164 | 0.0052 | 0.024 | (-0.306, -0.022) | * |
| 5 - 10 y | -0.040 | 0.0012 | 0.264 | (-0.109, 0.030) | |
| 0 - 5 y | -0.128 | 0.0027 | 0.013 | (-0.229, -0.027) | * |
| < 1 y | -0.044 | 0.0012 | 0.192 | (-0.110, 0.022) | |
| > 20 y | -0.031 | 0.0144 | 0.796 | (-0.267, 0.205) | |
| **Race (Ref: American Indian)** | | | | | |
| Asian | 0.329 | 0.0266 | 0.043 | ( 0.011, 0.648) | * |
| Black or African American | 0.262 | 0.0269 | 0.109 | (-0.058, 0.583) | |
| Hispanic or Latino or Spanish Origin | 0.262 | 0.0303 | 0.132 | (-0.079, 0.602) | |
| Other | 0.330 | 0.0269 | 0.044 | ( 0.009, 0.650) | * |
| White | 0.256 | 0.0269 | 0.119 | (-0.066, 0.578) | |
| **Gender (Ref: Female)** | | | | | |
| Male | 0.049 | 0.0007 | 0.067 | (-0.003, 0.102) | |
| Other | -0.252 | 0.0086 | 0.007 | (-0.434, -0.070) | ** |
| **Age (Ref: 18-24)** | | | | | |
| 25 - 34 | 0.065 | 0.0010 | 0.045 | ( 0.002, 0.128) | * |
| 35 - 44 | 0.110 | 0.0030 | 0.045 | ( 0.002, 0.218) | * |
| 45 - 54 | -0.157 | 0.0062 | 0.048 | (-0.312, -0.002) | * |
| 55 - 64 | 0.123 | 0.0110 | 0.245 | (-0.084, 0.329) | |
| 65 or older | 0.206 | 0.0645 | 0.417 | (-0.292, 0.704) | |
| Under 18 | 0.576 | 0.0740 | 0.034 | ( 0.042, 1.109) | * |
| **Other Covariates** | | | | | |
| HAI | 0.004 | 0.0000 | 0.282 | (-0.003, 0.010) | |
| XAT | 0.002 | 0.0000 | 0.478 | (-0.004, 0.009) | |
| AOT | 0.007 | 0.0000 | 0.006 | ( 0.002, 0.012) | ** |
| CRT | -0.044 | 0.0001 | <0.001 | (-0.066, -0.022) | *** |

**STable 26| Linear Mixed Model Results on Diagnostic Performance Across AI Explanations and Their Quality in PCP (Target Outcome: Accuracy)**

| Variables | β | $\sigma^2$ | p-value | 95% CI for β | Sig. Level |
|---|---|---|---|---|---|
| **Intercept** | -0.109 | 0.0844 | 0.708 | (-0.678, 0.461) | |
| **AI Correctness (Ref: AI Correct)** | | | | | |
| AI Incorrect | -0.271 | 0.0078 | 0.002 | (-0.443, -0.098) | *** |
| **XAI Quality (Ref: High)** | | | | | |
| Low | 0.008 | 0.0055 | 0.914 | (-0.138, 0.154) | |
| **Explanation (Ref: CBIR)** | | | | | |
| GradCAM | 0.551 | 0.026 | 0.001 | (0.235, 0.867) | ** |
| LLM | -0.050 | 0.0067 | 0.540 | (-0.210, 0.110) | |
| **Race (Ref: American Indian)** | | | | | |
| Asian | 0.386 | 0.0422 | 0.060 | (-0.016, 0.789) | |
| Black/African American | 0.443 | 0.0507 | 0.049 | (0.002, 0.884) | * |
| Hispanic/Latino/Spanish Origin | 0.508 | 0.0560 | 0.032 | (0.045, 0.972) | * |
| Other | 0.384 | 0.0483 | 0.080 | (-0.046, 0.815) | |
| White | 0.444 | 0.0509 | 0.049 | (0.002, 0.886) | * |
| **Gender (Ref: Female)** | | | | | |
| Male | 0.026 | 0.0035 | 0.656 | (-0.090, 0.143) | |
| Other | 0.039 | 0.0176 | 0.769 | (-0.221, 0.299) | |
| **Age (Ref: 18-24)** | | | | | |
| 25 - 34 | -0.071 | 0.0080 | 0.426 | (-0.246, 0.104) | |
| 35 - 44 | -0.040 | 0.0155 | 0.747 | (-0.285, 0.204) | |
| 45 - 54 | 0.010 | 0.0255 | 0.951 | (-0.303, 0.323) | |
| 55 - 64 | 0.131 | 0.0787 | 0.642 | (-0.419, 0.680) | |
| 65 or older | 0.062 | 0.1063 | 0.85 | (-0.577, 0.701) | |
| Under 18 | 0.059 | 0.0718 | 0.826 | (-0.466, 0.584) | |
| **Skin Disease Knowledge (Ref: Less Knowledgeable)** | | | | | |
| More Knowledgeable | -0.006 | 0.0025 | 0.902 | (-0.104, 0.092) | |
| **Year of Medical Experience (Ref: 1-3 y)** | | | | | |
| 10 - 20 y | 0.003 | 0.0130 | 0.981 | (-0.221, 0.226) | |
| 5 - 10 y | 0.007 | 0.0043 | 0.914 | (-0.121, 0.135) | |
| 0 - 5 y | 0.024 | 0.0106 | 0.816 | (-0.178, 0.225) | |
| < 1 y | -0.048 | 0.0065 | 0.55 | (-0.205, 0.110) | |
| > 20 y | -0.306 | 0.0646 | 0.228 | (-0.804, 0.192) | |
| **Interaction Terms** | | | | | |
| XAI quality Low : explanation GradCAM | 0.481 | 0.0642 | 0.058 | (-0.015, 0.978) | |
| XAI quality Low : explanation LLM | 0.054 | 0.0221 | 0.716 | (-0.237, 0.346) | |
| **Other Covariates** | | | | | |
| HAI | 0.013 | <0.0001 | 0.033 | (0.001, 0.024) | * |
| XAI | 0.006 | 0.0001 | 0.478 | (-0.010, 0.021) | |
| CRT | -0.036 | 0.0004 | 0.063 | (-0.073, 0.002) | |
| AOT | <0.001 | <0.0001 | 0.953 | (-0.011, 0.011) | |

**STable 27 | Post-hoc Pairwise EMMs Comparisons – XAI Quality (XAI Pairwise Comparison)**

| Condition | Variables | β | $\sigma^2$ | p-value | 95% CI for β | Sig. Level |
|---|---|---|---|---|---|---|
| AI Correct | **AI Explanation Quality High (LLM Compared with:)** | | | | | |
| | CBIR | -0.050 | 0.0067 | 0.540 | (-0.210, 0.110) | |
| | GradCAM | -0.602 | 0.0249 | <0.001 | (-0.910, -0.293) | *** |
| | **AI Explanation Quality Low (LLM Compared with:)** | | | | | |
| | CBIR | -0.117 | 0.0081 | 0.193 | (-0.293, 0.059) | |
| | GradCAM | -0.119 | 0.0059 | 0.119 | (-0.268, 0.031) | |
| AI Wrong | **AI Explanation Quality High (LLM Compared with:)** | | | | | |
| | CBIR | -0.003 | 0.0102 | 0.980 | (-0.201, 0.196) | |
| | GradCAM | -0.079 | 0.0318 | 0.660 | (-0.440, 0.271) | |
| | **AI Explanation Quality Low (LLM Compared with:)** | | | | | |
| | CBIR | -0.015 | 0.0079 | 0.863 | (-0.189, 0.159) | |
| | GradCAM | -0.023 | 0.0068 | 0.782 | (-0.185, 0.139) | |

## 6.6. Finding: Higher AI deference correlates with lower initial performance

**STable 28** investigates the diagnostic accuracy of the general public by deferential group. The non-deferential group demonstrated significantly higher initial diagnostic accuracy (Rd. 1) compared to the deferential group (p<0.001). AI assistance significantly improved accuracy for both groups (Rd. 2), but the benefit was greater for the deferential group, thus narrowing the performance gap (interaction effect: p<0.001).

**STable 29** investigates the Top-1 diagnostic accuracy of PCP by deferential group. Non-deferential PCPs showed significantly higher initial Top-1 accuracy compared to deferential PCPs (p<0.001). AI assistance significantly improved accuracy for all PCPs (p<0.001), with a slightly but significantly smaller gain for the non-deferential group (p=0.047) due to their higher baseline performance.

**STable 30** analyzes the factors associated with the deferential proportion among the general public. Older participants (55-64 years, p<0.001) and those with skin disease experience (p=0.001) showed significantly higher deference. Conversely, some minority race/ethnicity groups (Black or African American, Other) showed significantly lower deference than the American Indian reference group.

**STable 31** analyzes the factors associated with the deferential proportion among PCPs. Several race/ethnicity groups (Asian, Black/African American, Hispanic/Latino/Spanish Origin, Other, White) exhibited significantly higher deferential portions compared to the American Indian reference group. PCPs with more medical experience (45-54 years) showed a trend towards lower deference (p=0.094).

**STable 28 | Linear Mixed Model Results on Diagnostic Performance Across Deferential Group in General Public (Target Outcome: Accuracy)**

| Variables | β | $\sigma^2$ | p-value | 95% CI for β | Sig. Level |
|---|---|---|---|---|---|
| **Intercept** | 0.614 | 0.0008 | <0.001 | (0.557, 0.670) | *** |
| **Deferential Group (Ref: Deferential)** | | | | | |
| Non-Deferential | 0.082 | 0.0003 | <0.001 | (0.051, 0.114) | *** |
| **Decision (Ref: Rd. 1 without AI)** | | | | | |
| Rd. 2 with AI Assistance | 0.093 | 0.0001 | <0.001 | (0.074, 0.111) | *** |
| **Age (Ref: 18-24)** | | | | | |
| 25 - 34 | -0.004 | 0.0004 | 0.832 | (-0.042, 0.034) | |
| 35 - 44 | -0.003 | 0.0004 | 0.873 | (-0.045, 0.038) | |
| 45 - 54 | 0.039 | 0.0010 | 0.306 | (-0.035, 0.113) | |
| 55 - 64 | 0.078 | 0.0022 | 0.100 | (-0.015, 0.170) | |
| 65 or older | 0.056 | 0.0081 | 0.530 | (-0.120, 0.233) | |
| **Gender (Ref: Female)** | | | | | |
| Male | -0.056 | 0.0005 | 0.005 | (-0.087, -0.024) | ** |
| Other | -0.001 | 0.0006 | 0.989 | (-0.049, 0.047) | |
| **Race (Ref: American Indian)** | | | | | |
| Asian | 0.072 | 0.0014 | 0.049 | (0.000, 0.144) | * |
| Black/African American | 0.024 | 0.0013 | 0.504 | (-0.046, 0.094) | |
| Hispanic/Latino/Spanish Origin | 0.076 | 0.0014 | 0.046 | (0.001, 0.151) | * |
| Native Hawaiian/Other Pacific Islander | 0.022 | 0.0036 | 0.717 | (-0.096, 0.140) | |
| Other | 0.028 | 0.0014 | 0.444 | (-0.044, 0.100) | |
| White | 0.033 | 0.0006 | 0.161 | (-0.013, 0.079) | |
| **Skin Disease Experience (Ref: No)** | | | | | |
| Yes | 0.009 | 0.0002 | 0.574 | (-0.021, 0.039) | |
| **Interaction Terms** | | | | | |
| Deferential Group Non-Deferential : Decision Rd. 2 | -0.055 | 0.0002 | <0.001 | (-0.079, -0.030) | *** |
| **Other Covariates** | | | | | |
| HAI Collaboration Experience | 0.007 | <0.0001 | <0.001 | (0.004, 0.011) | *** |
| **Post-hoc Pairwise EMMs Comparisons – Decision Rounds (Deferential vs. Non-Deferential)** | | | | | |
| Rd. 1 (Deferential vs. Non-Deferential) | 0.0823 | 0.0003 | <0.001 | (0.051, 0.114) | *** |
| Rd. 2 (Deferential vs. Non-Deferential) | 0.0276 | 0.0003 | 0.083 | (-0.004, 0.059) | |

**STable 29 | Linear Mixed Model Results on Diagnostic Performance Across Deferential Group in PCP (Target Outcome: Top-1 Accuracy)**

| Variables | β | $\sigma^2$ | p-value | 95% CI for β | Sig. Level |
|---|---|---|---|---|---|
| **Intercept** | -0.067 | 0.0529 | 0.77 | (-0.517, 0.383) | |
| **Decision (Ref: Rd. 1 without AI)** | | | | | |
| Rd. 2 with AI Assistance | 0.235 | 0.0004 | <0.001 | (0.187, 0.284) | *** |
| **Deferential Group (Ref: Deferential)** | | | | | |
| Non-Deferential | 0.194 | 0.0035 | 0.001 | (0.079, 0.310) | ** |
| **Race (Ref: American Indian)** | | | | | |
| Asian | 0.208 | 0.0130 | 0.253 | (0.061, 0.355) | |
| Black/African American | 0.199 | 0.0353 | 0.289 | (-0.035, 0.433) | |
| Hispanic/Latino/Spanish Origin | 0.170 | 0.0420 | 0.407 | (-0.232, 0.573) | |
| Other | 0.210 | 0.0324 | 0.260 | (-0.155, 0.575) | |
| White | 0.206 | 0.0350 | 0.272 | (-0.161, 0.573) | |
| **Gender (Ref: Female)** | | | | | |
| Male | -0.055 | 0.0190 | 0.145 | (-0.279, 0.383) | |
| Other | -0.035 | 0.0144 | 0.768 | (-0.271, 0.200) | |
| **Age (Ref: 18-24)** | | | | | |
| 25 - 34 | -0.098 | 0.0052 | 0.175 | (-0.238, 0.043) | |
| 35 - 44 | -0.164 | 0.0052 | 0.085 | (-0.350, 0.023) | |
| 45 - 54 | -0.185 | 0.0164 | 0.149 | (-0.436, 0.066) | |
| 55 - 64 | 0.011 | 0.0052 | 0.963 | (-0.435, 0.456) | |
| 65 or older | 0.097 | 0.0718 | 0.718 | (-0.429, 0.623) | |
| Under 18 | -0.138 | 0.0595 | 0.570 | (-0.614, 0.338) | |
| **Skin Disease Knowledge (Ref: Less Knowledgeable)** | | | | | |
| More Knowledgeable | 0.034 | 0.0012 | 0.335 | (-0.335, 0.104) | |
| **Year of Medical Experience (Ref: 1-3 y)** | | | | | |
| 10 - 20 y | 0.046 | 0.0072 | 0.591 | (-0.122, 0.213) | |
| 5 - 10 y | 0.033 | 0.0072 | 0.475 | (-0.058, 0.124) | |
| 0 - 5 y | 0.073 | 0.0024 | 0.749 | (-0.054, 0.200) | |
| < 1 y | 0.045 | 0.0032 | 0.436 | (-0.068, 0.157) | |
| > 20 y | -0.115 | 0.0420 | 0.569 | (-0.512, 0.282) | |
| **Interaction Terms** | | | | | |
| Decision Rd. 2 : Deferential Group Non-Deferential | -0.118 | 0.0035 | 0.047 | (-0.234, -0.002) | * |
| **Other Covariates** | | | | | |
| HAI | 0.010 | <0.0001 | 0.038 | (0.001, 0.018) | * |
| XAI | 0.005 | <0.0001 | 0.298 | (-0.005, 0.015) | |
| CRT | 0.008 | <0.0001 | 0.591 | (-0.022, 0.039) | |
| AOT | -0.003 | <0.0001 | 0.482 | (-0.011, 0.005) | |
| | | | | | |
| **Post-hoc Pairwise EMMs Comparisons – Decision Rounds (Deferential vs. Non-Deferential)** | | | | | |
| Rd. 1 (Deferential vs. Non-Deferential) | 0.194 | 0.0035 | < 0.001 | (0.079, 0.310) | *** |
| Rd. 2 (Deferential vs. Non-Deferential) | 0.076 | 0.0035 | 0.196 | (-0.039, 0.192) | |

**STable 30 | Linear Mixed Model Results on Differential Portion Across XAI Explanations among General Public (Target Outcome: Deferential Portion)**

| Variables | β | $\sigma^2$ | p-value | 95% CI for β | Sig. Level |
|---|---|---|---|---|---|
| **Intercept** | 0.310 | 0.0066 | <0.001 | (0.151, 0.469) | *** |
| **Explanation Group (Ref: LLM)** | | | | | |
| Basic AI | -0.071 | 0.0027 | 0.169 | (-0.173, 0.030) | |
| CBIR | -0.040 | 0.0027 | 0.433 | (-0.141, 0.060) | |
| GradCAM | -0.074 | 0.0029 | 0.173 | (-0.181, 0.032) | |
| **Race (Ref: American Indian)** | | | | | |
| Asian | -0.006 | 0.0081 | 0.953 | (-0.201, 0.189) | |
| Black or African American | -0.287 | 0.0090 | 0.003 | (-0.474, -0.100) | ** |
| Hispanic/Latino/Spanish Origin | -0.099 | 0.0104 | 0.333 | (-0.298, 0.101) | |
| Native Hawaiian/Other Pacific Islander | -0.176 | 0.0250 | 0.474 | (-0.437, 0.084) | |
| Other | -0.276 | 0.0384 | 0.005 | (-0.469, -0.084) | ** |
| White | -0.090 | 0.0040 | 0.152 | (-0.213, 0.033) | |
| **Age (Ref: 18-24)** | | | | | |
| 25 - 34 | 0.067 | 0.0026 | 0.192 | (-0.034, 0.167) | |
| 35 - 44 | 0.095 | 0.0032 | 0.094 | (-0.016, 0.207) | |
| 45 - 54 | -0.139 | 0.0102 | 0.170 | (-0.337, 0.059) | |
| 55 - 64 | 0.596 | 0.0154 | <0.001 | (0.352, 0.839) | *** |
| 65 or older | 0.127 | 0.0581 | 0.597 | (-0.345, 0.599) | |
| **Gender (Ref: Female)** | | | | | |
| Male | -0.001 | 0.0014 | 0.985 | (-0.075, 0.074) | |
| Other | 0.269 | 0.0408 | 0.182 | (-0.126, 0.665) | |
| **Skin Disease Experience (Ref: No)** | | | | | |
| Yes | 0.135 | 0.0185 | 0.001 | (0.055, 0.215) | ** |
| **Other Covariates** | | | | | |
| HAI Collaboration Experience | 0.028 | <0.0001 | <0.001 | (0.019, 0.037) | *** |

**STable 31 | Linear Mixed Model Results on Differential Portion Across XAI Explanations among PCPs (Target Outcome: Deferential Portion)**

| Variables | β | $\sigma^2$ | p-value | 95% CI for β | Sig. Level |
|---|---|---|---|---|---|
| **Intercept** | 0.154 | 0.126 | 0.655 | (-0.521, 0.829) | |
| **Explanation Group (Ref: LLM)** | | | | | |
| Basic AI | 0.082 | 0.0055 | 0.268 | (-0.063, 0.228) | |
| CBIR | 0.046 | 0.0071 | 0.587 | (-0.119, 0.210) | |
| GradCAM | -0.053 | 0.0058 | 0.490 | (-0.202, 0.097) | |
| **Race (Ref: American Indian)** | | | | | |
| Asian | 0.939 | 0.0708 | <0.001 | (0.413, 1.465) | *** |
| Black/African American | 0.722 | 0.0784 | 0.010 | (0.174, 1.271) | ** |
| Hispanic/Latino/Spanish Origin | 1.158 | 0.0894 | <0.001 | (0.572, 1.745) | *** |
| Other | 0.851 | 0.0767 | 0.002 | (0.309, 1.394) | ** |
| White | 0.759 | 0.0784 | 0.007 | (0.211, 1.308) | ** |
| **Gender (Ref: Female)** | | | | | |
| Male | 0.110 | 0.0121 | 0.061 | (-0.005, 0.226) | |
| Other | 0.231 | 0.0338 | 0.209 | (-0.129, 0.590) | |
| **Age (Ref: 18-24)** | | | | | |
| 25 - 34 | 0.183 | 0.0357 | 0.092 | (-0.030, 0.396) | |
| 35 - 44 | 0.035 | 0.0216 | 0.813 | (-0.254, 0.324) | |
| 45 - 54 | -0.329 | 0.0384 | 0.094 | (-0.714, 0.056) | |
| 55 - 64 | -0.185 | 0.1260 | 0.601 | (-0.878, 0.508) | |
| 65 or older | -0.284 | 0.1689 | 0.489 | (-1.090, 0.521) | |
| Under 18 | 1.189 | 0.1347 | 0.001 | (0.471, 1.908) | ** |
| **Skin Disease Knowledge (Ref: Less Knowledgeable)** | | | | | |
| More Knowledgeable | -0.082 | 0.0231 | 0.141 | (-0.191, 0.027) | |
| **Year of Medical Experience (Ref: 1-3 y)** | | | | | |
| 10 - 20 y | -0.093 | 0.0174 | 0.482 | (-0.352, 0.166) | |
| 5 - 10 y | -0.084 | 0.0135 | 0.237 | (-0.223, 0.055) | |
| 0 - 5 y | -0.135 | 0.0128 | 0.232 | (-0.357, 0.086) | |
| < 1 y | -0.168 | 0.0350 | 0.054 | (-0.338, 0.003) | |
| > 20 y | 0.537 | 0.0980 | 0.091 | (-0.085, 1.160) | |
| **Other Covariates** | | | | | |
| HAI | 0.005 | <0.0001 | 0.441 | (-0.008, 0.018) | |
| XAT | -0.012 | <0.0001 | 0.152 | (-0.027, 0.004) | |
| CRT | -0.068 | 0.0005 | 0.004 | (-0.113, -0.022) | ** |
| AOT | 0.003 | <0.0001 | 0.595 | (-0.009, 0.015) | |

### 6.7. Finding: Putting AI before human decisions amplifies deference

For the general public in Study 1, **STable 32** examines the effect of HAI paradigm (AI-First vs. Human-First) on diagnostic accuracy. The Human-First paradigm significantly lowered initial (Rd. 1) accuracy compared to the AI-First paradigm (p<0.001), but this performance gap was eliminated after receiving AI assistance (Rd. 2), indicating a significant positive effect of the Human-First paradigm on performance gain (interaction effect: p<0.001). **STable 33** analyzes the diagnostic performance across deferential groups under the AI-First paradigm. The non-deferential group showed a higher trend of accuracy in Round 2 compared to the deferential group (p=0.049). **STable 36** explores the deferential proportion across HAI paradigms and XAI methods. Participants in the AI-First paradigm showed a higher deferential proportion compared to the Human-First paradigm, although the difference was not statistically significant. Several minority race/ethnicity groups (Black or African American, Native Hawaiian/Other Pacific Islander, Other, White) exhibited significantly lower deferential portions compared to the American Indian reference group.

For the PCPs in Study 2, **STable 34** investigates the effect of HAI paradigm on Top-1 diagnostic accuracy. The Human-First paradigm resulted in significantly lower initial (Rd. 1) Top-1 accuracy compared to the AI-First paradigm (p<0.001). However, the Human-First paradigm showed a significantly larger improvement in accuracy after AI assistance (Rd. 2), effectively eliminating the performance gap (interaction effect: p<0.001). **STable 35** analyzes the diagnostic performance across deferential groups under the AI-First paradigm. The non-deferential group maintained a significantly higher accuracy in Round 2 compared to the deferential group (p=0.041). **STable 37** explores the deferential proportion across HAI paradigms and XAI methods. Unlike the general public, no significant difference in deferential proportion was found between the Human-First and AI-First paradigms across XAI groups. However, several race/ethnicity groups (Asian, Hispanic or Latino or Spanish Origin) showed significantly higher deferential portions compared to the American Indian reference group.

**STable 32 | Linear Mixed Model Results on Diagnostic Performance Across HAI Paradigm in General Public (Target Outcome: Accuracy, AI-First vs. Human-First)**

| Variables | β | $\sigma^2$ | p-value | 95% CI for β | Sig. Level |
|---|---|---|---|---|---|
| **Intercept** | 0.737 | 0.0006 | <0.001 | (0.690, 0.783) | *** |
| **Decision (Ref: Rd. 1 without AI)** | | | | | |
| Rd. 2 with AI Assistance | 0.002 | <0.0001 | 0.699 | (-0.009, 0.013) | |
| **HAI Paradigm (Ref: AI First)** | | | | | |
| Human First | -0.054 | 0.0001 | <0.001 | (-0.076, -0.033) | *** |
| **Race (Ref: American Indian)** | | | | | |
| Asian | 0.053 | 0.0008 | 0.006 | (-0.002, 0.107) | |
| Black/African American | 0.004 | 0.0007 | 0.867 | (-0.046, 0.055) | |
| Hispanic/Latino/Spanish Origin | 0.042 | 0.0010 | 0.174 | (-0.039, 0.122) | |
| Native Hawaiian/Other Pacific Islander | -0.019 | 0.0014 | 0.620 | (-0.094, 0.056) | |
| Other | 0.037 | 0.0014 | 0.174 | (-0.016, 0.090) | |
| White | 0.030 | 0.0004 | 0.106 | (-0.006, 0.067) | |
| **Age (Ref: 18-24)** | | | | | |
| 25 - 34 | -0.011 | 0.0002 | 0.460 | (-0.042, 0.019) | |
| 35 - 44 | -0.011 | 0.0003 | 0.477 | (-0.042, 0.020) | |
| 45 - 54 | 0.044 | 0.0007 | 0.090 | (-0.007, 0.096) | |
| 55 - 64 | 0.035 | 0.0016 | 0.379 | (-0.043, 0.112) | |
| 65 or older | -0.031 | 0.0039 | 0.637 | (-0.158, 0.097) | |
| **Gender (Ref: Female)** | | | | | |
| Male | -0.034 | 0.0006 | 0.001 | (-0.054, -0.013) | ** |
| Other | 0.019 | 0.0042 | 0.772 | (-0.109, 0.146) | |
| **Skin Disease Experience (Ref: No)** | | | | | |
| Yes | -0.008 | 0.0001 | 0.446 | (-0.030, 0.013) | |
| **Interaction Terms** | | | | | |
| Decision Rd. 2 : HAI Paradigm Human First | 0.059 | <0.0001 | <0.001 | (0.045, 0.074) | *** |
| **Other Covariates** | | | | | |
| HAI Collaboration Experience | 0.005 | <0.0001 | < 0.001 | (0.002, 0.007) | *** |
| Decision Time (in minute) | 0.001 | <0.0001 | 0.909 | (-0.014, 0.016) | |
| | | | | | |
| **Post-hoc Pairwise EMMs Comparisons – Decision Round (Human-First vs. AI-First)** | | | | | |
| Round 1 (Human-First vs. AI-First) | -0.054 | 0.0001 | <0.001 | (-0.076, -0.033) | *** |
| Round 2 (Human-First vs. AI-First) | 0.005 | 0.0001 | 0.650 | (-0.017, 0.026) | |

**STable 33 | Linear Mixed Model Results on Diagnostic Performance Across Deferential Group in General Public (Target Outcome: Accuracy, AI First)**

| Variables | β | $\sigma^2$ | p-value | 95% CI for β | Sig. Level |
|---|---|---|---|---|---|
| **Intercept** | 0.761 | 0.0014 | <0.001 | (0.687, 0.835) | *** |
| **Deferential Group (Ref: Deferential)** | | | | | |
| Non-Deferential | 0.015 | 0.0003 | 0.335 | (-0.016, 0.046) | |
| **Decision (Ref: Rd. 1 without AI)** | | | | | |
| Rd. 2 with AI Assistance | -0.006 | 0.0000 | 0.214 | (-0.015, 0.003) | |
| **Gender (Ref: Female)** | | | | | |
| Male | -0.027 | 0.0003 | 0.082 | (-0.058, 0.004) | |
| Other | 0.123 | 0.0177 | 0.356 | (-0.139, 0.385) | |
| **Race (Ref: American Indian)** | | | | | |
| Asian | 0.004 | 0.0019 | 0.933 | (-0.083, 0.090) | |
| Black/African American | -0.036 | 0.0014 | 0.354 | (-0.113, 0.040) | |
| Hispanic/Latino/Spanish Origin | -0.033 | 0.0027 | 0.528 | (-0.134, 0.069) | |
| Native Hawaiian/Other Pacific Islander | -0.070 | 0.0025 | 0.184 | (-0.174, 0.033) | |
| Other | 0.013 | 0.0018 | 0.760 | (-0.060, 0.085) | |
| White | 0.008 | 0.0011 | 0.811 | (-0.054, 0.069) | |
| **Age (Ref: 18-24)** | | | | | |
| 25 - 34 | -0.015 | 0.0005 | 0.514 | (-0.059, 0.029) | |
| 35 - 44 | -0.012 | 0.0005 | 0.596 | (-0.058, 0.033) | |
| 45 - 54 | 0.032 | 0.0014 | 0.400 | (-0.042, 0.105) | |
| 55 - 64 | 0.028 | 0.0059 | 0.713 | (-0.123, 0.180) | |
| 65 or older | -0.117 | 0.0151 | 0.211 | (-0.300, 0.066) | |
| **Skin Disease Experience (Ref: No)** | | | | | |
| Yes | -0.017 | 0.0003 | 0.276 | (-0.049, 0.014) | |
| **Interaction Terms** | | | | | |
| Deferential Group Non-Deferential : Decision Rd. 2 | 0.016 | 0.0000 | 0.012 | (0.003, 0.028) | * |
| **Other Covariates** | | | | | |
| HAI Collaboration Experience | 0.004 | 0.0000 | 0.015 | (0.001, 0.008) | * |
| Decision Time (in minute) | 0.000 | 0.0000 | 0.990 | (-0.000, 0.001) | |
| | | | | | |
| **Post-hoc Pairwise EMMs Comparisons – Decision Round (Non-Deferential vs. Deferential)** | | | | | |
| Round 1 (Non-Deferential vs. Deferential) | 0.015 | 0.0002 | 0.335 | (-0.016, 0.046) | |
| Round 2 (Non-Deferential vs. Deferential) | 0.031 | 0.0002 | 0.049 | (0.000, 0.062) | * |

**STable 34 | Linear Mixed Model Results on Diagnostic Performance Across HAI Paradigm in PCP (Target Outcome: Top-1 Accuracy, AI-First vs. Human-First)**

| Variables | β | $\sigma^2$ | p-value | 95% CI for β | Sig. Level |
|---|---|---|---|---|---|
| **Intercept** | 0.389 | 0.0384 | 0.047 | (0.005, 0.774) | * |
| **Decision (Ref: Rd. 1 without AI)** | | | | | |
| Rd. 2 with AI Assistance | 0.023 | 0.0005 | 0.468 | (-0.038, 0.084) | |
| **HAI Paradigm (Ref: AI First)** | | | | | |
| Human First | -0.226 | 0.0013 | <0.001 | (-0.297, -0.154) | *** |
| **Race (Ref: American Indian)** | | | | | |
| Asian | -0.150 | 0.0210 | 0.302 | (-0.435, 0.135) | |
| Black/African American | -0.083 | 0.0231 | 0.586 | (-0.380, 0.215) | |
| Hispanic/Latino/Spanish Origin | -0.098 | 0.0266 | 0.550 | (-0.418, 0.223) | |
| Other | -0.125 | 0.0237 | 0.403 | (-0.419, 0.169) | |
| White | -0.117 | 0.0222 | 0.433 | (-0.409, 0.175) | |
| **Gender (Ref: Female)** | | | | | |
| Male | -0.013 | 0.0014 | 0.695 | (-0.080, 0.054) | |
| Other | 0.019 | 0.0135 | 0.874 | (-0.210, 0.247) | |
| **Age (Ref: 18-24)** | | | | | |
| 25 - 34 | -0.046 | 0.0053 | 0.533 | (-0.190, 0.098) | |
| 35 - 44 | -0.083 | 0.0086 | 0.376 | (-0.266, 0.101) | |
| 45 - 54 | -0.061 | 0.0135 | 0.603 | (-0.289, 0.168) | |
| 55 - 64 | 0.081 | 0.0313 | 0.646 | (-0.260, 0.428) | |
| 65 or older | 0.076 | 0.0484 | 0.719 | (-0.302, 0.492) | |
| Under 18 | -0.440 | 0.0590 | 0.070 | (-0.917, 0.037) | |
| **Skin Disease Knowledge (Ref: Less Knowledgeable)** | | | | | |
| More Knowledgeable | 0.052 | 0.0011 | 0.109 | (-0.012, 0.117) | |
| **Year of Medical Experience (Ref: 1-3 y)** | | | | | |
| 10 - 20 y | 0.031 | 0.0067 | 0.702 | (-0.129, 0.191) | |
| 5 - 10 y | 0.011 | 0.0018 | 0.791 | (-0.073, 0.096) | |
| 0 - 5 y | 0.016 | 0.0039 | 0.790 | (-0.104, 0.137) | |
| < 1 y | 0.062 | 0.0031 | 0.222 | (-0.038, 0.162) | |
| > 20 y | -0.065 | 0.0266 | 0.688 | (-0.385, 0.254) | |
| **Interaction Terms** | | | | | |
| Decision Rd. 2 : HAI Paradigm Human First | 0.205 | 0.0004 | <0.001 | (0.145, 0.266) | *** |
| **Other Covariates** | | | | | |
| HAI | 0.010 | <0.0001 | 0.010 | (0.002, 0.018) | * |
| XAI | 0.004 | <0.0001 | 0.330 | (-0.004, 0.013) | |
| CRT | 0.012 | 0.0002 | 0.430 | (-0.017, 0.040) | |
| AOT | -0.001 | <0.0001 | 0.871 | (-0.008, 0.007) | |
| Decision Time (in minute) | <0.001 | <0.0001 | 0.303 | (-0.000, 0.001) | |
| | | | | | |
| **Post-hoc Pairwise EMMs Comparisons – Decision Round (Human-First vs. AI-First)** | | | | | |
| Round 1 (Human-First vs. AI-First) | -0.226 | 0.0013 | <0.001 | (-0.297, -0.154) | *** |
| Round 2 (Human-First vs. AI-First) | -0.020 | 0.0013 | 0.572 | (-0.091, 0.050) | |

**STable 35 | Linear Mixed Model Results on Diagnostic Performance Across Deferential Group in PCP (Target Outcome: Top-1 Accuracy, AI First)**

| **Variables** | **β** | **$\sigma^2$** | **p-value** | **95% CI for β** | **Sig. Level** |
|---|---|---|---|---|---|
| **Intercept** | 0.512 | 0.3950 | 0.194 | (-0.261, 1.286) | |
| **Deferential Group (Ref: Deferential)** | | | | | |
| Non-Deferential | 0.206 | 0.1060 | 0.053 | (-0.002, 0.415) | |
| **Decision (Ref: Rd. 1 without AI)** | | | | | |
| Rd. 2 with AI Assistance | -0.007 | 0.0020 | 0.001 | (-0.011,-0.003) | ** |
| **Race (Ref: American Indian)** | | | | | |
| Asian | -0.503 | 0.2830 | 0.075 | (-1.058, 0.051) | |
| Black/African American | -0.307 | 0.3060 | 0.316 | (-0.907, 0.293) | |
| Hispanic/Latino/Spanish Origin | -0.211 | 0.3200 | 0.510 | (-0.838, 0.416) | |
| Other | -0.456 | 0.2910 | 0.117 | (-1.026, 0.114) | |
| White | -0.424 | 0.2830 | 0.134 | (-0.978, 0.131) | |
| **Gender (Ref: Female)** | | | | | |
| Male | 0.032 | 0.0790 | 0.691 | (-0.124, 0.187) | |
| Other | 0.223 | 0.2700 | 0.410 | (-0.306, 0.751) | |
| **Age (Ref: 18-24)** | | | | | |
| 25 - 34 | 0.013 | 0.2010 | 0.948 | (-0.381, 0.407) | |
| 35 - 44 | -0.014 | 0.2360 | 0.952 | (-0.476, 0.448) | |
| 45 - 54 | 0.090 | 0.2990 | 0.765 | (-0.497, 0.676) | |
| 55 - 64 | 0.006 | 0.3980 | 0.987 | (-0.774, 0.787) | |
| 65 or older | -0.002 | 0.4470 | 0.997 | (-0.877, 0.874) | |
| **Skin Disease Knowledge (Ref: Less Knowledgeable)** | | | | | |
| More Knowledgeable | 0.053 | 0.0710 | 0.456 | (-0.086, 0.192) | |
| **Year of Medical Experience (Ref: 1-3 y)** | | | | | |
| 10 - 20 y | 0.100 | 0.2500 | 0.690 | (-0.390, 0.589) | |
| 5 - 10 y | -0.080 | 0.0960 | 0.403 | (-0.269, 0.108) | |
| 0 - 5 y | -0.022 | 0.1270 | 0.861 | (-0.270, 0.226) | |
| < 1 y | 0.009 | 0.1220 | 0.940 | (-0.229, 0.248) | |
| > 20 y | 0.094 | 0.3590 | 0.793 | (-0.609, 0.797) | |
| **Interaction Terms** | | | | | |
| Deferential Group Non-Deferential : Decision Rd. 2 | 0.012 | 0.0040 | 0.009 | ( 0.003, 0.020) | ** |
| **Other Covariates** | | | | | |
| HAI | 0.008 | 0.0090 | 0.383 | (-0.009, 0.025) | |
| XAI | 0.005 | 0.0100 | 0.576 | (-0.014, 0.024) | |
| CRT | 0.018 | 0.0370 | 0.634 | (-0.055, 0.090) | |
| AOT | 0.006 | 0.0090 | 0.491 | (-0.011, 0.023) | |
| Decision Time (in minute) | -0.006 | <0.0001 | <0.001 | (-0.008, -0.003) | *** |
| | | | | | |
| **Post-hoc Pairwise EMMs Comparisons – Decision Round (Non-Deferential vs. Deferential)** | | | | | |
| Round 1 (Non-Deferential vs. Deferential) | 0.206 | 0.0114 | 0.053 | (-0.002, 0.415) | |
| Round 2 (Non-Deferential vs. Deferential) | 0.218 | 0.0114 | 0.041 | (0.009, 0.426) | * |

**STable 36 | Linear Mixed Model Results on Deferential Proportion Across HAI Paradigms and XAI Methods in General Public (Target Outcome: Deferential Proportion)**

| Variables | β | $\sigma^2$ | p-value | 95% CI for β | Sig. Level |
|---|---|---|---|---|---|
| **Intercept** | 0.485 | 0.0053 | <0.001 | (0.341, 0.629) | *** |
| **HAI Paradigm (Ref: AI First)** | | | | | |
| Human First | -0.097 | 0.0034 | 0.170 | (-0.210, 0.017) | |
| **Explanation Group (Ref: LLM)** | | | | | |
| Basic AI | -0.048 | 0.0035 | 0.418 | (-0.163, 0.067) | |
| CBIR | -0.033 | 0.0031 | 0.551 | (-0.143, 0.076) | |
| GradCAM | -0.063 | 0.0031 | 0.264 | (-0.173, 0.047) | |
| **Race (Ref: American Indian)** | | | | | |
| Asian | -0.124 | 0.0154 | 0.102 | (-0.274, 0.025) | |
| Black or African American | -0.155 | 0.0112 | 0.027 | (-0.292, -0.017) | * |
| Hispanic or Latino or Spanish Origin | -0.111 | 0.0117 | 0.181 | (-0.274, 0.052) | |
| Native Hawaiian/Other Pacific Islander | -0.262 | 0.0270 | 0.012 | (-0.466, -0.057) | * |
| Other | -0.255 | 0.0106 | 0.001 | (-0.399, -0.111) | ** |
| White | -0.103 | 0.0026 | 0.043 | (-0.203, -0.003) | * |
| **Age (Ref: 18-24)** | | | | | |
| 25 - 34 | 0.077 | 0.0014 | 0.051 | (-0.000, 0.153) | |
| 35 - 44 | 0.104 | 0.0017 | 0.014 | (0.018, 0.187) | * |
| 45 - 54 | -0.053 | 0.0050 | 0.451 | (-0.192, 0.085) | |
| 55 - 64 | 0.575 | 0.0115 | <0.001 | (0.365, 0.785) | *** |
| 65 or older | -0.177 | 0.0310 | 0.314 | (-0.521, 0.167) | |
| **Gender (Ref: Female)** | | | | | |
| Male | -0.042 | 0.0006 | 0.142 | (-0.098, 0.014) | |
| Other | 0.036 | 0.0276 | 0.839 | (-0.310, 0.382) | |
| **Skin Disease Knowledge (Ref: No)** | | | | | |
| Yes | 0.047 | 0.0009 | 0.115 | (-0.011, 0.105) | |
| **Other Covariates** | | | | | |
| HAI | 0.020 | <0.0001 | <0.001 | (0.014, 0.027) | *** |
| Decision Time (in minute) | <0.001 | <0.0001 | 1.000 | (-0.000, 0.000) | |
| **Interaction Terms** | | | | | |
| HAI Paradigm Human First : Explanation Group Basic | -0.023 | 0.0062 | 0.769 | (-0.178, 0.131) | |
| HAI Paradigm Human First : Explanation Group CBIR | -0.025 | 0.0061 | 0.744 | (-0.173, 0.124) | |
| HAI Paradigm Human First : Explanation Group GradCAM | -0.078 | 0.0061 | 0.749 | (-0.178, 0.128) | |
| | | | | | |

| Post-hoc Pairwise EMMs Comparisons – Explanation Group (Human First vs. AI First) | | | | | |
|---|---|---|---|---|---|
| Basic AI (Human First vs. AI First) | -0.097 | 0.0034 | 0.096 | (-0.210, 0.171) | |
| CBIR (Human First vs. AI First) | -0.098 | 0.0029 | 0.067 | (-0.203, 0.007) | |
| GradCAM (Human First vs. AI First) | -0.099 | 0.0032 | 0.084 | (-0.210, 0.013) | |
| LLM (Human First vs. AI First) | -0.074 | 0.0029 | 0.170 | (-0.179, 0.032) | |

**STable 37 | Linear Mixed Model Results on Deferential Proportion Across HAI Paradigms and XAI Methods in PCPs (Target Outcome: Deferential Proportion)**

| Variables | β | $\sigma^2$ | p-value | 95% CI for β | Sig. Level |
|---|---|---|---|---|---|
| **Intercept** | 0.644 | 0.0713 | 0.016 | (0.121, 1.166) | * |
| **HAI Paradigm (Ref: AI First)** | | | | | |
| Human First | -0.121 | 0.0071 | 0.150 | (-0.286, 0.044) | |
| **Explanation Group (Ref: LLM)** | | | | | |
| Basic AI | 0.013 | 0.0083 | 0.895 | (-0.174, 0.199) | |
| CBIR | 0.069 | 0.0125 | 0.547 | (-0.155, 0.294) | |
| GradCAM | 0.014 | 0.0061 | 0.878 | (-0.166, 0.194) | |
| **Race (Ref: American Indian)** | | | | | |
| Asian | 0.412 | 0.0380 | 0.035 | (0.030, 0.794) | * |
| Black or African American | 0.248 | 0.0416 | 0.218 | (-0.147, 0.644) | |
| Hispanic or Latino or Spanish Origin | 0.537 | 0.0475 | 0.014 | (0.111, 0.964) | * |
| Other | 0.324 | 0.0488 | 0.107 | (-0.070, 0.718) | |
| White | 0.360 | 0.0392 | 0.069 | (-0.029, 0.748) | |
| **Age (Ref: 18-24)** | | | | | |
| 25 - 34 | 0.052 | 0.0077 | 0.594 | (-0.140, 0.245) | |
| 35 - 44 | -0.005 | 0.0156 | 0.970 | (-0.249, 0.240) | |
| 45 - 54 | -0.144 | 0.0237 | 0.349 | (-0.446, 0.158) | |
| 55 - 64 | -0.073 | 0.0571 | 0.759 | (-0.539, 0.393) | |
| 65 or older | 0.168 | 0.0718 | 0.558 | (-0.395, 0.731) | |
| **Gender (Ref: Female)** | | | | | |
| Male | -0.006 | 0.0023 | 0.816 | (-0.100, 0.087) | |
| Other | 0.193 | 0.0243 | 0.217 | (-0.130, 0.517) | |
| **Skin Disease Knowledge (Ref: Less Knowledgeable)** | | | | | |
| More Knowledgeable | -0.038 | 0.0031 | 0.384 | (-0.125, 0.048) | |
| **Other Covariates** | | | | | |
| HAI | -0.001 | <0.0001 | 0.297 | (-0.005, 0.016) | |
| XAI | 0.006 | <0.0001 | 0.853 | (-0.013, 0.011) | |
| CRT | -0.039 | 0.0004 | 0.049 | (-0.077, -0.000) | * |
| AOT | 0.003 | <0.0001 | 0.574 | (-0.007, 0.012) | |
| Decision Time (in minute) | <0.001 | <0.0001 | 1.000 | (0.000, 0.000) | |
| **Interaction Terms** | | | | | |
| HAI Paradigm Human First : Explanation Group Basic | -0.023 | 0.0146 | 0.306 | (-0.113, 0.360) | |
| HAI Paradigm Human First : Explanation Group CBIR | 0.031 | 0.0146 | 0.824 | (-0.244, 0.306) | |
| HAI Paradigm Human First : Explanation Group GradCAM | 0.022 | 0.0142 | 0.850 | (-0.202, 0.245) | |
| | | | | | |
| **Post-hoc Pairwise EMMs Comparisons – Explanation Group (Human First vs. AI First)** | | | | | |
| Basic AI (Human First vs. AI First) | 0.002 | 0.0076 | 0.978 | (-0.168, 0.173) | |
| CBIR (Human First vs. AI First) | -0.090 | 0.0121 | 0.412 | (-0.305, 0.125) | |
| GradCAM (Human First vs. AI First) | -0.099 | 0.0060 | 0.198 | (-0.251, 0.052) | |
| LLM (Human First vs. AI First) | -0.121 | 0.0071 | 0.150 | (-0.286, 0.044) | |